\documentclass[a4paper,11pt]{article}
\pdfoutput=1 
\usepackage{jheppub} 
\usepackage[numbers]{natbib}
\usepackage[T1]{fontenc} 
\usepackage[tight]{subfigure}
\usepackage{hyperref}
\usepackage{bbold}
\usepackage{float}
\usepackage{feynman}
\usepackage{yhmath}
\usepackage{xcolor}
\usepackage{adjustbox}
\usepackage{blindtext}
\usepackage{multirow}
\usepackage{multicol}
\usepackage{verbatim}
\usepackage{amsmath}
\usepackage{relsize}
\usepackage{bm, nicefrac}
\usepackage{cleveref}
\usepackage[normalem]{ulem}
\usepackage{booktabs}
\usepackage{array}

\newcolumntype{R}[1]{>{\raggedleft\arraybackslash}p{#1}}
\newcolumntype{L}[1]{>{\raggedright\arraybackslash}p{#1}}
\newcolumntype{C}[1]{>{\centering\arraybackslash}p{#1}}

\newcommand{\beq}{\begin{equation}}
\newcommand{\eeq}{\end{equation}}
\newcommand{\beqn}{\begin{eqnarray}}
\newcommand{\eeqn}{\end{eqnarray}}

\newcommand{\madgraphnlos}{{\sc\small MG5\_aMC@NLO}\xspace}
\newcommand{\mocanlo}{{\sc\small MoCaNLO}\xspace}
\newcommand{\powheg}{{\sc\small PowHeg-Box-Res}\xspace}

\newcommand{\Pp}{\ensuremath{\text{p}}}
\newcommand{\Pe}{\ensuremath{\text{e}}\xspace}

\newcommand{\Pt}{\ensuremath{\text{t}}\xspace}

\newcommand{\PW}{\ensuremath{\text{W}}\xspace}
\newcommand{\PZ}{\ensuremath{\text{Z}}\xspace}

\newcommand{\GeV}{\ensuremath{\,\text{GeV}}\xspace}
\newcommand{\Mt}{\ensuremath{m_\Pt}\xspace}

\newcommand{\MW}{\ensuremath{M_\PW}\xspace}

\newcommand{\MZ}{\ensuremath{M_\PZ}\xspace}
\newcommand{\GZ}{\ensuremath{\Gamma_\PZ}\xspace}

\newcommand{\GW}{\ensuremath{\Gamma_\PW}\xspace}

\newcommand{\GF}{\ensuremath{G_\mu}}
\newcommand{\alphas}{\ensuremath{\alpha_\text{s}}\xspace}

\newcommand{\rd}{\mathrm d}
\newcommand{\mc}{\mathcal}

\title{Triple-gauge couplings in LHC diboson production: a SMEFT view from every angle}
\author[a]{Hesham El Faham,}
\author[b]{Giovanni Pelliccioli,}
\author[a]{Eleni Vryonidou\,}

\affiliation[a]{Department of Physics and Astronomy, University of Manchester, Oxford Road, Manchester M13~9PL, United Kingdom}
\affiliation[b]{Max-Planck-Institut f\"ur Physik, Boltzmannstra{\ss}e 8, 85748 Garching, Germany}

\emailAdd{hesham.elfaham@manchester.ac.uk}
\emailAdd{gpellicc@mpp.mpg.de}
\emailAdd{eleni.vryonidou@manchester.ac.uk}

\preprint{
\begin{flushright}
{COMETA-2024-08}\\
{MPP-2024-102}
\end{flushright}
}

\abstract{This study explores fully leptonic WZ and WW production at the LHC within the SMEFT framework at NLO in QCD, focusing on both CP-even and CP-odd triple-gauge-coupling dimension-six operators. We investigate the off-shell processes, contrasting our findings in inclusive setups with those in the presence of realistic fiducial selections. Alongside the conventional kinematic observables, we examine polarisation-sensitive observables and angular coefficients. Moreover, we assess potential SMEFT effects on asymmetry observables. Through a sensitivity analysis, we identify critical LHC observables that are particularly sensitive to SMEFT-induced modifications, thereby shedding light on potential avenues for new physics searches in diboson production at the LHC.}
\begin{document}
\maketitle

\section{Introduction}\label{sec:intro}
In the absence of experimental evidence pointing at the existence of new light degrees of freedom, a new era of searches for indirect signs of New Physics through a campaign of precision measurements is underway. The Large Hadron Collider (LHC) offers a great testing ground, as the LHC experiments perform a plethora of increasingly precise and differential measurements. 

The Standard Model Effective Field Theory (SMEFT) offers a model-independent framework to parametrise deviations from the Standard Model (SM) and interpret experimental data to reveal signs of New Physics. The SMEFT introduces a tower of higher-dimensional operators modifying the interactions of SM particles. These higher-dimension operators manifest themselves by modifying the rates and differential distributions of various physical observables at the LHC. 

A class of particularly interesting processes in the search for deviations from the SM is electroweak (EW) boson pair production. Observed for the first time in high-energy colliders in lepton collisions at LEP, they offer a window of opportunity to understand the nature of EW symmetry breaking.
The luminosities that will become available with the Run-3 and High-Lumi LHC stages render diboson production one of the most promising channels for precise and highly differential measurements. In the fully leptonic decay channel, gauge-boson pair production offers especially clean signatures. 
The production of $\PW^\pm\PZ$ and $\PW^+\PW^-$ pairs represents the simplest class of processes providing access to the self-interaction of gauge bosons, as it is sensitive to the triple-gauge coupling. 

The importance of these processes in the LHC precision programme has motivated much theoretical progress regarding precision computations within the SM. The SM off-shell modelling at fixed order is known up to next-to-next-to-leading order (NNLO) QCD \cite{Grazzini:2016ctr,Grazzini:2017ckn} and next-to-leading order (NLO) EW corrections \cite{Biedermann:2016guo,Biedermann:2016yvs,Biedermann:2017oae, Grazzini:2019jkl}. The matching to parton-shower (PS) and hadronisation effects has been achieved, including both NNLO QCD \cite{Lombardi:2021rvg,Lindert:2022qdd,Gavardi:2023aco} and NLO EW corrections \cite{Chiesa:2020ttl,Lindert:2022qdd}. Transverse-momentum resummation \cite{Grazzini:2015wpa, Kallweit:2020gva,Campbell:2022uzw}, as well as jet-veto resummation \cite{Dawson:2016ysj,Campbell:2023cha,Gavardi:2023aco} in QCD are also known to high accuracy. Predictions for intermediate polarised bosons in the pole \cite{Stuart:1991xk,Aeppli:1993cb,Aeppli:1993rs,Denner:2000bj} or narrow-width approximation \cite{Richardson:2001df,Uhlemann:2008pm,Artoisenet:2012st} are also known up to (N)NLO in QCD \cite{Denner:2020bcz,Denner:2020eck,Poncelet:2021jmj} and NLO EW \cite{Le:2022lrp,Le:2022ppa,Denner:2023ehn,Dao:2023pkl} accuracy, including also NLO matching to PS \cite{Hoppe:2023uux,Pelliccioli:2023zpd}.

The combination of the high degree of precision in the SM computations and the precise measurements available offer an excellent opportunity for SMEFT interpretations of the diboson processes. Several studies exist where diboson production is explored with the SMEFT 
\cite{Falkowski:2015jaa,Falkowski:2016cxu,Helset:2017mlf,Baglio:2017bfe,Azatov:2017kzw,Franceschini:2017xkh,Chiesa:2018lcs,Baglio:2018bkm,Liu:2018pkg,Grojean:2018dqj,Azatov:2019xxn,Baglio:2019uty,Baglio:2020oqu,Ellis:2020unq,Degrande:2021zpv,Degrande:2023iob,Aoude:2023hxv,Degrande:2024bmd}. These studies highlight not only the prospects of probing triple gauge couplings in diboson processes but also the possibility of extracting information on the couplings of fermions to EW gauge bosons, complementary to that obtained by considering EW precision observables (EWPOs). 

An interesting observation from SMEFT studies of diboson production is the suppressed interference between the amplitudes for the SM and the new TGC interaction arising from 
\begin{equation}
   \epsilon_{ijk} W_{\mu\nu}^{i} W^{j,\nu\rho} W^{k,\mu}_{\rho},
   \end{equation}
due to helicity selection rules \cite{Azatov:2016sqh}, as different helicity configurations dominate the SM and SMEFT LO amplitudes. This non-interference is evident in inclusive observables and has, in turn, motivated several phenomenological studies exploring {\it{interference-resurrection}} observables. A partial restoration of SMEFT--SM interference at dimension-six in diboson production can be achieved with the differential description of chirality-sensitive azimuthal angles associated with the decay products \cite{Azatov:2017kzw,Panico:2017frx,Franceschini:2017xkh,Azatov:2019xxn}. An enhancement of the dimension-six interference can also be fulfilled when considering the production of a boson pair in association with QCD jets \cite{Azatov:2017kzw} or computing NLO QCD corrections to inclusive boson-pair production \cite{Panico:2017frx,Azatov:2019xxn,Franceschini:2017xkh}. This is possible as the $2 \to 3$ amplitudes do not exhibit the same helicity selection rules. It has also been noted that fiducial cuts, as typically applied to the leptons originating from the gauge boson decays, also resurrect the interference \cite{Azatov:2019xxn}.
   A partial enhancement of the interference is also given by the off-shell modelling of weak-boson production and decay, compared to the
on-shell modelling and the narrow-width approximation \cite{Helset:2017mlf}.

Given that chirality-sensitive observables resurrect the interference, investigating the polarisation of intermediate EW bosons
could unveil further effects coming from the \sloppy dimension-six interference as well. Achieving this task relies either on direct Monte Carlo (MC) simulations of separately polarised signals (for WZ production see for example refs.~\cite{Denner:2020eck,Le:2022ppa,Le:2022lrp,Dao:2023pkl,Hoppe:2023uux,Pelliccioli:2023zpd}), or projecting decay-angle distributions onto suitable combinations of rank-2 spherical harmonics to extract polarisation and spin-correlation coefficients. The latter strategy is known to be reliable only in the absence of fiducial cuts on the decay products \cite{Bern:2011ie,Stirling:2012zt,Belyaev:2013nla} and of radiative corrections that may distort the two-body structure of the LO decay of EW bosons \cite{Denner:2020eck,Baglio:2018rcu}.

In this work we aim to fully explore the impact of new CP-even and CP-odd triple gauge interactions, as predicted within the SMEFT in WZ and WW production. We focus on different observables, inclusive and in the presence of fiducial cuts, to establish the magnitude of the interference between the SM and EFT in each case. Our predictions include NLO corrections and off-shell effects to compute the EFT contributions reliably. We consider both standard observables, such as invariant masses and transverse momenta, as well as angular observables, which are sensitive to the polarisation of the EW gauge bosons. We also compute the impact of new interactions on asymmetries defined for either the gauge bosons or their leptonic decay products. These are also sensitive to the polarisation of the gauge bosons whilst benefiting from reduced systematic uncertainties due to their definition as ratios of cross-sections. We aim to determine which classes of observables offer the best sensitivity to new interactions by performing a fit using existing measurements from ATLAS and CMS. We also explore the impact of NLO QCD corrections and quadratic EFT contributions on the constraints set on the Wilson coefficients (WCs). 

This paper is organised as follows. In \Cref{sec:setup}, we define our setup, whilst in \Cref{sec:WZ,sec:WW}, we show our results for WZ and WW production, respectively. We study the boost asymmetries in \Cref{sec:asym}. In \Cref{sec:sens_study}, we present our sensitivity study, and in \Cref{sec:conclusions}, we draw our conclusions. 

\section{Theoretical framework and computational setup}
\label{sec:setup}
The dimension-six effective Lagrangian relevant for diboson production, \(\mathcal{L}_{VV}\), parametrising modifications to the SM vertices, can be schematically expressed as,
\begin{equation}
    \mathcal{L}_{VV} \supset \mathcal{L}_{3V} + \mathcal{L}_{Vq\bar{q}}\,.
\end{equation}
The first term on the right-hand side encapsulates interactions involving three EW gauge bosons (\(V=\PW,\PZ,\gamma\)). In contrast, the second term addresses interactions between an EW gauge boson and a quark-antiquark pair (\(q\bar{q}\)). The \(\mathcal{L}_{3V}\) component can be defined as:
\begin{align}
\mathcal{L}_{3V} &= \,\,\,\,i e \,\left(W^{+}_{\mu\nu} W^-_\mu - W^{-}_{\mu\nu} W^+_\mu\right) A_{\nu} \nonumber\\
&\quad + i e \left[\left(1 + \delta \kappa_\gamma\right) A_{\mu\nu} W^+_\mu W^{-}_{\nu} + \tilde{\kappa}_\gamma \tilde{A}_{\mu\nu} W^+_\mu W^{-}_{\nu}\right] \nonumber \\
&\quad + i g c_\theta \left[ (1 + \delta g_{1,z}) (W^{+}_{\mu\nu} W^{-}_{\mu} - W^{-}_{\mu\nu} W^+_\mu) Z_\nu \right. \nonumber \\ 
&\quad \left. + (1 + \delta \kappa_z) Z_{\mu\nu} W^+_\mu W^{-}_\nu + \tilde{\kappa}_z \tilde{Z}_{\mu\nu} W^+_\mu W^{-}_{\nu} \right] \nonumber \\
&\quad + \frac{i e}{2 \MW^2} \left[ \lambda_\gamma W^{+}_{\mu\nu} W^{-}_{\nu\rho} A_{\rho\mu} + \tilde{\lambda}_\gamma W^{+}_{\mu\nu} W^{-}_{\nu\rho} \tilde{A}_{\rho\mu} \right] \nonumber \\
&\quad + \frac{i g c_\theta}{2 \MW^2} \left[ \lambda_zW^{+}_{\mu\nu} W^{-}_{\nu\rho} Z_{\rho\mu} + \tilde{\lambda}_z W^{+}_{\mu\nu} W^{-}_{\nu\rho} \tilde{Z}_{\rho\mu} \right]\,,
\end{align}
where \(\delta g_{1,z}\), the CP-conserving \(\delta \kappa_{z/\gamma}\), and \(\lambda_{z/\gamma}\), along with their CP-violating counterparts (\(x \to \tilde{x}\)), represent the TGCs. While sensitivity to \(Vq\bar{q}\) interactions also arises through the EWPOs measured at LEP and SLD \cite{Altarelli:1991fk, ALEPH:2005ab}, \(\mathcal{L}_{3V}\) is specifically probed through diboson production. Although LEP provided some constraints, LHC diboson measurements are crucial in precisely determining the parameters entering \(\mathcal{L}_{3V}\).

Before focusing on the impact of LHC diboson measurements on probing TGCs, it is crucial to stress that diboson processes can offer sensitivity to \(Vq\bar{q}\) corrections, as extensively discussed in Ref.~\cite{Grojean:2018dqj}. This study
finds that LHC data can constrain modifications to \(Vq\bar{q}\) couplings beyond the bounds set by LEP-1 data. The study also assesses the impact of non-vanishing \(Vq\bar{q}\) corrections on determining TGCs. Under the Minimal Flavour Violation (MFV) and Flavour Universality (FU) assumptions, the stability of a TGC fit is not guaranteed when profiling over \(Vq\bar{q}\) corrections. Such non-trivial correlations have been shown to affect \(g_{1,z}\) and \(\kappa_{z}\), while \(\lambda_z\) remains independent of the inclusion of \(Vq\bar{q}\) corrections. For this reason, we conservatively focus on the TGC modifications directly related to \(\lambda_z\) and \(\tilde{\lambda}_z\). These are respectively mapped to SMEFT in the Warsaw basis \cite{Grzadkowski:2010es} through the WCs \(c_{W}\) and \(c_{\tilde{W}}\) as follows:
\begin{equation}\label{couplings}
    \lambda_{z} = -c_{W} \frac{v}{\Lambda^2} \frac{3}{2} g, \hspace{1cm} \tilde{\lambda}_{z} = -c_{\tilde{W}} \frac{v}{\Lambda^2} \frac{3}{2} g,
\end{equation}
and are associated with the following CP-even ($O_{3W}$) and CP-odd ($O_{3\widetilde{W}}$) dimension-six operators:
\begin{equation}\label{operators}
   \epsilon_{ijk} W_{\mu\nu}^{i} W^{j,\nu\rho} W^{k,\mu}_{\rho}, \hspace{1cm} \epsilon_{ijk} \tilde{W}_{\mu\nu}^{i} W^{j,\nu\rho} W^{k,\mu}_{\rho},
\end{equation}
where \(\tilde{W}_{\mu\nu}\) represents the dual field strength tensor. 

The findings of Ref.~\cite{Grojean:2018dqj} are also confirmed by the recent global SMEFT fit of Ref.~\cite{Celada:2024mcf}, which shows that individual and marginalised bounds on the CP-even Wilson coefficient $c_{W}$ are almost identical, with the complete sensitivity being originated from diboson production at the LHC. This justifies our choice of focusing on TGCs of Eq.~\ref{couplings} rather than considering other modifications that are better constrained through different processes. 
In addition to not considering $\mathcal{L}_{Vq\bar{q}}$, we also note that deformations in the lepton sector (entering through the decays of the EW bosons) are disregarded in our analysis due to the stringent bounds already imposed by LEP data~\cite{Celada:2024mcf}. Additionally, as our focus is on \(\lambda_z\) and \(\tilde{\lambda}_z\), the $\PW$-boson mass shifts are also ignored since such modifications are expected to be negligible in diboson production. 

Before proceeding to our results, we mention that the impact of SMEFT on diboson processes at dimension-eight has been recently explored in Ref.~\cite{Degrande:2023iob}. This study highlights dimension-eight SMEFT operators that exhibit interference contributions comparable to those from dimension-six. However, dedicated collider simulations are necessary to accurately assess the effect of these contributions on constraining SMEFT, as has already been undertaken in Ref.~\cite{corbett:2023qtg}. Therein, the authors extend the parameterisation presented in this section to include the relevant $\mathcal{O}$(\(1/\Lambda^4\)) terms. As well-known for some time~\cite{PhysRevD.48.2182}, dimension-eight operators decorrelate the photon and Z-boson TGCs leading to \(\lambda_{\gamma} \neq \lambda_{z}\). The analysis in Ref.~\cite{corbett:2023qtg} concludes that the impact of said contribution is minimal, \emph{i.e.} the bounds originating from the squared dimension-six terms are barely altered by including the dimension-8 contributions. In contrast, dimension-eight contributions relevant for \(g_{1,z}\) and \(\kappa_{z}\) were shown to be non-negligible (see Table IV of Ref.~\cite{corbett:2023qtg}). In summary, focusing on \(\lambda_z\) and \(\tilde{\lambda}_z\) further simplifies our analysis since, although dimension-eight contributions can not be neglected a-priori, the relevant dimension-eight effects are not expected to alter the constraining power induced by dimension-six quadratic terms.

\subsection{Numerical input and validation}
\label{sec:input}
We consider the inclusive production of \(\PW^+\PZ\) and \(\PW^+\PW^-\) pairs at the LHC@13TeV, in the fully leptonic decay channels, namely
\beq
\Pp\Pp \to \Pe^+ \nu_{\Pe} \hspace{0.1cm} \mu^+\mu^- + X\,,\qquad 
\Pp\Pp \to \Pe^+ \nu_{\Pe} \hspace{0.1cm} \mu^-\bar{\nu}_\mu + X\,,
\eeq
at NLO QCD accuracy in the SM and in the presence of dimension-six SMEFT operators introduced in Eq.~\ref{operators}.

We carry out the calculations in the $G_\mu$~scheme \cite{Denner:2000bj,Dittmaier:2001ay}, which is a suitable choice for SMEFT analyses, according to the recent recommendations of the LHC EFT WG \cite{Brivio:2021yjb}.
The Fermi-constant value is set to $\GF = 1.16638\cdot10^{-5} \GeV^{-2}$.
We treat unstable particles in the complex-mass scheme \cite{Denner:2005fg,Denner:2006ic,Denner:2019vbn}.
The pole masses and widths of the gauge bosons ($M_V,\Gamma_V$, $V=\PW,\PZ$) are derived, according to Ref.~\cite{Bardin:1988xt}, from the corresponding on-shell values \cite{ParticleDataGroup:2018ovx},
\begin{align}
\MW^{\rm os} ={}& 80.379\GeV\,,&  \GW^{\rm os}   ={}& 2.085\GeV\,, \nonumber\\
\MZ^{\rm os} ={}& 91.1876\GeV\,,&   \GZ^{\rm os} ={}& 2.4952\GeV\,,
\end{align}
The pole mass and width of the top quark (only entering the bottom-induced $\PW\PW$ production) read,
\begin{align}
\Mt ={}& 172\GeV\,,&  \Gamma_{\rm t}  ={}& 1.47\GeV\,.
\end{align}
We perform the calculations in the five-flavour scheme ($N_{f}=5$).
No Higgs-boson physical parameters enter the NLO QCD calculation.
The NNPDF3.1 PDF set \cite{Ball:2017nwa}, computed at NLO with $\alphas(\MZ)=0.118$ (\texttt{NNPDF31\_nlo\_as\_0118}), is given as an input to all MC simulations through the LHAPDF interface \cite{Buckley:2014ana}.
A fixed-scale choice is made for the renormalisation ($\mu_{\rm R}$) and factorisation scale ($\mu_{\rm F}$), 
as done in recent SM polarisation studies \cite{Denner:2020bcz,Denner:2020eck,Poncelet:2021jmj,Le:2022lrp,Le:2022ppa,Denner:2023ehn,Dao:2023pkl,Hoppe:2023uux,Pelliccioli:2023zpd}, 
\beq\label{eq:fixedscale}
\mu_{\rm F}=\mu_{\rm R}=\frac{\MW +\MZ}2\,\quad (\textrm{for }\PW\PZ)\,, \,\qquad \mu_{\rm F}=\mu_{\rm R}=\MW\,\,\quad (\textrm{for }\PW\PW) \,,
\eeq
We have checked (see \Cref{sec:dyn_sect}) that a dynamical choice leads to similar results to the fixed-scale ones, both at the integrated level and in differential distributions.

The SM calculation of both processes have been performed independently at NLO QCD  accuracy with \madgraphnlos \cite{Alwall:2014hca}, \mocanlo \cite{Denner:2020bcz,Denner:2020eck} and \powheg \cite{Chiesa:2020ttl,Pelliccioli:2023zpd}, finding agreement within integration uncertainties both at inclusive and at differential level. The SMEFT results have been extracted using SMEFT\@NLO \cite{Degrande:2020evl} within \madgraphnlos. The model has also been extended to include the CP-odd triple gauge operator. The SMEFT calculations have been validated through a comparison with the results of Ref.~\cite{Chiesa:2018lcs}. 

\section{WZ production}
\label{sec:WZ}
The ATLAS and CMS collaborations have already investigated the WZ channel in terms of polarisations and spin correlations \cite{ATLAS:2019bsc,CMS:2021icx,ATLAS:2022oge}, owing to the good signal purity and the possibility of accessing the complete final state up to a single-neutrino reconstruction.
At LO, the initial state is strictly $qq'$ induced, while at NLO QCD, the additional $qg,\bar{q}g$ partonic processes open up. 
It is well known \cite{Rubin:2010xp,Grazzini:2019jkl} that the NLO QCD corrections to this process in the SM are very large and driven by hard real radiation, owing to the kinematic enhancement of phase-space regions with one hard EW boson recoiling against the system of a jet and a soft EW boson, as well as to the enhancement from the gluon luminosity in the proton. The NLO EW corrections to fully leptonic WZ are at the few-percent level \cite{Biedermann:2017oae,Grazzini:2019jkl} and grow negatively in the tails of transverse-momentum distributions, where large EW Sudakov logarithms of virtual origin appear \cite{Denner:2000jv,Accomando:2004de}.
The SMEFT effects have been investigated in WZ production, with a special focus on the well-known \emph{resurrection} of dimension-six interference
\cite{Panico:2017frx,Azatov:2017kzw,Franceschini:2017xkh,Azatov:2019xxn} but also on dimension-eight operators \cite{Chiesa:2018lcs,Baglio:2019uty} and the matching to various beyond-the-SM scenarios \cite{Liu:2018pkg,Grojean:2018dqj}.

\subsection{Setup}\label{sec:WZsetup}
We consider an inclusive and a fiducial setup for $\PW^+\PZ$ production. The inclusive one features uniquely a cut on the invariant mass of the same-flavour, opposite-sign leptons,
\beq\label{eq:mllcut}
81\GeV<M_{\mu^+\mu^-}<101\GeV\,.
\eeq
The fiducial selections mimic those of Refs.~\cite{ATLAS:2019bsc,ATLAS:2022oge}:
\begin{align}
 & 81\GeV<M_{\mu^+\mu^-}<101\GeV\,, \nonumber\\
 & p_{\rm T,\,\Pe^+}>20\GeV\,,\qquad p_{\rm T,\,\mu^\pm}>15\GeV\nonumber\\
 &M_{\rm T,\,\PW} > 30\GeV\,,\qquad |y_\ell|<2.5 \nonumber\\
 &\Delta R_{\mu^+\mu^-}>0.2\,,\qquad\Delta R_{\mu^\pm\Pe^+}>0.3, \label{eq:fidSetup}
\end{align}
where 
\beq
M_{\rm T,\,\PW} = \sqrt{2\,p_{\rm T,\,\Pe^+}\,p_{\rm T,\,mis}\left(1-\cos\Delta\phi_{\Pe^+,\rm mis}\right)}. 
\eeq
No veto is imposed on possible additional hadronic activity.
Because of the presence of a single neutrino in the considered final state, it is possible to employ standard on-shell requirements on the $\PW$ boson to access the complete final-state kinematics. We apply the reconstruction technique used in Ref.~\cite{ATLAS:2019bsc}.

\subsection{Selection-cut effects}
\begin{figure}
  \centering
  \subfigure[Inclusive\label{fig:m4l_1}]{\includegraphics[scale=0.4]{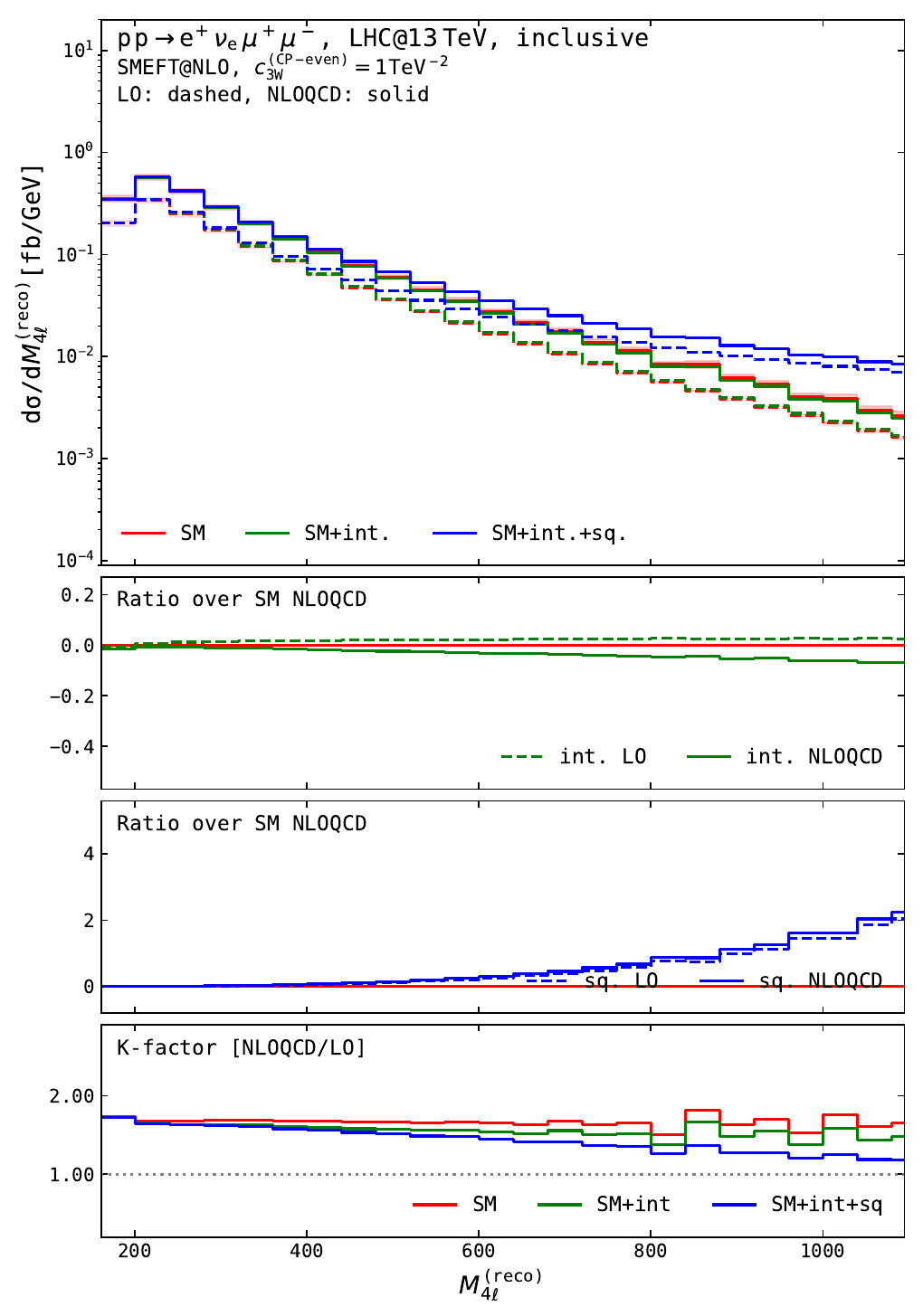}}
  \subfigure[ATLAS fiducial\label{fig:m4l_2}]{\includegraphics[scale=0.4]{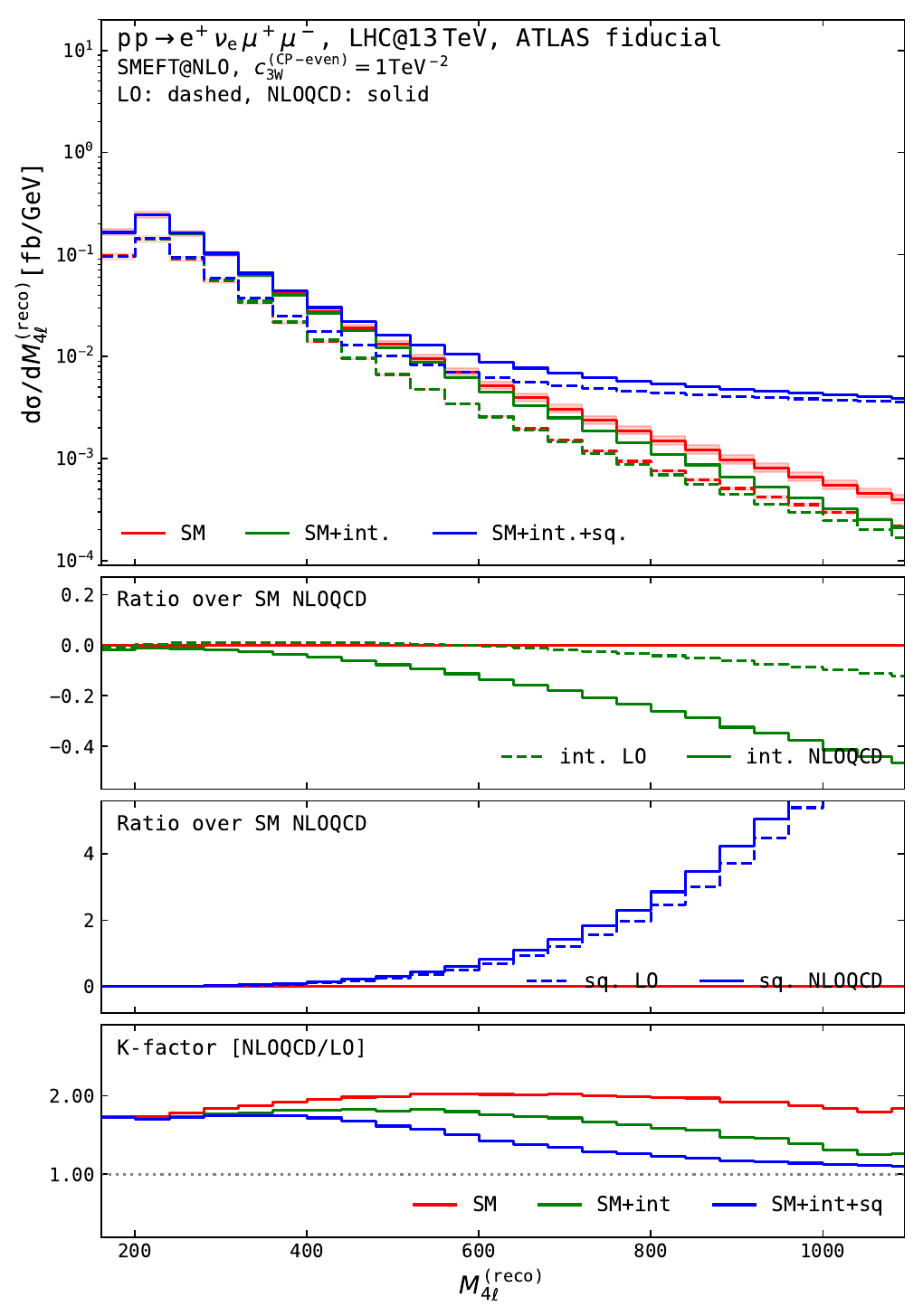}}
  \caption{Distributions in the reconstructed four-lepton invariant mass in W$^+$Z production at the LHC@13TeV, for the inclusive (left) and fiducial ATLAS setup (right): effect of the inclusion of the CP-even operator ${O}_{3{W}}$ with WCs set to $1{\rm TeV}^{-2}$. 
  Main panels: absolute differential cross-sections at LO (dashed) and NLO QCD (solid), for the SM (red), for the sum of the SM and the SMEFT linear term (green), and also including the quadratic SMEFT term (blue). First/second inset: ratio of the linear/quadratic SMEFT term (at LO and NLO QCD) over the NLO QCD SM cross-section. Third inset: QCD K-factors (NLO QCD over LO).
  }\label{fig:m4lrec}
\end{figure}

\begin{figure}[h]
  \centering
  \subfigure[Inclusive\label{fig:ptZ_1}]{\includegraphics[scale=0.4]{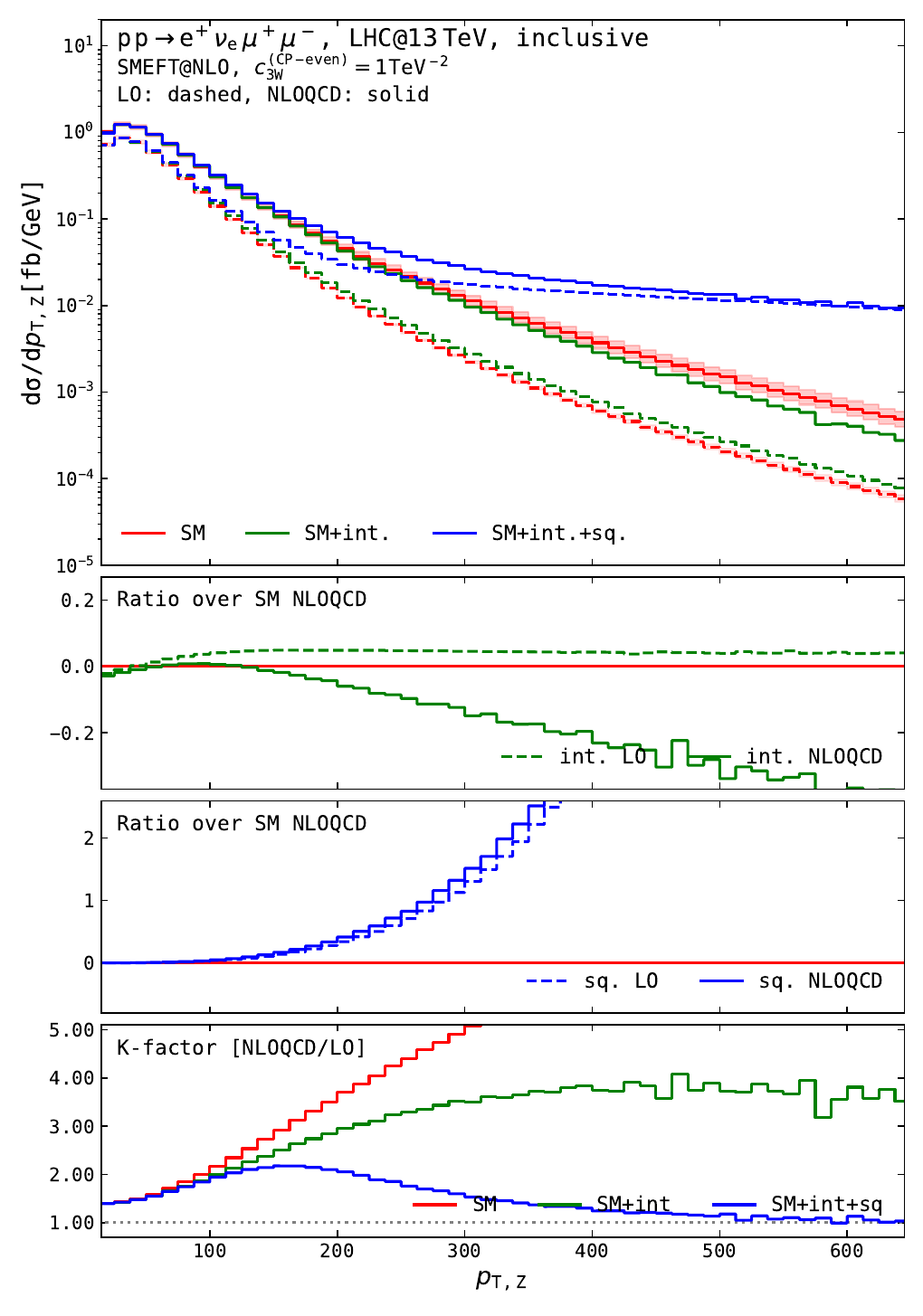}}
  \subfigure[ATLAS fiducial\label{fig:ptZ_2}]{\includegraphics[scale=0.4]{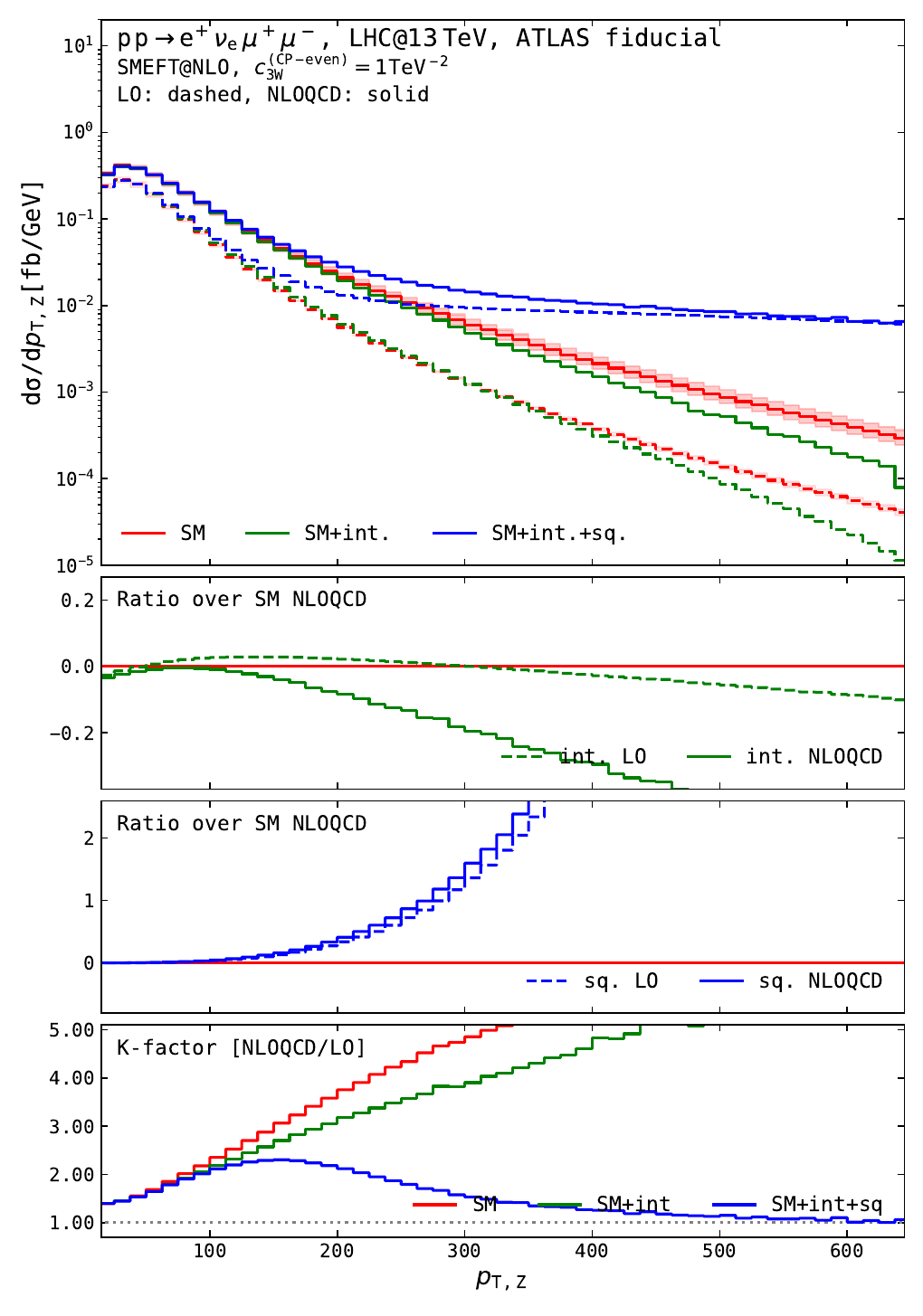}}
  \caption{Distributions in the $\PZ$-boson transverse momentum in W$^+$Z production at the LHC@13TeV, for the inclusive (left) and fiducial ATLAS setup (right): effect of the inclusion of the CP-even operator ${O}_{3{W}}$ with WCs set to $1~{\rm TeV}^{-2}$. Same structure as \Cref{fig:m4lrec}.
 }\label{fig:ptZ}
\end{figure}

The integrated cross-sections in the ATLAS fiducial setup detailed in \Cref{sec:WZsetup} are shown in \Cref{tab:sigmasWZ}, compared with those obtained in the inclusive setup.
\begin{table}[t]
    \centering
    \begin{tabular}{ccccc}
\hline\\[-0.5cm]
  accuracy                 &   $\sigma_{\rm LO}$             &   $\sigma_{\rm NLO}$        &   $\sigma_{\rm LO}$        &   $\sigma_{\rm NLO}$       \\[0.1cm]
\hline\\[-0.4cm]
  setup                &  inclusive &  inclusive &  ATLAS fiducial & ATLAS fiducial   \\[0.1cm]
\hline\\[-0.4cm]
 SM          &    $ 58.421(5) ^{+   3.6 \%}_{ -4.6 \%}$  &   $   98.37(2) ^{+   5.3 \%}_{ -5.1 \%}$   & $19.691(3) ^{+   4.8 \%}_{ -6.0 \%}$& $35.34(1) ^{+   5.6 \%}_{ -5.6 \%}$\\
 CP-even int &    $ 1.0783(6) ^{+   3.1 \%}_{ -4.1 \%}  $  &   $    -1.418(2) _{-22.1 \%}^{  +26.2 \%}   $   & $0.0980(4) ^{+   9.6 \%}_{ -11.7 \%}$& $-0.997(2) _{-13.6 \%}^{+16.4 \%}$\\
 CP-even sq  &    $ 11.972(4) ^{+   6.0 \%}_{ -5.3 \%} $  &   $   12.468(9) ^{+   0.9 \%}_{ -1.1 \%}       $   & $6.192(3) ^{+   5.3 \%}_{ -4.8 \%}$& $6.582(6) ^{+   1.6 \%}_{ -1.6 \%}$\\
 CP-odd int &  $-0.1475(5) _{  -5.9 \%}^{ + 4.8 \%}$ &  $-0.182(3) _{  -4.0 \%}^{ + 3.4 \%}$    &   $-0.0601(3) _{  -6.6 \%}^{ + 5.4 \%}$  & $-0.059(1) _{  -1.0 \%}^{ + 1.1 \%} $\\
 CP-odd sq  &  $12.248(6) ^{+   5.7 \%}_{ -5.0 \%} $    &  $12.78(2) ^{+   0.8 \%}_{ -1.1 \%}$    &  $6.310(4) ^{+   5.1 \%}_{ -4.6 \%}$  & $6.71(1) ^{+   1.5 \%}_{ -1.5 \%} $\\[0.1cm] 
\hline
\end{tabular}
    \caption{Integrated cross-sections (in fb) for the SM and dimension-six SMEFT linear (int) and quadratic (sq) contributions in $\PW^+\PZ$ production at the LHC@13TeV. Two setups are considered: inclusive (Eq.~\ref{eq:mllcut}) and ATLAS fiducial (Eq.~\ref{eq:fidSetup}). The QCD uncertainties from 9-point scale variations are shown in percentages. MC uncertainties on central values are shown in parentheses.}
    \label{tab:sigmasWZ}
\end{table}
The SM cross-section receives large QCD corrections (about $+80\%$).
The contribution of the SMEFT--SM interference term gets a sign flip when including NLO QCD corrections and a sizeable increase of the QCD-scale uncertainty. The squared SMEFT cross-section is instead characterised by a moderate increase ($+6\%$) due to QCD radiative corrections and a reduction in the NLO scale uncertainty. The rather small QCD corrections in the squared SMEFT term is due to the absence of large Sudakov logarithms from soft-boson radiation off quark lines, which instead dominate the SM cross-section at NLO \cite{Rubin:2010xp,Grazzini:2019jkl}. 
This effect will be further appreciated in the differential distributions of \Cref{fig:m4lrec,fig:ptZ}.

It is instructive to compare the fiducial cross-sections with the inclusive ones,
computed in the presence of the sole cut shown in Eq.~\ref{eq:mllcut}.
The application of fiducial cuts increases the relative size of QCD corrections compared to LO in the SM (from 68\% to 79\%) and the squared SMEFT term (from 4\% to 6\%).  

The fiducial cross-sections confirm the literature results showing the \emph{resurrection} of SMEFT--SM interference owing to the inclusion of additional QCD radiation \cite{Azatov:2017kzw,Azatov:2019xxn} 
In particular, the linear SMEFT term amounts to 0.5\% of the SM cross-section at LO, while, up to the sign flip, almost 3\% at NLO QCD accuracy.
 It is worth noticing that the interference resurrection is more evident in the presence of selection cuts than in the inclusive setup. 
This stems from a delicate interplay between the negative shift in the EFT contribution due to QCD corrections and the application of selection cuts on the decay products, which introduce shape distortion, preventing the complete integration over the decay angles of the leptons. 

Despite the CP invariance of the SM (no quark-family mixing is assumed),
the linear CP-odd contribution does not vanish ($-0.2\%$ of the SM at NLO QCD). This can be traced back to the inclusion of complete spin correlations between production and decay sub-processes and non-resonant topologies. 
We have checked analytically that when considering the $2\rightarrow 2$ process ($\Pp\Pp\rightarrow \PW^{+}\PZ$) with EW-boson widths set to zero, the interference between SM and CP-odd EFT amplitudes vanishes, whilst this does not hold anymore for the $2\rightarrow 4$ process. 

We continue by exploring the importance of fiducial cuts employed by the experimental analyses for the relative impact of the EFT operators on various observables. We first show in \Cref{fig:m4lrec} the reconstructed four-lepton invariant mass in $\PW^+\PZ$ production at the inclusive level and with ATLAS fiducial cuts as defined in \Cref{sec:WZsetup}. Generally, the deviation from the SM due to the triple-gauge operator is more pronounced in the high energy tails of the differential distributions. This is expected due to the energy growth of the corresponding EFT amplitude. We find that the impact of the EFT contributions is significantly enhanced for the fiducial distribution, both at the interference and squared levels. In the same plots, we also explore the impact of NLO QCD corrections, which are found to vary between the SM and EFT and their interference. K-factors are modified in the presence of fiducial cuts and depend on the kinematic region, i.e. K-factors are not flat.  

Similar observations can be made for the $\PZ$-boson transverse momentum, where fiducial cuts again enhance the impact of the EFT. In particular, for this observable, we notice the large QCD K-factors that arise in the SM \cite{Rubin:2010xp,Grazzini:2019jkl}. These large K-factors are well understood to originate from Sudakov logarithms from the emission of a relatively soft gauge boson from a quark line. We notice, though, that for the quadratic EFT contribution, the K-factor is significantly smaller, tending to unity at high transverse momentum. This is because the topology of the EFT diagram only allows a limited set of kinematic configurations that are Sudakov-logarithm enhanced. 
When only including the EFT linear term, the QCD K-factor stabilises for $p_{\rm T,Z}>400\GeV$ in the inclusive setup, while in the presence of fiducial cuts, the radiative corrections keep growing even in this boosted regime.

As expected from the helicity amplitudes in the SM and EFT, interference contributions are minimal at LO but grow after including QCD corrections, leading to a change of sign in the most populated phase space regions as also expected from the integrated cross-sections shown in \Cref{tab:sigmasWZ}. In the high transverse-momentum region, the NLO contribution is at least one order of magnitude larger than that of the LO. Nevertheless, for the choice of the coefficient shown, the EFT contribution is dominated by the quadratic term. 

\begin{figure}
  \centering
  \subfigure[Inclusive\label{fig:dphi_1}]{\includegraphics[scale=0.43]{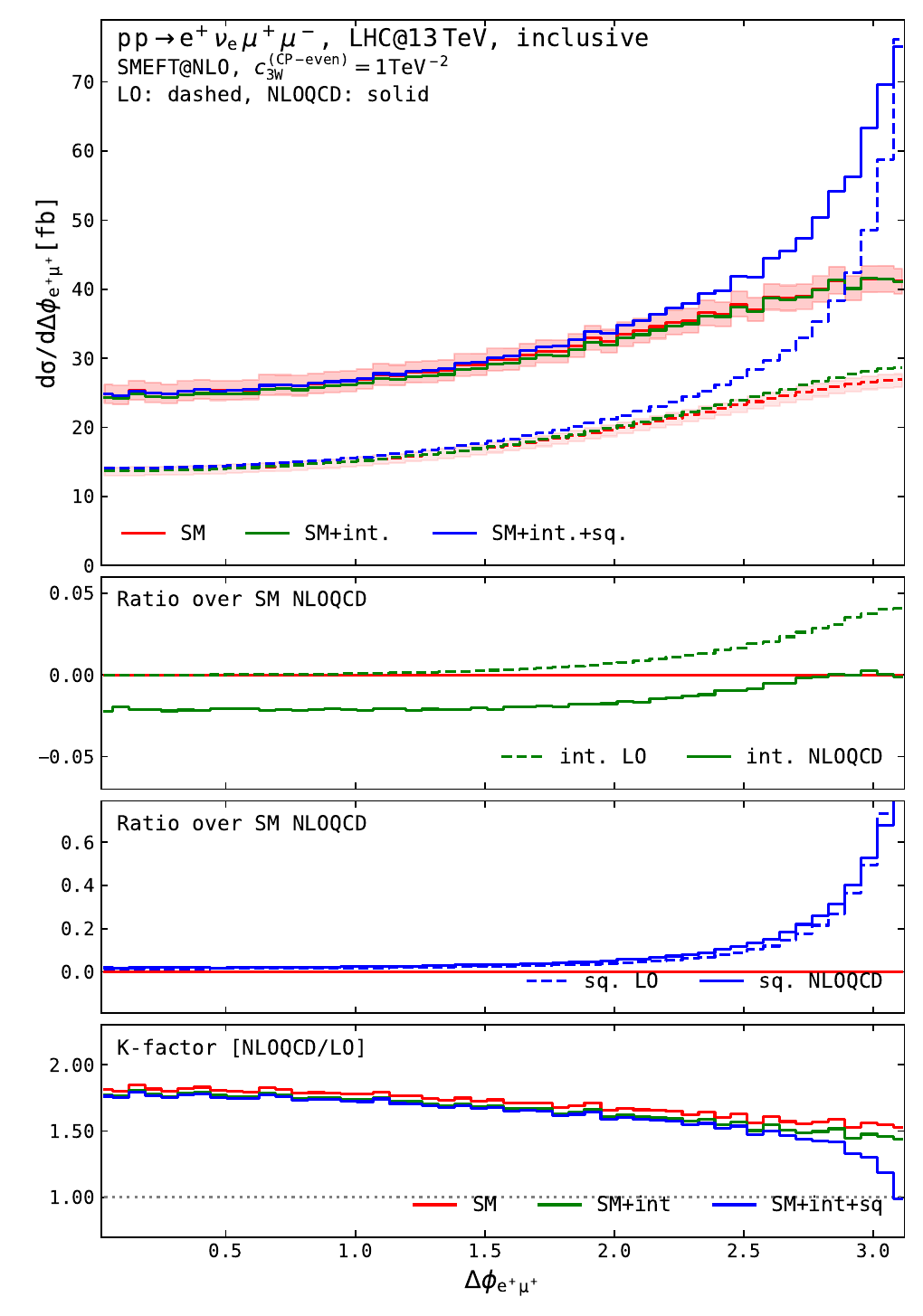}}
  \subfigure[ATLAS fiducial\label{fig:dphi_2}]{\includegraphics[scale=0.43]{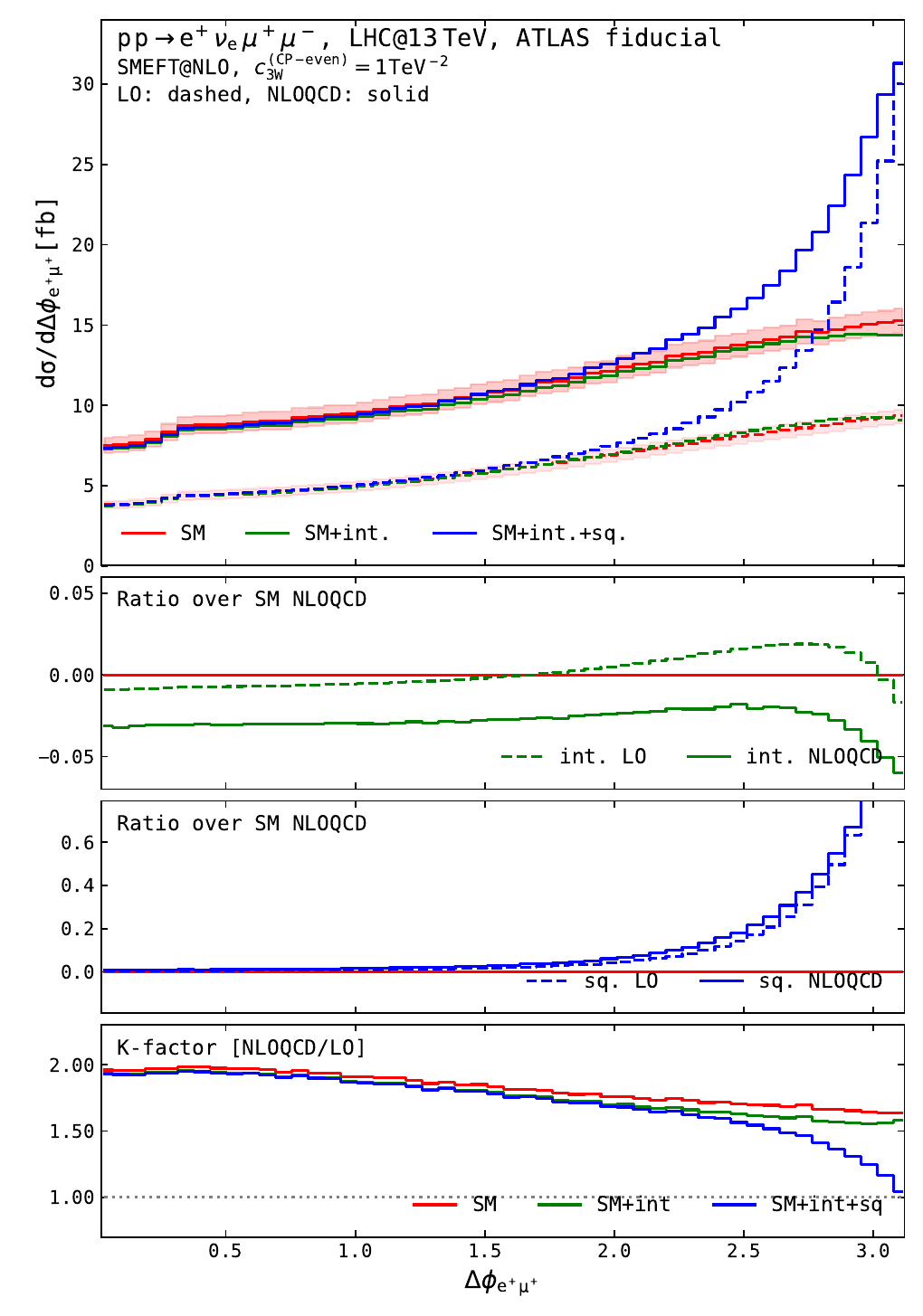}}
  \caption{Distributions in the azimuthal-angle separation between the positron and the antimuon in W$^+$Z production at the LHC@13TeV, for the inclusive (left) and fiducial ATLAS setup (right): effect of the inclusion of the CP-even operator ${O}_{3{W}}$, with Wilson coefficient set to $1{\rm TeV}^{-2}$. 
   Same structure as \Cref{fig:m4lrec}.
  }\label{fig:dphi}
\end{figure}

Finally, we also consider the impact of fiducial cuts on the azimuthal separation between the positron and antimuon in \Cref{fig:dphi}. We find that the quadratic contributions dominate in the large-separation region. Applying the fiducial cuts does not significantly alter the shape of the distribution. For the interference, the cuts push the distribution towards more negative values in the region $\Delta \phi\sim \pi$, which is correlated with a boosted kinematics of the EW bosons (similar effects in the tail of the $p_{\rm T,Z}$ spectrum in \Cref{fig:ptZ}), 
but the interference contribution remains below 5\% of the SM. QCD corrections have a non-trivial impact on the interference contribution, whilst, for the squared contribution, the K-factor is close to unity, as evidenced by driving the overall K-factor to 1 in the large-separation region where quadratic contributions dominate.  

The CP-odd effects induced by the dimension-six operator $O_{3\widetilde{W}}$ lead to negligible impacts relative to the SM (almost vanishing linear term, flat correction from the squared term) for the considered observables in Figures~\ref{fig:m4lrec}--\ref{fig:dphi}, therefore no results are shown. To increase the sensitivity to linear CP-odd effects, it is necessary to measure suitable angular variables described in the following section. The squared CP-odd term leads to similar effects found for the CP-even one in the considered observables, both in size and the distortion of SM distribution shapes.

\subsection{Polarisation-sensitive observables}
As discussed in \Cref{sec:intro}, inclusive observables exhibit a small interference, owing to the nature of the SM and EFT amplitudes. 
In the previous sections, we have demonstrated the revival of dimension-six interference through NLO corrections and fiducial cuts. Now, we turn our attention to the decay angular observables, which serve as crucial tools for directly probing interference contributions.
In an inclusive setup (no selections on decay leptons) and considering MC-truth momenta (no neutrino reconstruction), the differential two-body decay rate of an EW boson reads,
\beqn\label{eq:angular}
\frac1{\sigma}\frac{\rd\sigma}{\rd\cos\theta^* \,\rd\phi^*} 
&=&
\frac{3}{16\pi}
\bigg[
1+\cos^2\theta^*  + A_0\frac{1-3\cos^2\theta^* }2 + A_1 \sin2\theta^* \cos\phi^* \nonumber\\
&&\hspace*{0.8cm}+\frac12 A_2 \sin^2\theta^* \cos2\phi^*  + A_3 \sin\theta^* \cos\phi^* 
+ A_4\cos\theta^*  \nonumber\\
&&\hspace*{0.8cm}+ A_5 \sin\theta^* \sin\phi^* 
+ \; A_6  \sin2\theta^* \sin\phi\;+  A_7 \sin^2\theta^* \sin2\phi^* \bigg]\,,
\eeqn
where $\phi^*$ and $\theta^*$ are, respectively, the azimuthal and polar angles of one of the decay leptons, computed in the EW-boson rest frame, with respect to the boson flight direction in the diboson CM frame. This choice gives access to polarisation coefficients defined in the diboson CM frame, which is typically used in experimental analyses \cite{ATLAS:2019bsc,ATLAS:2022oge,ATLAS:2024qbd} and considered as the most natural frame for diboson studies 
\cite{Denner:2020eck,Le:2022lrp,Le:2022ppa,Denner:2023ehn,Dao:2023pkl,Hoppe:2023uux,Pelliccioli:2023zpd}. To determine the coordinate system fully \cite{Boudjema:2009fz}, we fix the zero-axis for the azimuthal decay angle to the Z-boson decay plane in the diboson CM frame.

Upon integrating Eq.~\ref{eq:angular} over the azimuthal decay angle, the decay rate takes the form: 
\beqn\label{eq:costhetaA0A4}
\frac{1}{\sigma} \frac{\rd \sigma}{\rd\, \cos\theta^*} &=&
\frac38 \left(1+\cos ^2\theta^*+ A_0 \frac{1-3 \cos ^2\theta^*}2+A_4 \cos\theta^* \right).
\eeqn
Through linear combinations of the coefficients $A_0,\,A_4$, this expression can be re-written in terms of polarisation fractions:
\beqn
\frac{1}{\sigma} \frac{\rd \sigma}{\rd\, \textrm{cos}\theta^*} 
&=&
\frac38\bigg[
\,\,2\,f_{\rm 0}\,\sin^2\theta^*\nonumber\\
&&\hspace*{0.5cm}+f_{\rm L}\,\left(1+\cos^2\theta^{*}-2\,c_{\rm \tiny LR}\,\cos\theta^*\right) \,\nonumber\\
&&\hspace*{0.5cm}+f_{\rm R}\,\left(1+\cos^2\theta^{*}+2\,c_{\rm LR}\,\cos\theta^*\right) \, \bigg], \label{eq:polfrac}
\eeqn
where $f_{\rm 0}, f_{\rm R}, f_{\rm L}$ are the longitudinal, right-handed, and left-handed polarisation fractions, respectively. The coefficient $c_{\rm LR}$ parametrises the relative balance of the left-chirality and right-chirality coupling of the EW boson to massless leptons, 
\beq\label{eq:CLR}
c_{\rm LR} = \frac{|g_{\rm L}|^2-|g_{\rm R}|^2}{|g_{\rm L}|^2+|g_{\rm R}|^2}\,.
\eeq
The bosonic SMEFT operators we consider in this work do not modify by any means the coupling of EW bosons to fermions; therefore the $c_{\rm LR}$ values are the SM ones, computed in the $G_\mu$  scheme,
\beq
c_{\rm LR}^{\rm (W)} = 1\,,\qquad c_{\rm LR}^{\rm (Z)} = \frac{1-4\sin^2\theta_{\rm w}}{1-4\sin^2\theta_{\rm w}+8\sin^4\theta_{\rm w}}\,\approx 0.215\,,\quad \sin^2\theta_{\rm w} = 1-\frac{\MW^2}{\MZ^2}\,.
\eeq

\subsubsection{Inclusive angular coefficients and polarisation fractions}\label{sec:angcoeff}
Using the expressions in Eqs.~\ref{eq:angular}--\ref{eq:polfrac}, we can extract the angular coefficients as well as the polarisation fractions of the gauge bosons in the process by means of suitable projections on spherical harmonics up to rank 2. This approach has also been utilised for Higgs decays and Higgs-strahlung \cite{Banerjee:2019twi,Banerjee:2020vtm} in the context of SMEFT.
The values of the angular coefficients extracted from the angular distributions in the inclusive setup are shown in \Cref{tab:Acoeff_phi} for the SM and in the presence of the CP-even and CP-odd operator coefficients. 
\begin{table}[t]
    \centering
    \begin{tabular}{lcccc}
\hline
    & $A_0$    & $A_4$ & $A_2$    & $A_7$ \\
\hline
& \multicolumn{4}{c}{$\PW^+$ boson} \\
\hline
SM                  & 0.367(1)& -0.381(1) &   -0.223(7) &  -0.001(3) \\
SM+int, CP-even     & 0.378(1)& -0.379(1) &   -0.103(6) &  -0.003(4) \\
SM+int+sq, CP-even  & 0.339(1)& -0.336(1)&    -0.089(7) &  -0.003(3) \\
SM+int, CP-odd      & 0.367(1)& -0.381(1) &   -0.219(5) &   0.061(2) \\
SM+int+sq, CP-odd   & 0.332(2)& -0.338(2) &   -0.196(4) &   0.053(2) \\
\hline
& \multicolumn{4}{c}{$\PZ$ boson} \\
\hline
SM                    & 0.358(1)  & -0.0357(4) & -0.148(6)  &  -0.002(3)\\
SM+int, CP-even       & 0.370(1) & -0.0377(5) & -0.025(7) & 0.001(3)\\
SM+int+sq, CP-even    & 0.332(1) & -0.0332(6) & -0.021(6) & -0.001(2)\\
SM+int, CP-odd        & 0.357(1)  & -0.0375(4)& -0.151(4)& 0.062(2) \\
SM+int+sq, CP-odd     & 0.323(1)  & -0.0329(5)& -0.135(4) & 0.055(2)\\
\hline
    \end{tabular}
    \caption{Angular coefficients of the $\PW^+$ and $\PZ$ bosons in W$^+$Z production at the LHC@13TeV, extracted at NLO QCD in the inclusive setup (see Eq.~\ref{eq:mllcut}) from decay angular distributions: effect of the inclusion of the CP-even and CP-odd operators (${O}_{3{W}}$ and $O_{3\widetilde{W}}$, respectively). The WCs are set to $1{\rm TeV}^{-2}$. The MC uncertainties are shown in parentheses.}
    \label{tab:Acoeff_phi}
\end{table}

The $A_0$ and $A_4$ coefficients are extracted from Eq.~\ref{eq:costhetaA0A4} and they are directly related to the 
longitudinal- and transverse-polarisation fractions via Eq.~\ref{eq:polfrac}, namely,
\beq\label{eq:Atof_corresp}
A_0 = 2\,f_0\,,\qquad A_4 = 2\,c_{\rm LR}\,(f_{\rm R}-f_{\rm L})\,.
\eeq
The linear SMEFT effect of CP-even coefficient mildly changes the $\PW^+$-boson polarisation fractions, while the corresponding CP-odd effects vanish, as they only affect off-diagonal entries of the spin-density matrix, \emph{i.e.} azimuthal-dependent angular coefficients \cite{Azatov:2017kzw,Azatov:2019xxn,Degrande:2021zpv}.
The quadratic SMEFT effects (both CP-even and CP-odd) change the W-boson longitudinal fraction from 18\% to 16\% and conversely increase its right-handed fraction from 31\% to 33\%, having a negligible effect on the dominant left-handed contribution (50\%).
A similar picture is found for the longitudinal-polarisation content of the $\PZ$ boson. At variance with the $\PW$ boson, the $\PZ$ boson is characterised by an almost equal content of left- and right-handed polarisations, resulting in a much smaller $A_4$ coefficient. This comes from the mixed right- and left-chirality of the $\PZ$ coupling to fermions. The considered SMEFT operators have mild effects on the left--right-polarisation balance of the $\PZ$ boson.

The coefficients $A_5, \,A_7$ are parity odd and in the SM, they only take non-vanishing values when including
higher-order QCD corrections in the production process \cite{Bern:2011ie,Baglio:2018rcu}. The impact of the $A_7$ contribution is typically negligible for practical applications; the one of $A_5$ is slightly larger only if computing forward-backwards asymmetries in the W-boson rapidity \cite{Bern:2011ie}. As expected, the CP-even operator does not modify $A_7$, whose value significantly rises in the presence of the CP-odd operator.
The two parity-even coefficients $A_2$ and $A_3$ control the left-right and longitudinal-transverse interference terms, respectively.
The SM suppression of the longitudinal mode in inclusive WZ production \cite{Denner:2020eck,Le:2022lrp} makes the longitudinal-transverse interference term (and therefore the $A_3$ value) small. 
In the presence of the EFT coefficients, we find that the value of the $A_2$ coefficient is significantly reduced by both the interference and quadratic contributions of the CP-even Wilson coefficient. The CP-odd contribution only modifies $A_2$ at the quadratic level, but, as expected, it gives rise to a more considerable parity odd $A_7$ coefficient.  

It is worth recalling that the parity-odd $A_5$ coefficient is suppressed in the SM, as it only takes non-vanishing values from absorptive loop contributions of NNLO QCD corrections \cite{Hagiwara:2006qe,Bern:2011ie}. The picture is not changed by including SMEFT effects (not even CP-odd ones) at the considered accuracy. This holds for the $\PW$ and the $\PZ$ boson.
Similarly, the parity-even $A_3$ coefficient comes from off-diagonal components of the spin-density matrix (for on-shell bosons) and, therefore, is small in the SM. SMEFT effects are also negligible. Additionally, the $A_3$ coefficient is sensitive to the off-shell-ness of the lepton pairs \cite{Baglio:2019uty}. We do not show any result for $A_3$ or $A_5$ in this context.

\begin{figure}
  \centering
  \subfigure[ \label{fig:ppol1}]{\includegraphics[scale=0.4]{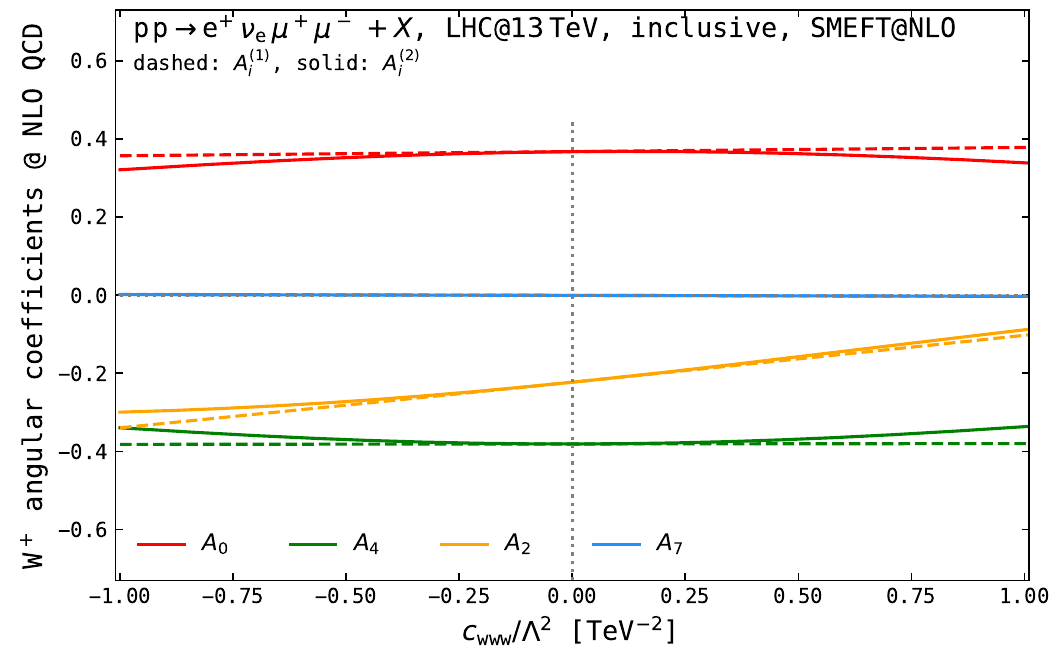}}
  \subfigure[ \label{fig:ppol2}]{\includegraphics[scale=0.4]{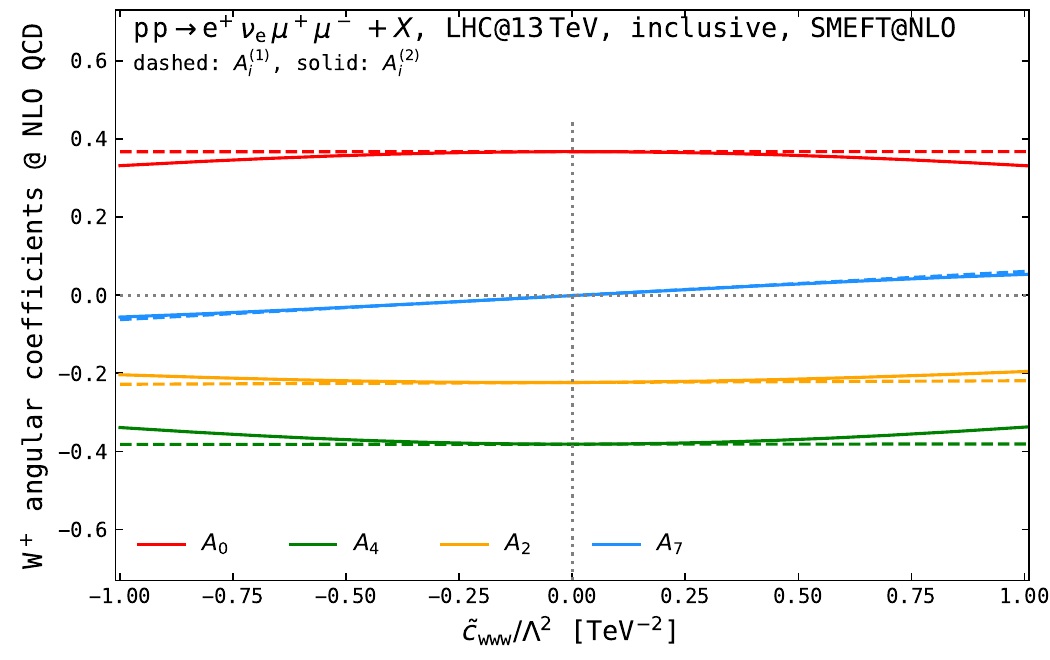}}
  \subfigure[ \label{fig:ppol1z}]{\includegraphics[scale=0.4]{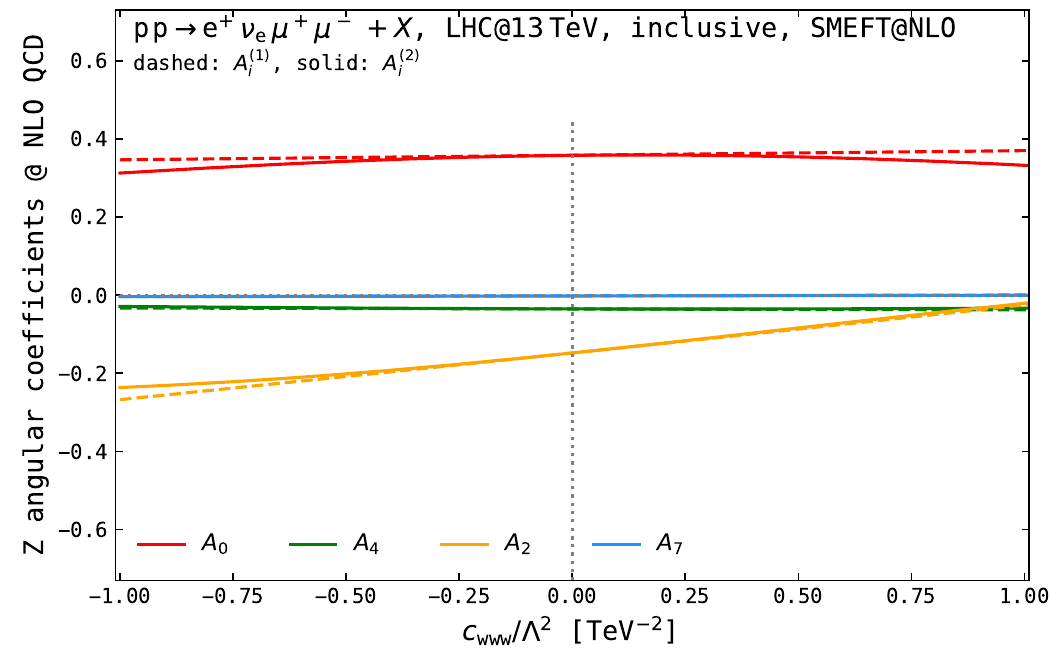}}
  \subfigure[ \label{fig:ppol2z}]{\includegraphics[scale=0.4]{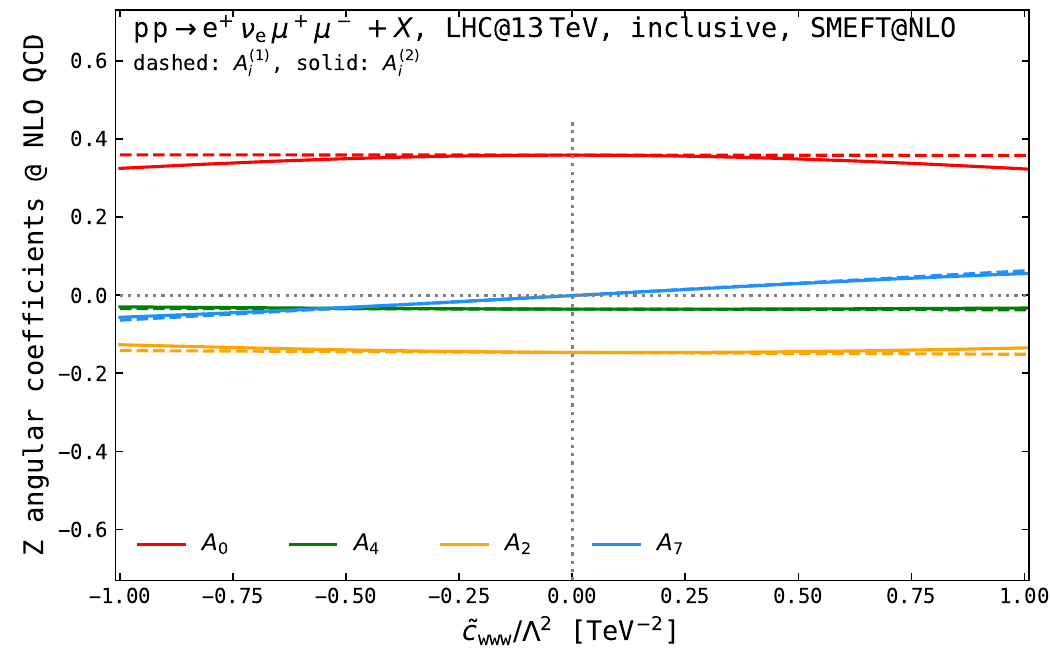}}
  \caption{Dependence of the angular coefficients $A_0,\,A_2,\,A_4,\,A_7$ of the W (top) and Z boson (bottom) in W$^+$Z production at the LHC@13TeV on the Wilson coefficient value for the CP-even operator ${O}_{3{W}}$ (left) and of the CP-odd one $O_{3\widetilde{W}}$ (right). The inclusive setup (see Eq.~\ref{eq:mllcut}) is understood. The dashed and solid curves correspond respectively to the $A^{(1)}$ and $A^{(2)}$ SMEFT parametrisations of the coefficients defined in Eq.~\ref{eq:coef_par}. \label{fig:polWSMEFT}}
\end{figure}

We have also studied the angular coefficients for changing values of the WCs to obtain a complete and more general picture.
For a given coefficient $A_i$, we consider two parametrisations that rely on the expansion in the SMEFT series up to the linear and quadratic term, respectively:
\beqn\label{eq:coef_par}
  A^{(1)}_i(\lambda) &=& \frac{  A_i^{\rm SM} + \lambda \,{  A_i^{\rm int}}\,\kappa^{\rm int}}{1 + \lambda \kappa^{\rm int}   }\,,\nonumber\\
  A^{(2)}_i(\lambda) &=& \frac{  A_i^{\rm SM} + \lambda \,{  A_i^{\rm int}}\,\kappa^{\rm int} +  { \lambda^2 A_i^{\rm sq}} \kappa^{\rm sq}}{1 + \lambda \kappa^{\rm int} +  \lambda^2 \kappa^{\rm sq}}\,,
\eeqn
where,
\beq\label{eq:kappadef}
\kappa^{\rm int}=\frac{\sigma^{\rm int}}{\sigma^{\rm SM}}\,,\qquad\kappa^{\rm sq}=\frac{\sigma^{\rm sq}}{\sigma^{\rm SM}}\,,\qquad \lambda=\frac{c_{\rm www}}{\Lambda^2} \textrm{ or } \frac{\tilde{c}_{\rm www}}{\Lambda^2}\,.
\eeq
The results are shown in \Cref{fig:polWSMEFT} for the four considered coefficients.
The plots confirm that the EFT contribution from the CP-even triple-gauge operator leads to a slightly smaller longitudinal-polarisation fraction for non-zero Wilson-coefficient values. The CP-odd operator does not modify the polarisation fractions, hence the $A_0, A_4$ coefficients at the linear level.  
At the quadratic level (solid curves), both the CP-even and the CP-odd operators give similar distortions (up to a sign change) to the $A_0$ and $A_4$ coefficients of the W boson. 
In the considered inclusive setup, the size of the absolute values of $A_0$ and $A_4$ are very close to each other. This results in a negligible change of the dominant left-handed fraction. The coefficient 
$A_2$ is significantly modified by CP-even effects already at the linear level. The main impact of the CP-odd Wilson coefficient for both the W and Z boson is to increase the value of the parity odd $A_7$ coefficient, which is compatible with zero in the SM.  

As a last comment, we stress that the extraction of such coefficients through suitable spherical-harmonics projections (according to Eq.~\ref{eq:angular}) is well defined only in the absence of fiducial cuts on individual decay products, and assuming a perfect neutrino reconstruction, owing to the possibility to perform the complete integration over polar and azimuthal decay angle ranges. The situation is entirely different in a fiducial setup and in the presence of a realistic neutrino reconstruction, where the application of the same strategy as in the inclusive case may lead to unphysical results  \cite{Boudjema:2009fz,Stirling:2012zt,Ballestrero:2017bxn,Denner:2020bcz}.
To quantify the size of these effects, we consider the extraction of the $A_0$ coefficient of the $\PW^+$ boson via Eq.~\ref{eq:angular} in four scenarios: the inclusive setup without neutrino reconstruction (corresponding to the numerical values of \Cref{tab:Acoeff_phi}), the inclusive setup with neutrino reconstruction, the fiducial setup without and with neutrino reconstruction.
For this example, we only consider the effect of the $O_{3\PW}$ operator, but similar results follow for its CP-odd counterpart.
The numerical results are shown in \Cref{tab:A0_check}.
\begin{table}[t]
    \centering
    \begin{tabular}{lcccc}
\hline
setup                   & inclusive & inclusive & ATLAS fid. & ATLAS fid. \\
neutrino reco.          & no        & yes       & no             & yes            \\
\hline
& \multicolumn{4}{c}{$A_0$ ($\PW^+$ boson)} \\
\hline
SM                  & 0.367(1) & 0.515(4) & 0.930(2) & 0.983(3) \\
SM+int, CP-even     & 0.378(1) & 0.509(4) & 0.935(2) & 0.975(3) \\
SM+int+sq, CP-even  & 0.339(1) & 0.449(5) & 0.834(2) & 0.861(3) \\
\hline
& \multicolumn{4}{c}{$A_0$ ($\PZ$ boson)} \\
\hline
SM                  & 0.358(1) & 0.431(1) & 0.855(2) & 0.839(2) \\
SM+int, CP-even     & 0.370(1) & 0.440(1) & 0.861(2) & 0.845(3) \\
SM+int+sq, CP-even  & 0.332(1) & 0.394(2) & 0.776(2) & 0.761(3) \\
\hline
    \end{tabular}
    \caption{Effect of selection cuts and neutrino reconstruction on the $A_0$ coefficient in W$^+$Z production at the LHC@13TeV at NLO QCD, extracted through projection of Eq.~\ref{eq:angular} onto spherical harmonics. Only numerical values in the first column (inclusive, no neutrino reconstruction) are reliable for interpretation in terms of polarisation states. The results are shown for the SM and in the presence of the linear and squared term in the CP-even operator ${O}_{3{W}}$, with the Wilson coefficient set to $1{\rm TeV}^{-2}$. The MC uncertainties are shown in parentheses.}
    \label{tab:A0_check}
\end{table}
It is evident that the standard analytic extraction of the coefficients leads to unphysical results for both bosons when 
selection cuts are applied and/or neutrino reconstruction.
The $A_0$ coefficient is proportional to the longitudinal fraction of the gauge boson (see Eq.~\ref{eq:Atof_corresp}),
therefore estimating it at the LHC, \emph{i.e.} in the presence of cuts, is extremely important to constrain possible modification of the EW-symmetry-breaking mechanism due to new physics. This has to be done properly to avoid having spurious, unphysical effects. 
It can be appreciated in \Cref{tab:A0_check} how the standard analytic approach would lead to a longitudinal fraction of about 45\% in the presence of ATLAS fiducial selections. The effect of neutrino reconstruction is minor but non-negligible as well.
While a sound extraction may still be possible, though with large systematics, through extrapolation from the fiducial to the inclusive phase space, a better strategy is represented by the so-called polarisation-template method 
\cite{Ballestrero:2017bxn,BuarqueFranzosi:2019boy,Ballestrero:2019qoy,Ballestrero:2020qgv,Denner:2020bcz,Denner:2020eck,Poncelet:2021jmj,Le:2022lrp,Le:2022ppa,Denner:2023ehn,Dao:2023pkl,Hoppe:2023uux,Pelliccioli:2023zpd}.

\subsubsection{Angular distributions: inclusive and fiducial}
To extract more information than that obtained from the inclusive angular coefficients and polarisation fractions, we now explore the full angular distributions in the azimuthal and polar angles, both at the inclusive and fiducial (reconstruction) levels. 

\begin{figure}[h]
  \centering
  \subfigure[Inclusive, truth\label{fig:Tp_1}]{\includegraphics[scale=0.37,page=2]{fig/Wdecay_azimuth_pwg.pdf}}
  \subfigure[ATLAS fiducial, reconstructed\label{fig:Tp_4}]{\includegraphics[scale=0.37,page=1]{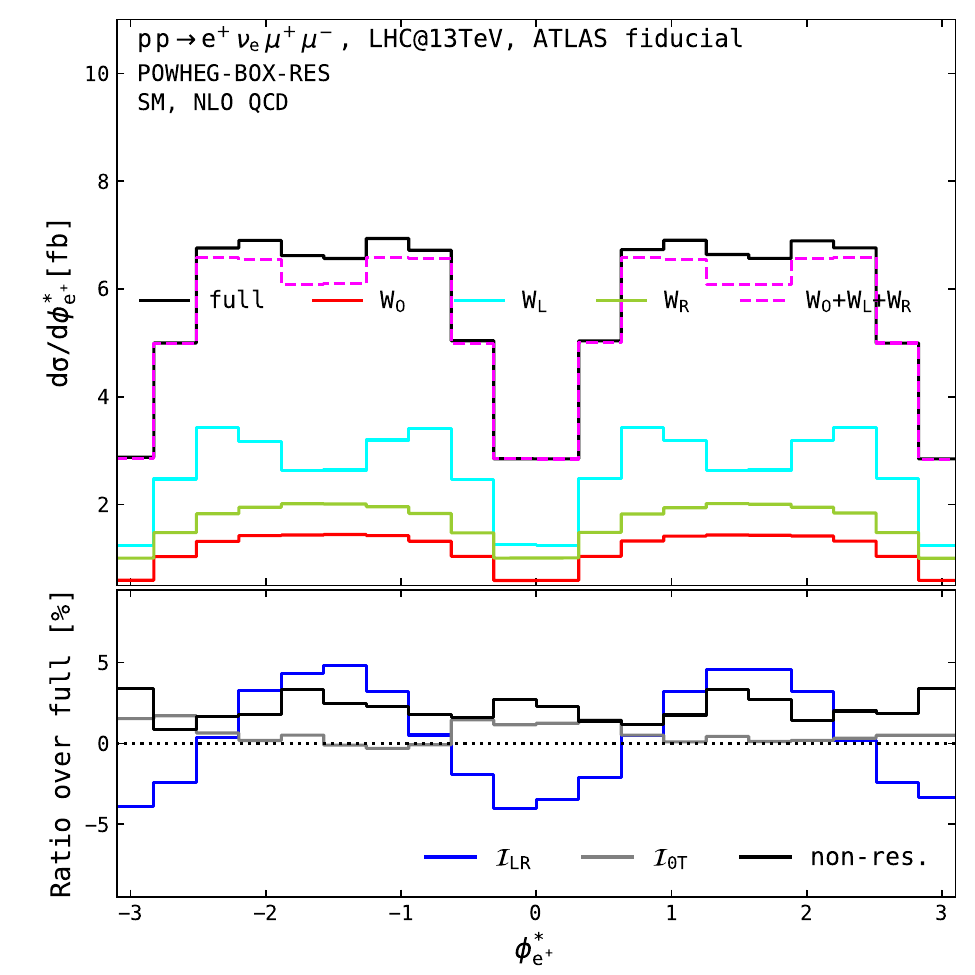}}
  \caption{Distributions in the azimuthal decay angle of the positron in the W-boson rest frame, for a longitudinal (0), left-handed (L) and right-handed (R) W boson, in W$^+$Z production at the LHC@13TeV.  The results are shown for an inclusive setup at the MC-truth level (left) and for the ATLAS fiducial setup after neutrino reconstruction (right).
  Polarised-boson signals have been computed in the SM at NLO QCD accuracy with \powheg \cite{Chiesa:2020ttl,Pelliccioli:2023zpd}.\label{fig:TransvCPhi}}
\end{figure}

We start by considering the distribution of the azimuthal decay angle of the positron in the W$^+$ rest frame ($\phi^*_{\rm e^+}$) in the SM. A sizeable left-right interference characterises this distribution. This can be appreciated in \Cref{fig:TransvCPhi}, where the SM distributions are shown. 
Simulating polarised signals at NLO QCD in the double-pole approximation with \powheg \cite{Pelliccioli:2023zpd,Chiesa:2020ttl}, the incoherent sum of left- and right-handed modes gives a flat distribution in $\phi^*_{\rm e^+}$. In contrast, the transverse one (coherent sum of the same modes) features a modulation proportional to $\cos2\phi^*_{\rm e^+}$, highlighting a sizeable contribution from left-right interference. Although the application of lepton cuts and neutrino reconstruction diminishes the sensitivity to this interference term, the effect is still accessible also with realistic setups (see \Cref{fig:Tp_4}). The left-right interference retains its $\cos2\phi^*_{\rm e^+}$ modulation shape, contributing up to 5\% of the total distribution.

\begin{figure}
  \centering
  \subfigure[Inclusive, truth\label{fig:pth_1}]{\includegraphics[scale=0.37]{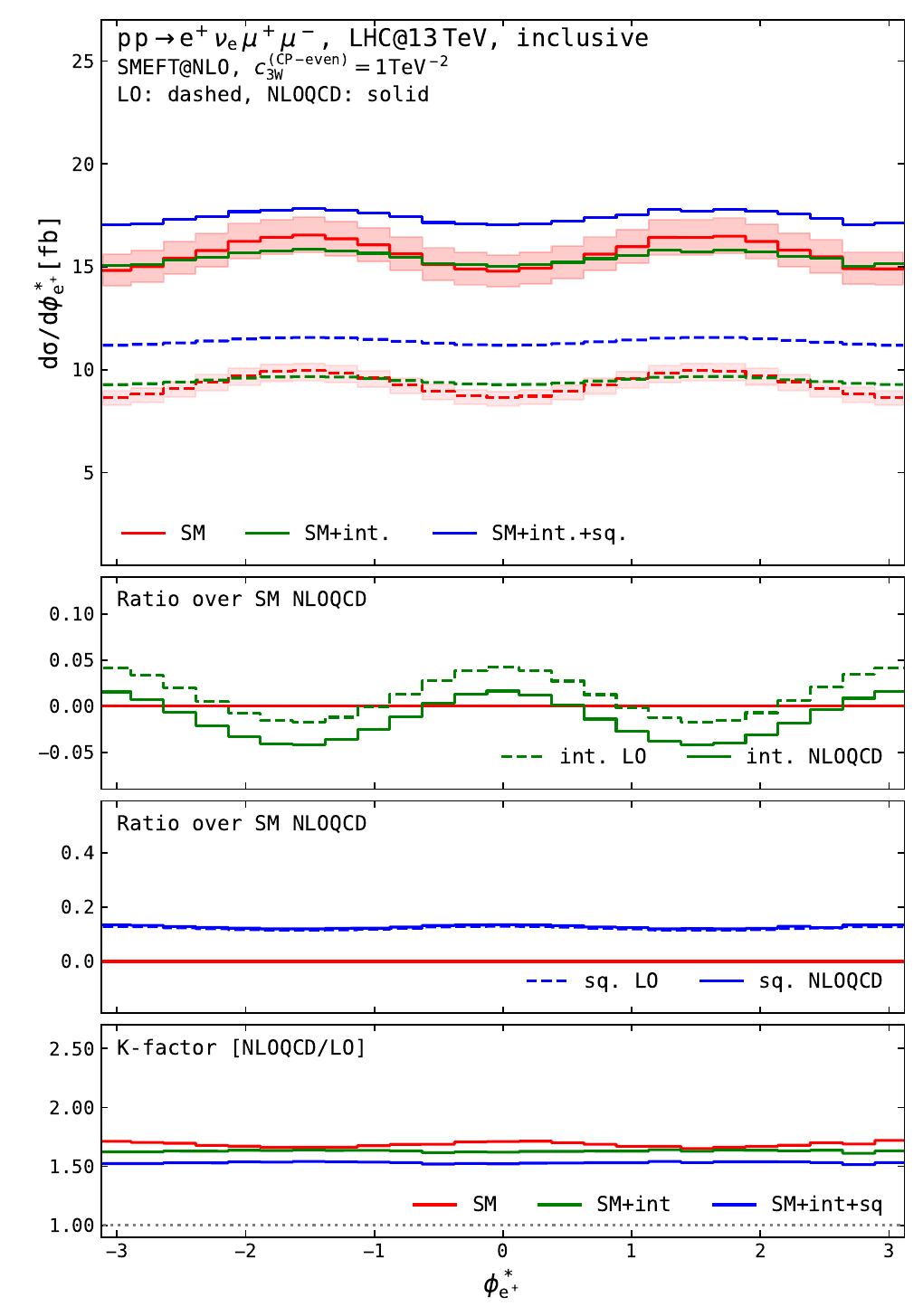}}
  \subfigure[ATLAS fiducial, reconstructed\label{fig:pth_4}]{\includegraphics[scale=0.37]{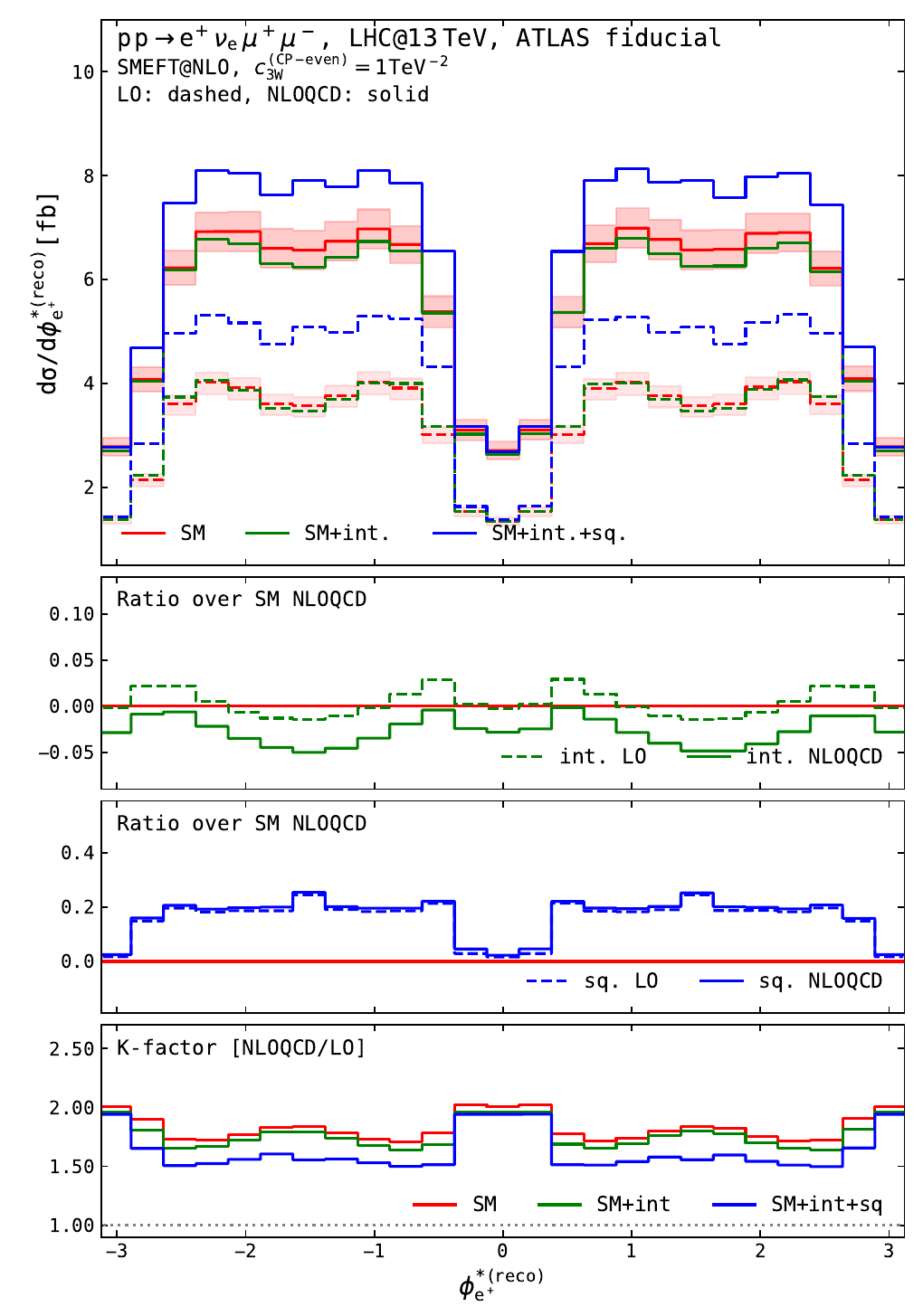}}
  \subfigure[Inclusive, truth\label{fig:PhiPos_CPodd_1}]{\includegraphics[scale=0.37,page=2]{fig/Wdec_cpodd_phi_st_ep.pdf}}
  \subfigure[ATLAS fiducial, reconstructed\label{fig:PhiPos_CPodd_2}]{\includegraphics[scale=0.37,page=1]{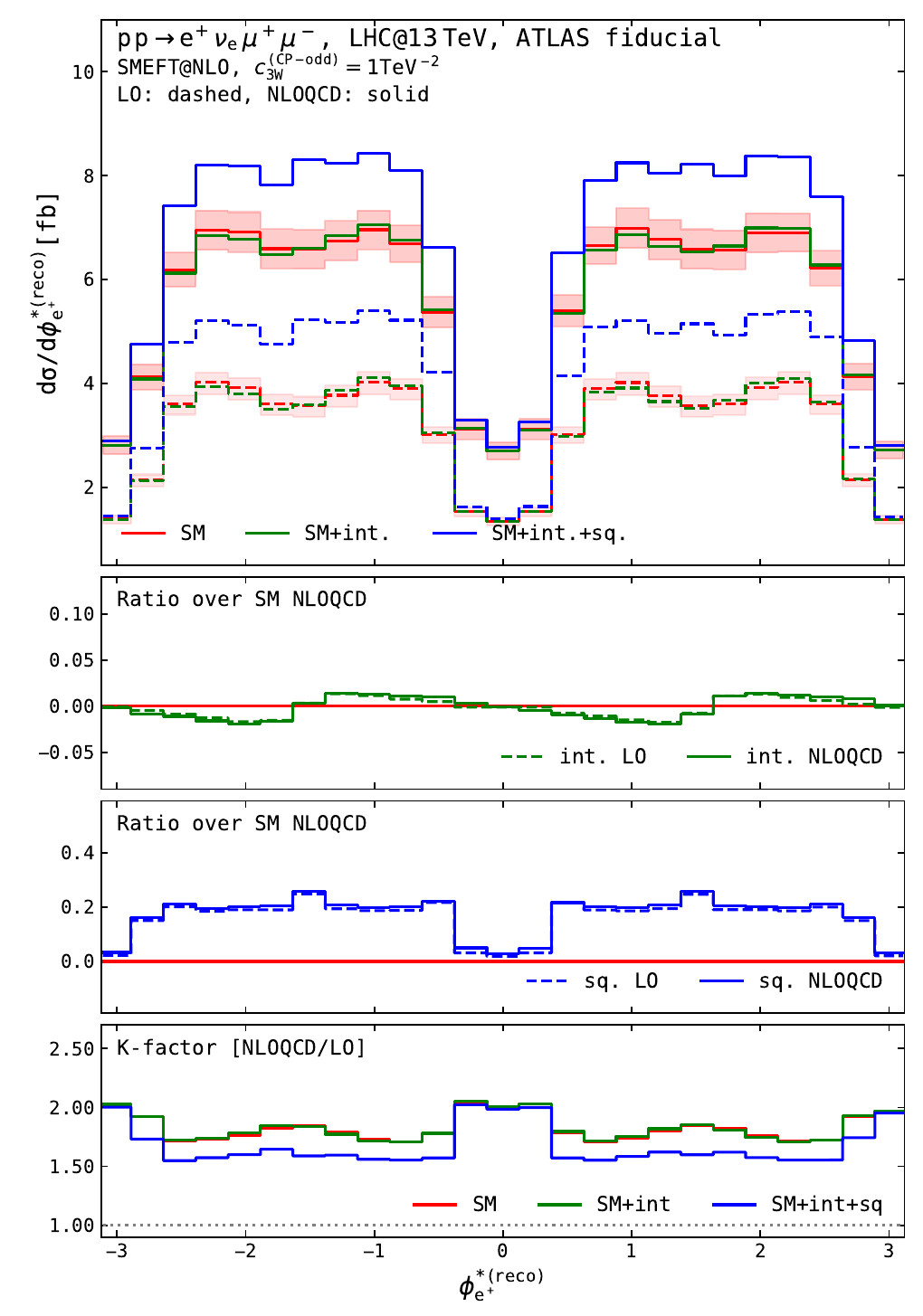}}
  \caption{Distributions in the azimuthal decay angle of the positron in the $\PW^+$ rest frame in W$^+$Z production at the LHC@13TeV, for the inclusive setup at MC-truth level (left) and for the fiducial ATLAS setup after neutrino reconstruction (right): effect of the inclusion of the CP-even operator ${O}_{3{W}}$ (a-b) and of the CP-odd one $O_{3\widetilde{W}}$ (c-d) with WCs set to $1{\rm TeV}^{-2}$. Same structure as \Cref{fig:m4lrec}.}\label{fig:Phie_CPev}
\end{figure}

The SMEFT distributions for the azimuthal decay angle of the positron in $\PW^+$ rest frame are shown in \Cref{fig:Phie_CPev}. We find that the ratio of the CP-even interference over the SM shows a modulation with the azimuthal angle both at LO and NLO. Interestingly, the modulation follows a cosine pattern for the CP-even coefficient ($\cos2\phi^*_{\Pe^+}$) and a sine ($\sin2\phi^*_{\Pe^+}$) one for the CP-odd as seen in the inclusive results. Therefore, these observables offer a way to distinguish linear CP-odd and CP-even contributions. Quadratic contributions follow the same shapes as the SM for CP-oven and CP-odd contributions.
Applying fiducial cuts has a non-trivial impact on the shapes of the distributions. The impact of the cuts affects the SM, interference and quadratic EFT contributions in different ways, resulting in different deviations from the SM. In particular, we notice that the fiducial cuts and neutrino reconstruction to the lepton kinematics lead to evident shape distortions compared to the inclusive MC-truth results.
Comparing \Cref{fig:pth_1,fig:PhiPos_CPodd_1} with \Cref{fig:pth_4,fig:PhiPos_CPodd_2}, we find that the fiducial selections (especially the transverse-momentum ones) deplete the distribution at the minima ($\phi^*_{\Pe}=0,\pm \pi$). At the same time, the neutrino reconstruction gives artificial spikes at $\phi^*_{\Pe^+}=\pm \pi/2$, which are clearly visible for the squared SMEFT term while still present but to a lesser extent in the SM and in the linear SMEFT term. In addition, the realistic effects change the sign of the NLO QCD linear SMEFT contribution at  $\phi^*_{\Pe^+}\approx 0, \pm \pi/2$, while a very similar shape to the inclusive one is found in the rest of the angular spectrum. 
While the CP-even operator only changes the amplitude of the SM $\cos2\phi^*_{\Pe^+}$ modulation, the linear contribution of the CP-odd operator introduces a sizeable modulation in $\sin2\phi^*_{\Pe^+}$ amounting to about $4\%$ (at the maxima and minima) of the central SM value at NLO QCD. The realistic effects in the ATLAS fiducial setup wash this modulation out, decreasing the modulation amplitude to $2\%$.
\begin{figure}
  \centering
  \subfigure[Inclusive, truth\label{fig:PhiMu_CPev_1}]{\includegraphics[scale=0.37]{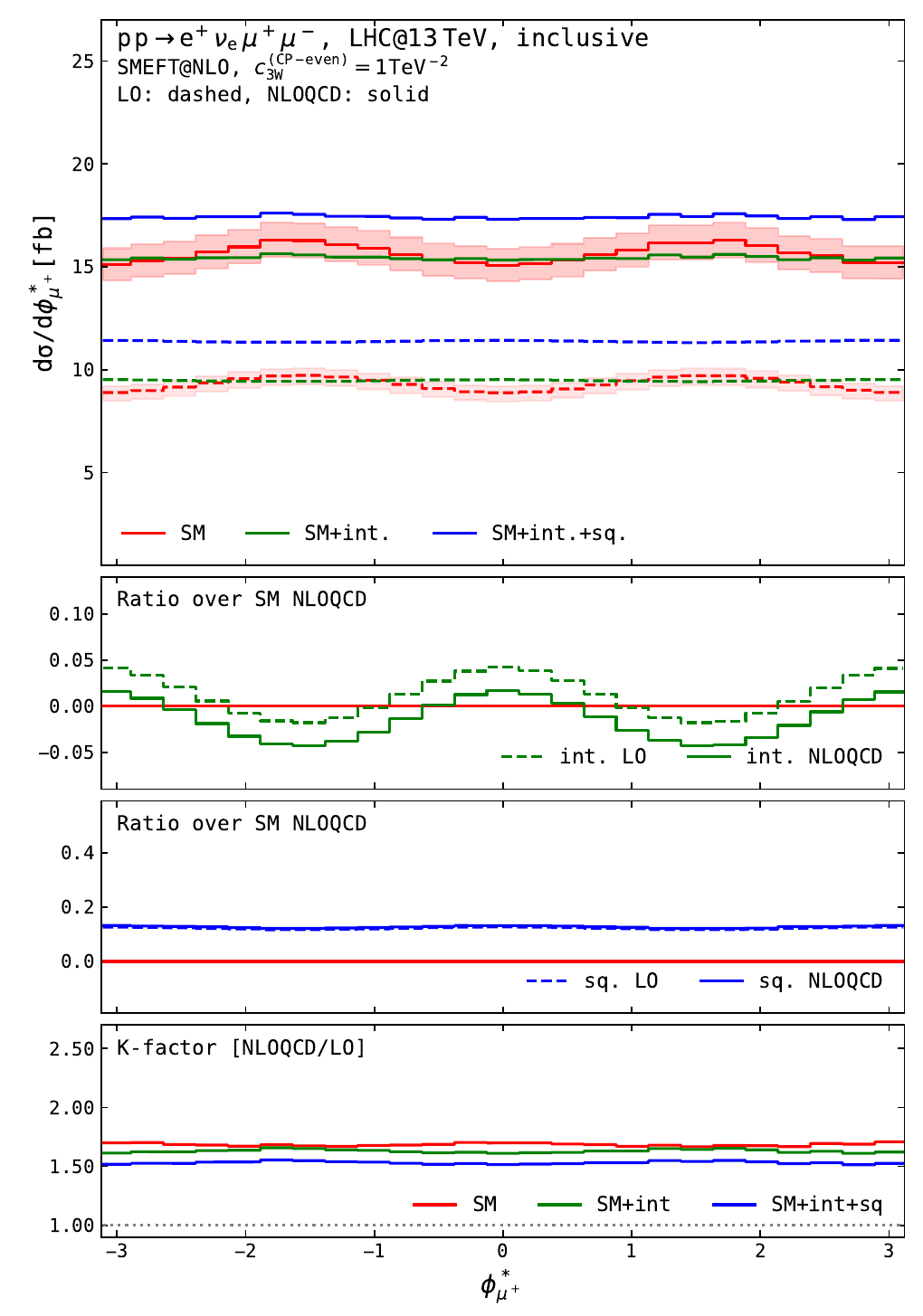}}
  \subfigure[ATLAS fiducial, reconstructed\label{fig:PhiMu_CPev_2}]{\includegraphics[scale=0.37]{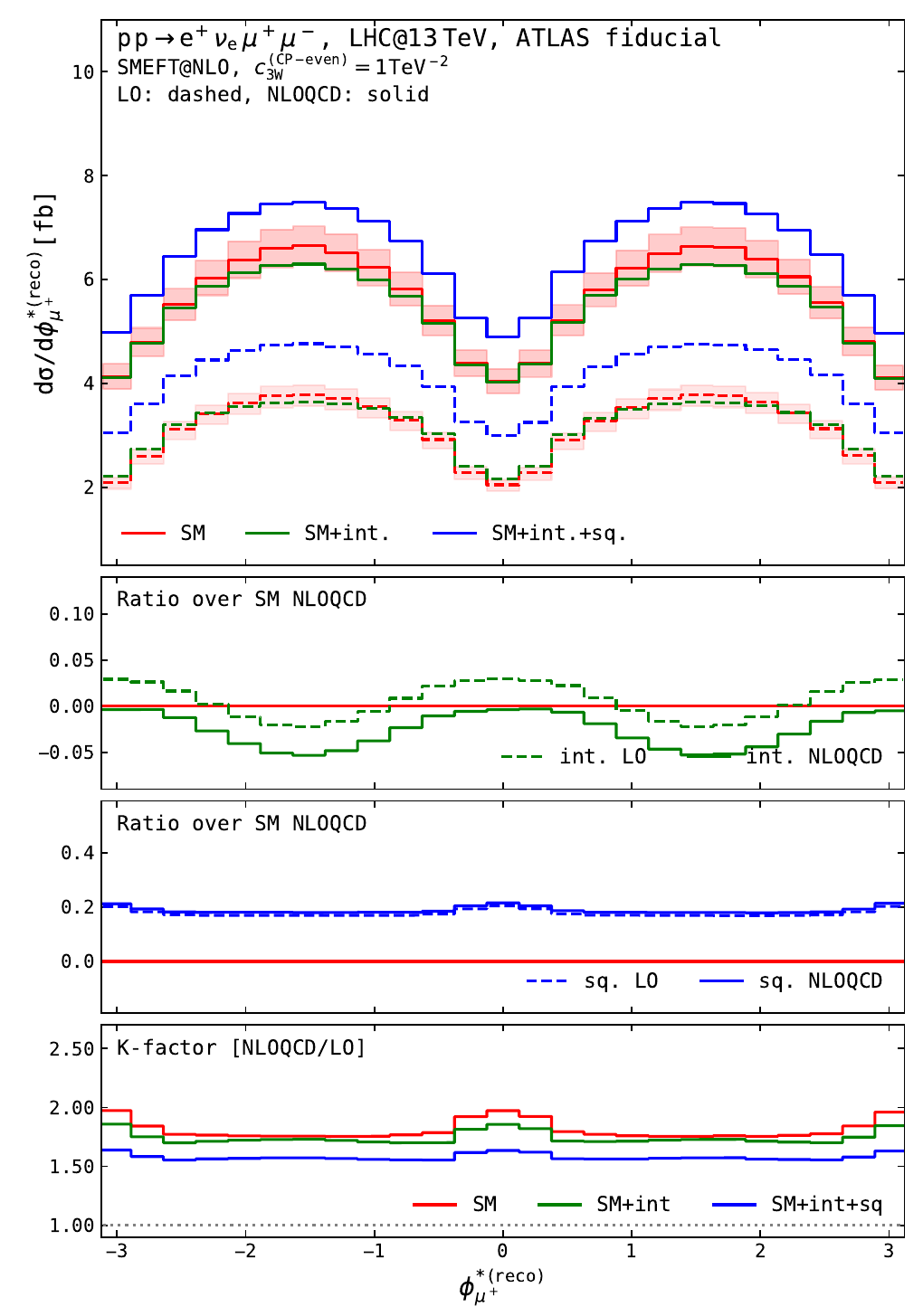}}
  \subfigure[Inclusive, truth\label{fig:PhiMu_CPodd_1}]{\includegraphics[scale=0.37,page=2]{fig/Wdec_cpodd_phi_st_mp.pdf}}
  \subfigure[ATLAS fiducial, reconstructed\label{fig:PhiMu_CPodd_2}]{\includegraphics[scale=0.37,page=1]{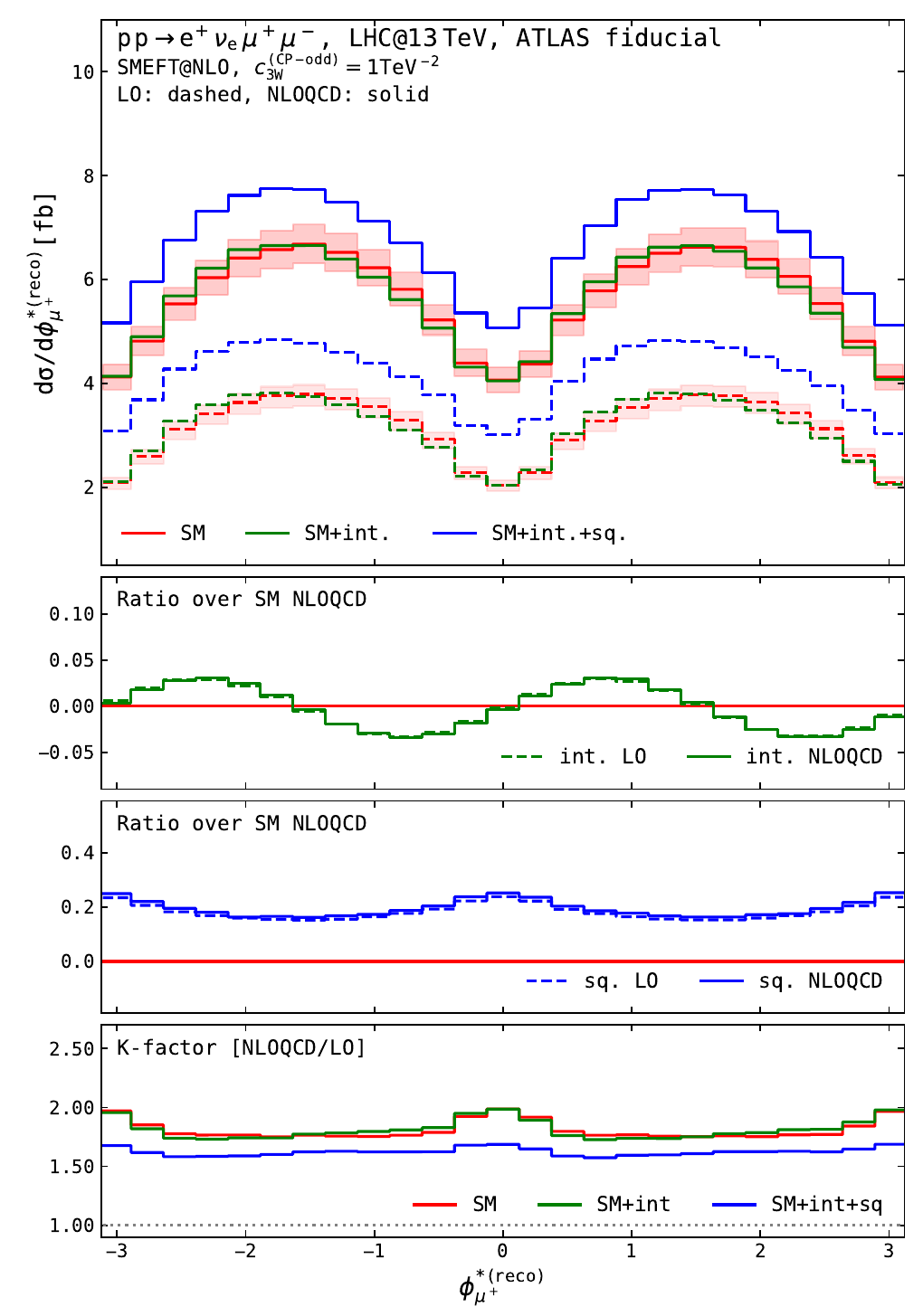}}
  \caption{Distributions in the azimuthal decay angle of the antimuon in the $\PZ$ rest frame in W$^+$Z production at the LHC@13TeV, for the inclusive setup at MC-truth level (left) and for the fiducial ATLAS setup after neutrino reconstruction (right): effect of the inclusion of the CP-even operator ${O}_{3{W}}$ (a-b) and of the CP-odd one $O_{3\widetilde{W}}$ (c-d), with WCs set to $1{\rm TeV}^{-2}$. Same structure as \Cref{fig:m4lrec}.\label{fig:PhiMu_CPev}}
\end{figure}
\begin{figure}
  \centering
  \subfigure[Inclusive, truth\label{fig:cth_1}]{\includegraphics[scale=0.37]{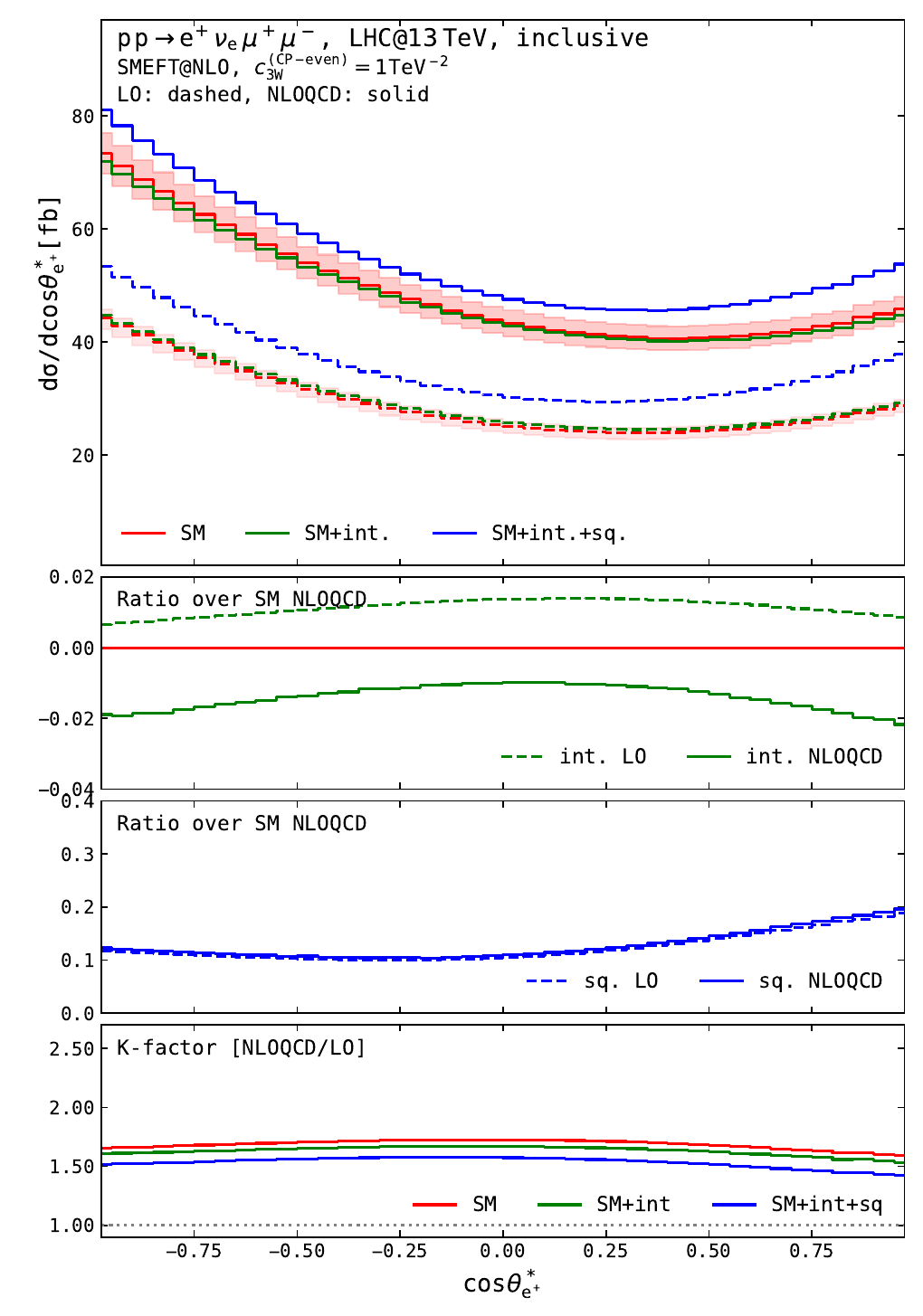}}
  \subfigure[ATLAS fiducial, reconstructed\label{fig:cth_4}]{\includegraphics[scale=0.37]{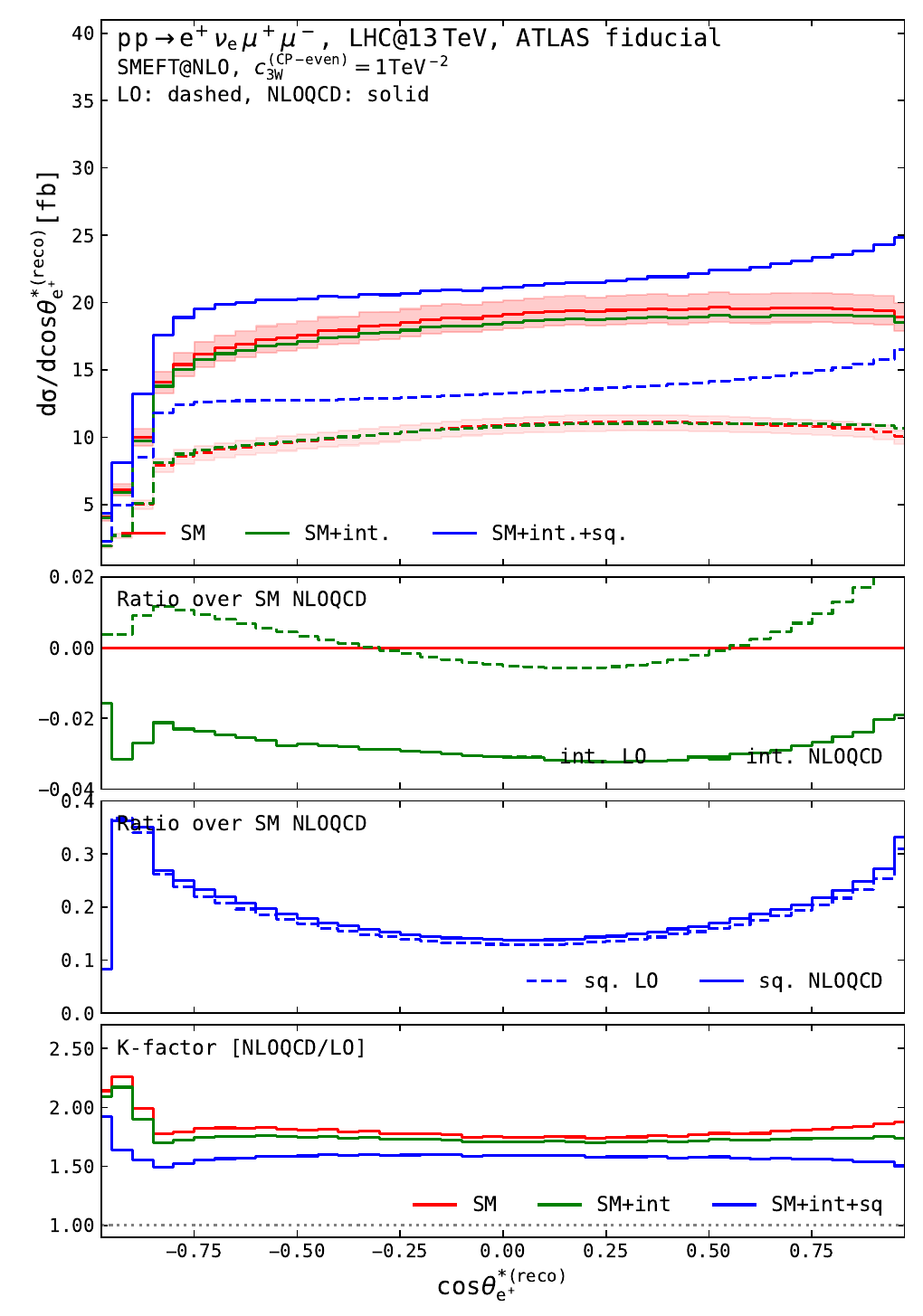}}
  \subfigure[Inclusive, truth\label{fig:cthODD_1}]{\includegraphics[scale=0.37]{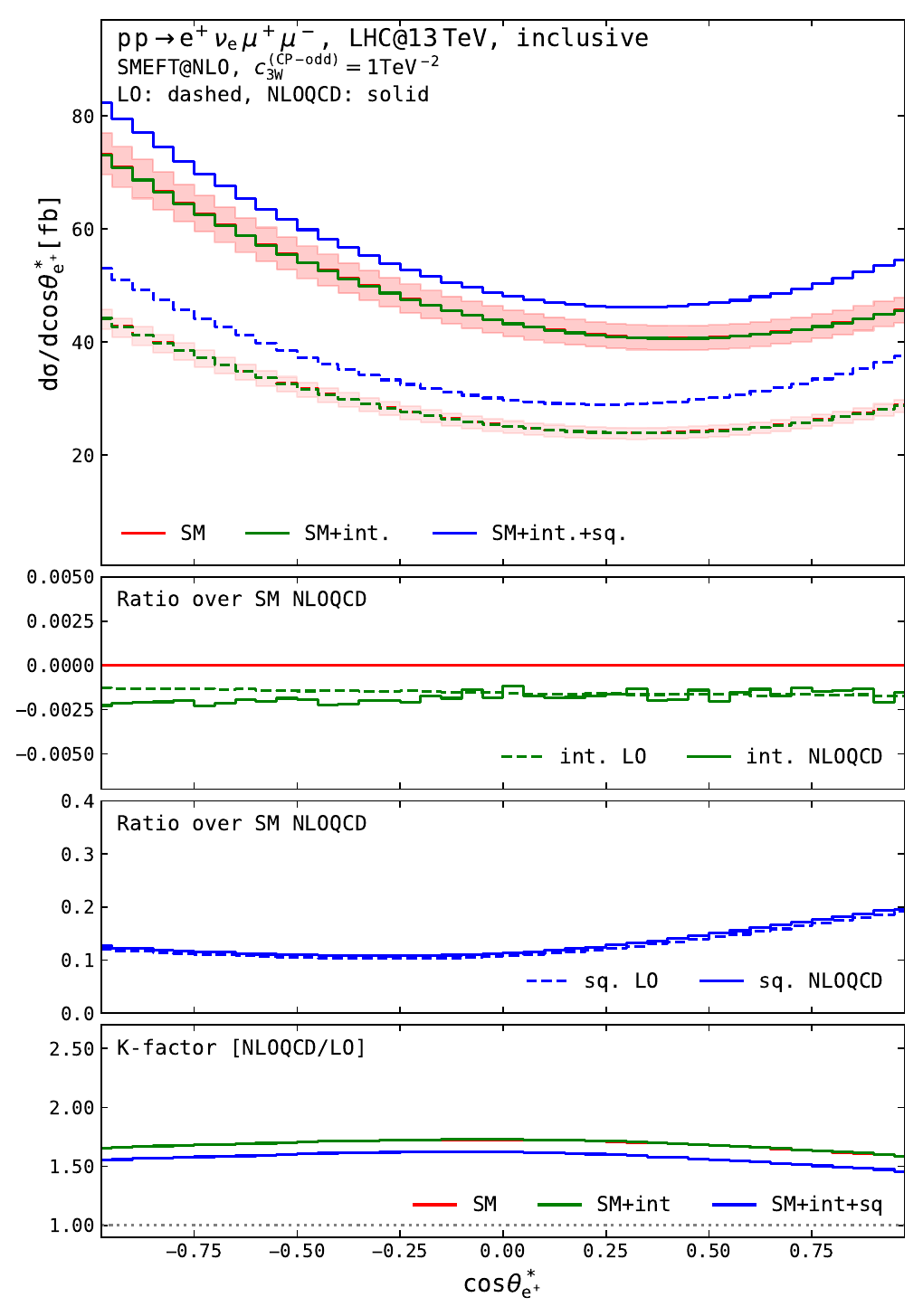}}
  \subfigure[ATLAS fiducial, reconstructed\label{fig:cthODD_4}]{\includegraphics[scale=0.37]{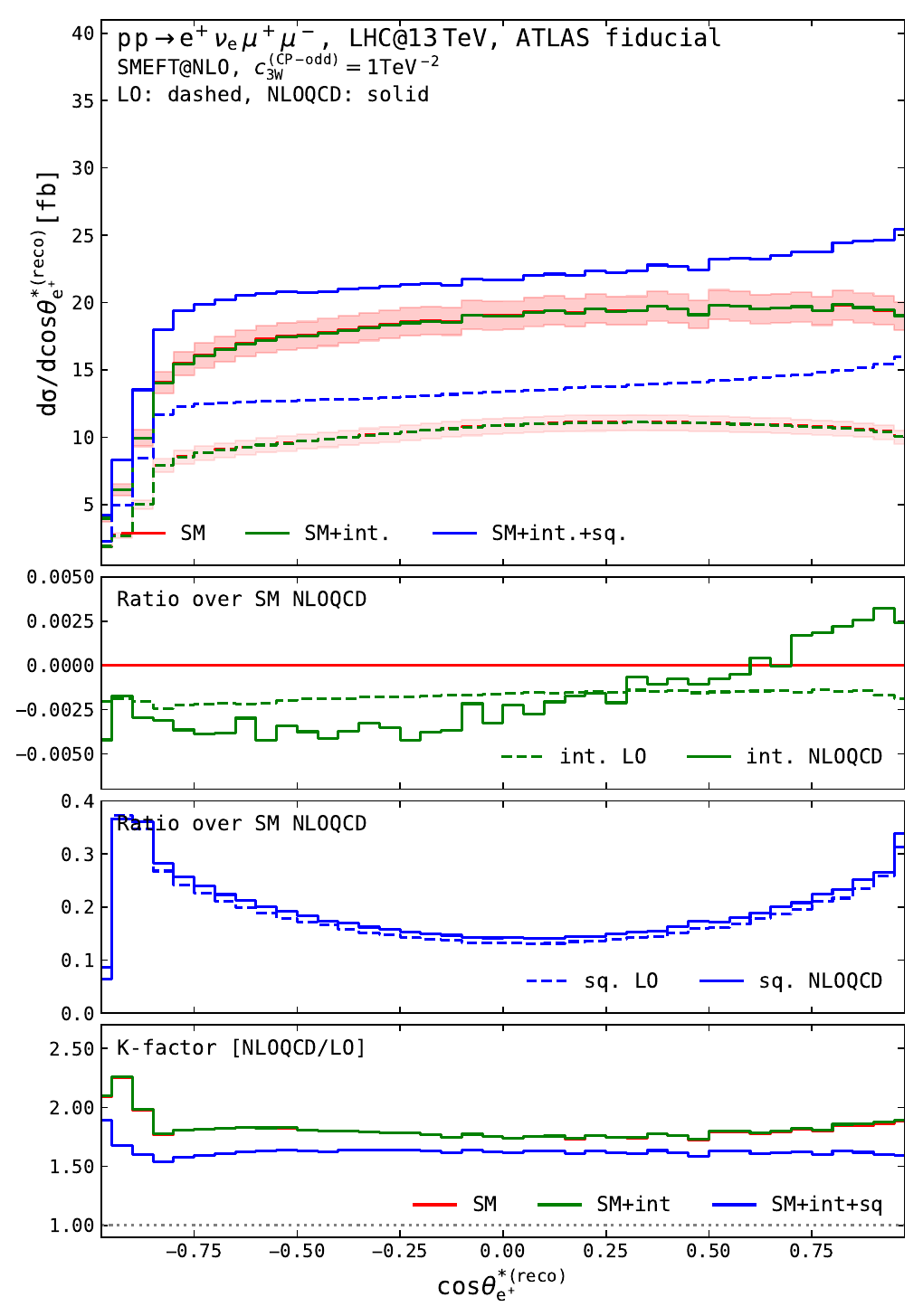}}
  \caption{Distributions in the polar decay angle of the positron in the $\PW^+$ rest frame in W$^+$Z production at the LHC@13TeV, for the inclusive setup at MC-truth level (left) and for the fiducial ATLAS setup after neutrino reconstruction (right): effect of the inclusion of the CP-even operator ${O}_{3{W}}$ (a-b) and of the CP-odd one $O_{3\widetilde{W}}$ (c-d), with WCs set to $1{\rm TeV}^{-2}$. Same structure as \Cref{fig:m4lrec}. 
  }\label{fig:cth}
\end{figure}

In \Cref{fig:PhiMu_CPev}, we consider the distribution of the azimuthal decay angle of the antimuon in the $\PZ$ rest frame in W$^+$Z, which is expected to be less affected by realistic cuts than the corresponding angle in the W-boson decay, since the neutrino reconstruction only affects the reconstruction of the diboson CM frame in this case. Regarding the CP-even effects relative to the SM, the situation is similar to the one found for $\phi^*_{\Pe^+}$ in the inclusive setup. In the fiducial setup, the artificial effect introduced by neutrino reconstruction at $\phi^*_{\Pe} \approx \pm\pi/2$ is now absent for $\phi^*_\mu$, while the effect of the transverse-momentum cuts is very similar to the one found for $\phi^*_{\Pe}$. The linear CP-even SMEFT term features the same dominant $\cos2\phi^{*}_\mu$ modulation, although with an opposite sign compared to the SM (in agreement with the decrease in the $A_2$ coefficient shown in \Cref{tab:Acoeff_phi}). 
This azimuthal observable is more interesting when looking into the linear contribution of the CP-odd operator.
The same $\sin2\phi^*$ modulation, already observed in \Cref{fig:Phie_CPev}, is also present in the Z-boson decay. At variance with the W-boson case, the contribution of the linear term relatively to the SM preserves the same shape and size when applying fiducial cuts and neutrino reconstruction. 
As expected from previous literature results \cite{Azatov:2017kzw,Panico:2017frx,Franceschini:2017xkh,Azatov:2019xxn,Banerjee:2019twi,Banerjee:2020vtm}, azimuthal-angle variables provide an optimal discrimination power to spot CP-odd effects. Our findings show that extracting such CP-odd effects from the $\phi^*_{\mu}$ distribution can be performed at the LHC in a rather cut-independent manner as the deviation from the SM persists in the fiducial cut setup.
The squared SMEFT contributions give a relatively flat enhancement to the SM distribution in the inclusive setup; both CP-even and CP-odd operators introduce a slight change in shape in the fiducial setup.
We also observe that K-factors are different between the SM and EFT contributions, and whilst flat in the inclusive contributions, they acquire a non-trivial dependence on $\phi^*_\mu$ once fiducial cuts are applied. 

The final distribution we consider in the $\PW\PZ$ process is the polar decay angle of the positron in the $\PW^+$ rest frame in \Cref{fig:cth}. We find that fiducial cuts remove events in the region of $\theta^*_{e^+}\sim \pi$. NLO corrections change the normalisation of the interference contribution, rendering it negative over the whole range of angles in the fiducial setup. The squared contribution dominates in the region of $\theta^*_{e^+}\sim 0$ and $\pi$, but the deviation is not as large as achieved in the high $p_T$ or high $m_{4l}$ distribution tails.
The CP-odd linear effects on this observable, considered in \Cref{fig:cthODD_1,fig:cthODD_4} are at the 0.2\% level, as already shown in the integrated cross-sections (see \Cref{tab:sigmasWZ}). While in the inclusive setup, its impact is flat over the whole angular range, in the fiducial setup, the CP-odd interference gives a slight distortion to the SM shape, 
driven mainly by the neutrino-reconstruction procedure.
The squared CP-odd effects are very similar to the CP-even ones discussed above.

\subsection{Dynamical-scale and off-shell effects}\label{sec:dyn_sect}
In the SM \cite{Grazzini:2019jkl}, a fixed-scale choice for inclusive diboson production may lead to shape distortions in the tails of energy-dependent distributions. 
Although the fixed-scale choice is well motivated for this work (see \Cref{sec:input}), we compare our default fixed-scale results with those relying on the following dynamical scale definition,
\beq\label{eq:dyndef}
\mu_0^{\rm dyn}\,=\, \sum_{i \in {\rm FS}} \frac{E_{\rm T,i}}2\,
\eeq where the sum runs over all final state particles, and $E_{\rm T,i}$ is the transverse energy of the final-state particle $i$. 
This choice leads to mildly smaller QCD corrections ($+65\%$) at the level of the fiducial cross-section compared to the fixed-scale case ($+73\%$), in the case where the linear contribution from the ${O}_{3{W}}$ operator is included. When the corresponding quadratic term is included, the dynamical and fixed scale give very similar QCD corrections ($+57\%$ and $+55\%$, respectively).

\begin{figure}
  \centering
  \includegraphics[scale=0.43]{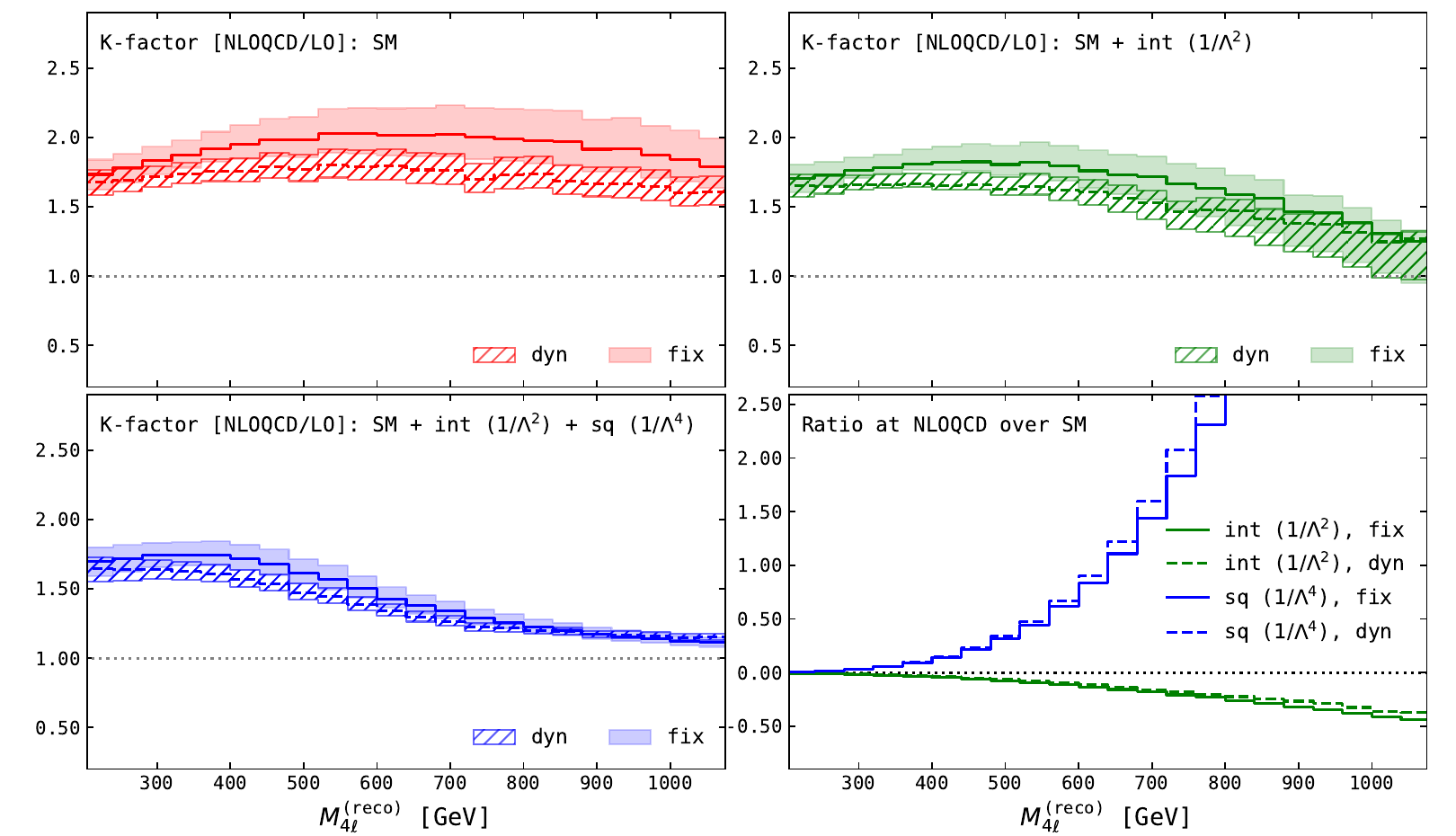}
  \caption{
  Differential K-factors and ratio plots in the reconstructed four-lepton invariant mass for off-shell W$^+$Z production at the LHC@13TeV in the fiducial ATLAS setup \cite{ATLAS:2022oge}. 
  The default fixed scale (``fix'', solid curves) of Eq.~\ref{eq:fixedscale} and the dynamical one (``dyn'', dashed curves) defined in Eq.~\ref{eq:dyndef} are compared. 
  NLO QCD K-factors are shown for the SM (red, top-left panel), including the linear SMEFT term (green, bottom-left panel), including the linear and squared SMEFT term (blue, top-right panel). 
  The ratios at NLO QCD of the linear (quadratic) SMEFT term over the SM are shown in green (blue) on the bottom-right panel.
  The CP-even operator ${O}_{3{W}}$ is considered, with its Wilson coefficient set to $1{\rm TeV}^{-2}$. 
  Uncertainty bands come from 9-point scale variations of the NLO QCD cross-section about the central scale, normalised to the central LO value. 
  }\label{fig:dynscale}
\end{figure}

Differential observables give a more complete picture of the scale-choice effects.
In \Cref{fig:dynscale}, we consider the reconstructed invariant mass of the diboson system and only focus on the effects of the ${O}_{3{W}}$ operator.
The QCD K-factors are flatter for the dynamical scale both in the SM and when including CP-even SMEFT effects. The choice of dynamical scale mildly reduces the QCD-scale uncertainties.
The contributions at NLO QCD of the linear and quadratic SMEFT terms relative to the SM are 
relatively independent of the scale choice, with some differences found in the distribution tail for the squared term. We note, however, that both scale choices agree within scale variations. 
 
It is also instructive to compare the full off-shell results with those obtained through the narrow-width approximation \cite{Artoisenet:2012st,BuarqueFranzosi:2019boy}. Notice that the narrow-width approximation is expected to reproduce off-shell calculations in kinematic regions which are dominated by resonant contributions, with discrepancies that are of order ${\mc O}(\Gamma_V/M_V)$ \cite{Denner:2000bj,Denner:2005fg,Aeppli:1993cb,Aeppli:1993rs,Uhlemann:2008pm}, \emph{i.e.} approximately 2\% in the case of EW bosons. 
The impact of the off-shell effects is found at the per cent level, both for integrated cross-sections and for most of the considered distributions. 
This shows that the modelling in the narrow-width approximation is not missing relevant physics effects, primarily because of the preservation of spin correlations \cite{Artoisenet:2012st,BuarqueFranzosi:2019boy}.
Somewhat larger off-shell effects would be found in more exclusive setups, in kinematic regions where non-resonant contributions are not any more suppressed compared to the dominant double-resonant topologies \cite{Biedermann:2017oae,Denner:2020bcz}. 
It is worth noticing that the operators considered in this work (${O}_{3{W}}$ and $O_{3\widetilde{W}}$) can only affect diagrams with the triple-gauge coupling, which are double-resonant diagrams. Therefore, non-trivial deviations from the SM can only be due to the interference between SM non-resonant diagrams and SMEFT resonant diagrams at dimension-six. 
Larger off-shell effects could be found in the presence of SMEFT operators involving fermions \cite{Helset:2017mlf}.

\section{WW  production}\label{sec:WW}
Compared to W$^\pm$Z, the W$^+$W$^-$ process is characterised by a larger production rate but features challenging complications, especially in the fully leptonic decay channel. First, the large top-quark backgrounds leading to the same final-state signature render separating the W$^+$W$^-$ signal rather intricate. Second, the presence of two neutrinos dramatically reduces the number of differential observables that can be extracted from the data, preventing access to a single-W rest frame.  
It is important to notice that $\PW^\pm\PZ$ and $\PW^+\PW^-$ share the same gauge structure at tree level in the SM, with triple-gauge-coupling contributions in $s$-channel diagrams, together with additional $t/u$-channel diagrams only involving gauge couplings of EW bosons to fermions. This motivates a number of SMEFT studies of $\PW^+\PW^-$ in combination with $\PW^\pm\PZ$ production \cite{Chiesa:2018lcs,Baglio:2019uty,Baglio:2020oqu}.

\subsection{Setup}\label{sec:WWsetup}
We consider three setups for the W$^+$W$^-$ process. 
The first one is fully inclusive, with no cut applied to the leptons and without any jet veto.
The second one is also inclusive of the lepton kinematics but features the jet vetoes applied in the ATLAS analysis of Ref.~\cite{ATLAS:2019rob}:
\begin{itemize}
    \item no b-tagged jets with $p_{\rm T,b}>20\GeV$ and $|\eta_{\rm b}|<2.5$\,,
    \item no light jets with $p_{\rm T,j}>35\GeV$ and $|\eta_{\rm j}|<4.5$\,,
\end{itemize}
which reduce the contamination from top-quark and QCD multi-jet backgrounds.
The third setup features the complete set of fiducial selections of Ref.~\cite{ATLAS:2019rob}, including the jet vetoes described above and the following cuts:
\begin{align}
 & p_{\rm T,\,\ell}>27\GeV\,,\quad p_{\rm T, miss}>20\GeV\,,\quad p_{\rm T, \Pe^+\mu^-}>30\GeV\nonumber\\ 
 &|y_{\mu^-}|<2.5\,,\qquad |y_{\Pe^+}|<2.47 \,,\qquad M_{\Pe^+\mu^-}>55\GeV \,.
\end{align}
The presence of two neutrinos prevents reconstructing individual $\PW$ bosons with standard techniques. Therefore, we limit our phenomenological analysis to differential observables accessible in the laboratory without any reconstruction methods.

\subsection{Selection-cut effects}
The potential presence of b-tagged jets in the final state at NLO in QCD leads to sizeable contamination of the $\PW^+\PW^-$ EW signal from the single-top background. This situation becomes even more severe at NNLO QCD, owing to the considerable production rate of top-antitop pairs at the LHC. 
The application of jet vetoes helps reduce the impact of such backgrounds. 
It is worth recalling that in this work, we only focus on dimension-six triple-gauge coupling operators; therefore, the single-top channels ($\rm g b \rightarrow t \PW^-,\,\rm g \bar{\rm b} \rightarrow \bar{\rm t} \PW^+,\,$) that open up at NLO QCD are not affected by the considered SMEFT effects.
The effect of jet vetoes and the interplay with anomalous couplings of weak bosons to fermions in $\PW^+\PW^-$ are known in the literature \cite{Baglio:2018bkm,Baglio:2019uty}.

In \Cref{tab:sigmasWW}, we show the integrated cross-sections for the SM and SMEFT contributions in the three setups described in \Cref{sec:WWsetup}, from the fully inclusive one (LO and NLO) to the ATLAS fiducial one.
\begin{table}[t]
    \centering
    \begin{tabular}{ccccc}
\hline\\[-0.5cm]
  accuracy                 &   $\sigma_{\rm LO}$             &   $\sigma_{\rm NLO}$        &   $\sigma_{\rm NLO}$        &   $\sigma_{\rm NLO}$       \\[0.1cm]
\hline\\[-0.4cm]
  setup                &  fully inclusive &  fully inclusive &  jet-vetoes only & ATLAS fiducial   \\[0.1cm]
\hline\\[-0.4cm]
 SM          &    $ 876.1(1)  ^{+ 5.0 \%} _{ -6.2 \%}$  &   $    1895.1(5) ^{+ 11 \%} _{ -11 \%} $   & $1004.5(3)  ^{+ 3.5 \%} _{ -4.4 \%}$& $187.8(2)  ^{+ 3.6 \%} _{ -4.8 \%}$\\
 CP-even int &    $ 2.29(1)  ^{+ 2.7 \%} _{ -3.6 \%}  $  &   $    -8.57(6)  ^{+ 17 \%} _{ -15 \%}   $   & $-0.43(2)  _{-66 \%} ^{ +76 \%}$& $-0.74(4)  _{-12 \%} ^{ +14\%}$\\
 CP-even sq  &    $ 38.20(5)  ^{+ 5.0 \%} _{ -4.4 \%} $  &   $   40.77(9)  ^{+ 1.3 \%} _{ -1.4 \%}      $   & $17.8(2)  ^{+ 19 \%} _{ -27 \%}$& $10.4(1)  ^{+ 27 \%} _{ -39 \%}$\\
 CP-odd int &    $ 0.04(5)  ^{+ 1.2 \%} _{ -2.0 \%}  $  &   $     0.01(5)  ^{+ 29\%} _{ -40\%}   $   & $ -0.10(9)  _{-25 \%} ^{+ 35 \%}$& $0.02(1)  ^{+ 20 \%} _{ -15 \%} $\\
 CP-odd sq  &    $ 21.60(8)  ^{+ 4.5 \%} _{ -4.0 \%} $  &   $   23.2(1)  ^{+ 1.2 \%} _{ -1.3 \%}      $   & $10.3(3)  ^{+ 18 \%} _{ -26 \%} $& $6.0(1) ^{+ 26 \%} _{ -38 \%}$\\[0.1cm]
\hline
\end{tabular}
    \caption{Integrated cross-sections (in fb) for the SM and dimension-six SMEFT contributions in $\PW^+\PW^-$ production at the LHC@13TeV. Three setups are used: fully inclusive (LO and NLO), inclusive with jet vetoes only, and ATLAS fiducial. The uncertainties from 9-point QCD-scale variations are shown in percentages. The MC uncertainties of central values are shown in parentheses.}
    \label{tab:sigmasWW}
\end{table}
In all setups, despite small differences, the linear SMEFT contribution of the CP-even operator is at the sub-percent level relative to the SM cross-section. 
The relative impact of squared dimension-six contributions ranges between 1\% and 5\% of the SM, with the size of CP-odd contributions roughly half of the corresponding CP-even ones.

{It is worth noting that the CP-odd EFT--SM interference is expected to vanish when considering the $2 \to 2$ scattering $q\bar{q}\rightarrow\PW^+\PW^-$, owing to its invariance under charge conjugation. This aspect makes the initial and final states CP eigenstates, as a consequence of the Optical \cite{Eden:1966dnq,PhysRev.66.163} and CPT theorem \cite{Luders:1957bpq}.
These theorems relate the original amplitude to its time-reversed counterpart with a minus sign induced by the CP weak phase. The absence of interference in the $2 \to 2$ process is due to the exact cancellation stemming from different $\PW$-boson helicity configurations. However, these arguments do not hold for the complete $2 \to 4$ scattering considered here. We have verified that the cancellations above do not occur when decays are included. Nonetheless, the integrated results show that the CP-odd linear contributions remain extremely small and are compatible with zero within numerical uncertainties at both LO and NLO QCD levels. Similar observations have been reported in Refs.~\cite{Degrande:2023iob,Degrande:2021zpv}.
}

While we have computed integrated cross-sections in all three setups described in the previous sub-section, \emph{i.e.} with and without jet vetoes, in the following, we show differential results only for the two setups characterised by the application of realistic ATLAS jet vetoes.

\begin{figure}
  \centering
  \subfigure[Inclusive, jet vetoes only\label{fig:WWeven_mll_inc}]{\includegraphics[scale=0.43]{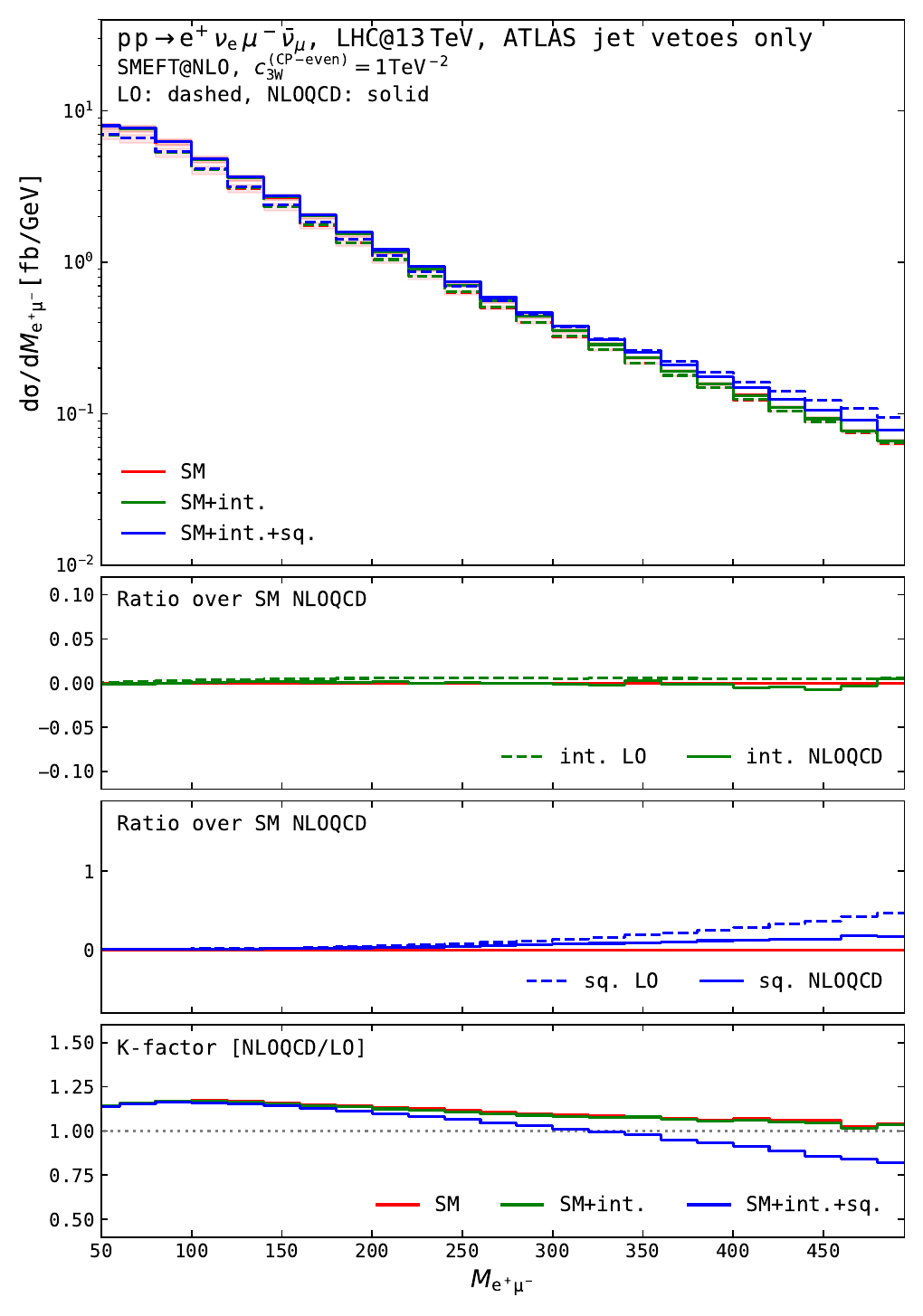}}
  \subfigure[ATLAS fiducial\label{fig:WWeven_mll_fid}]{\includegraphics[scale=0.43]{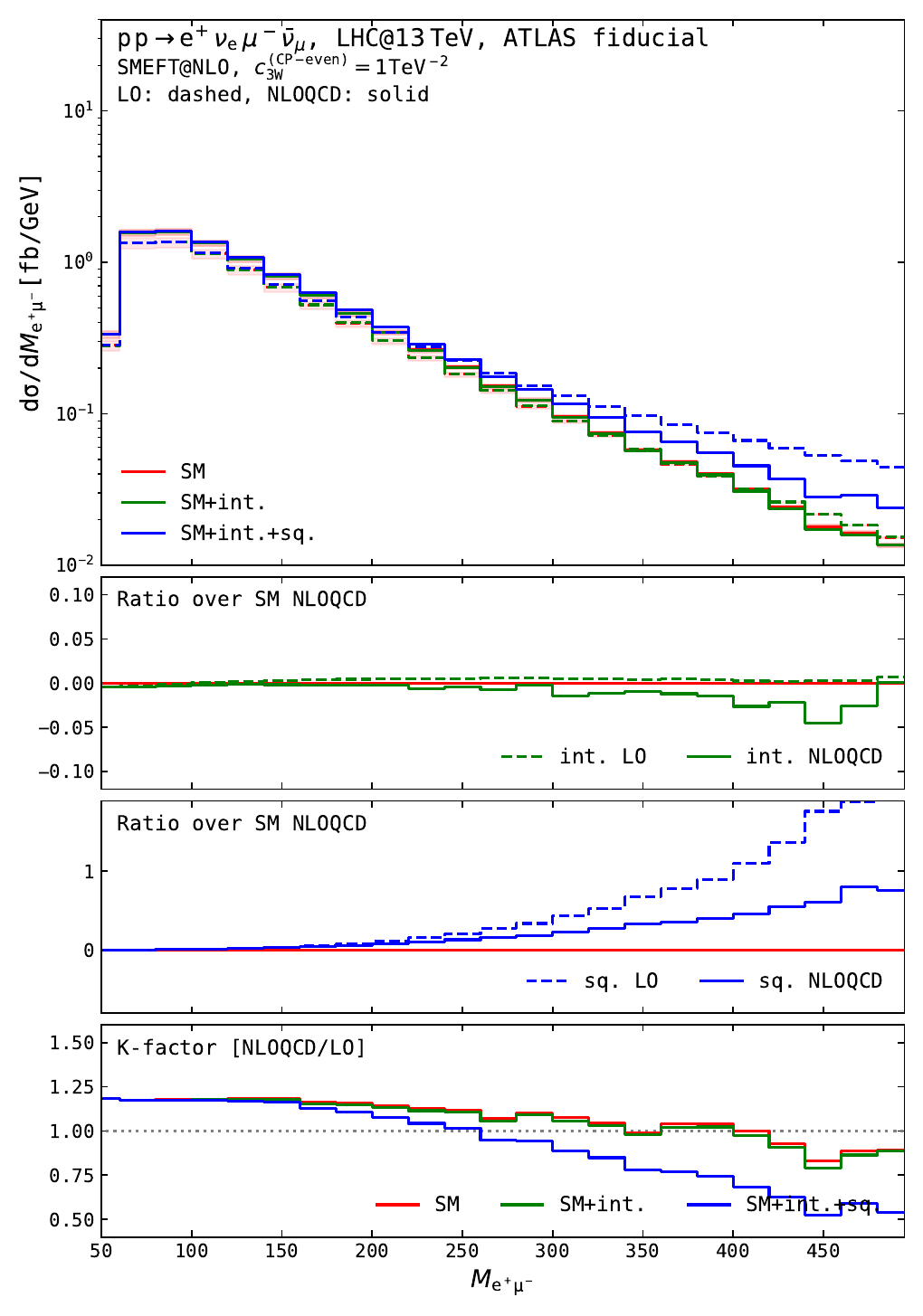}}
  \caption{Distributions in the invariant mass of the di-lepton system in $\PW^+\PW^-$ production at the LHC@13TeV, for the inclusive setup with jet vetoes only (left) and for the fiducial ATLAS setup (right): effect of the inclusion of the CP-even operator ${O}_{3{W}}$ with Wilson coefficient set to $1{\rm TeV}^{-2}$. Same structure as \Cref{fig:m4lrec}.
  }\label{fig:WWeven_mll}
\end{figure}

\begin{figure}
  \centering
  \subfigure[Inclusive, jet vetoes only\label{fig:WWeven_dphi_inc}]{\includegraphics[scale=0.43]{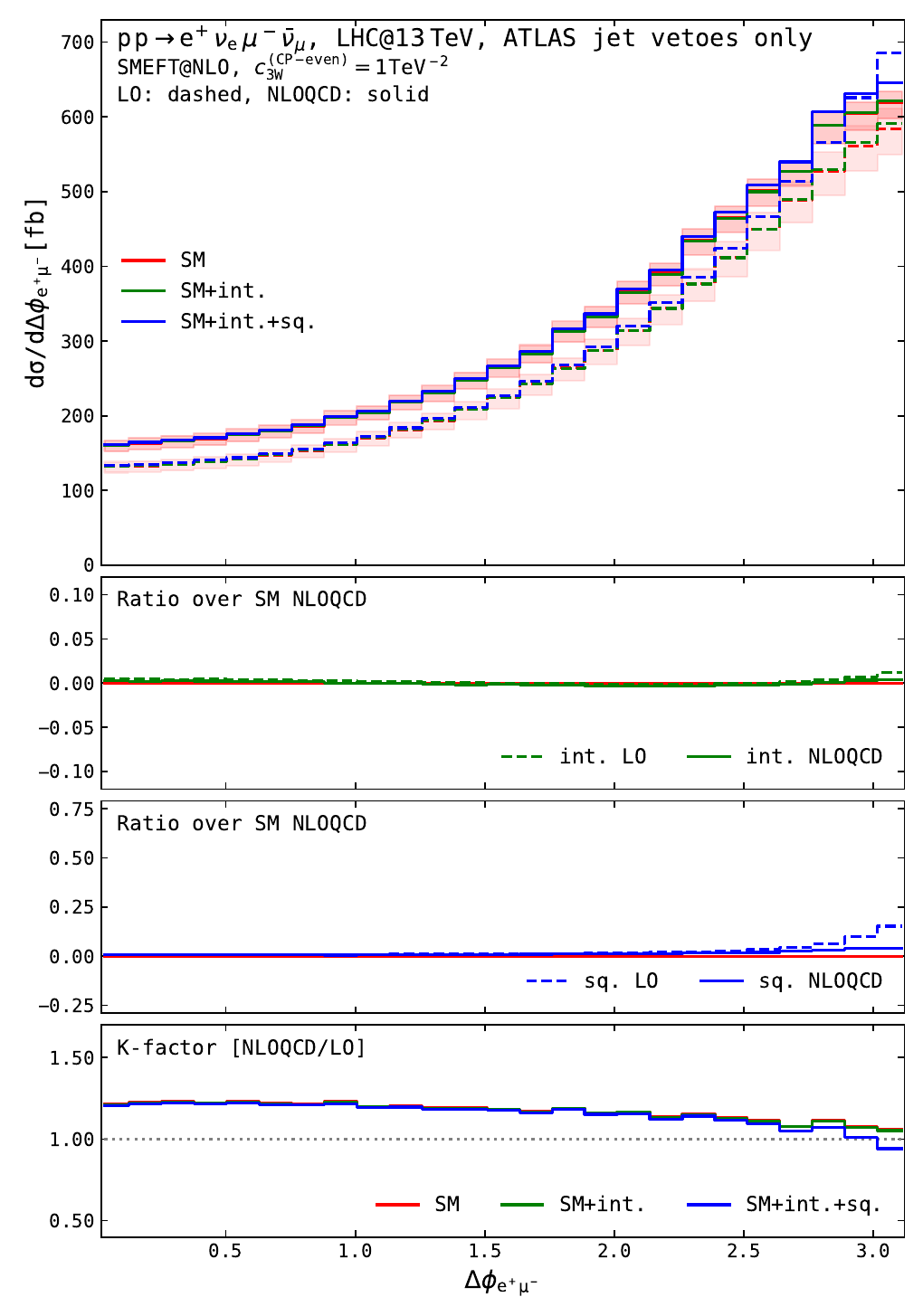}}
  \subfigure[ATLAS fiducial\label{fig:WWeven_dphi_fid}]{\includegraphics[scale=0.43]{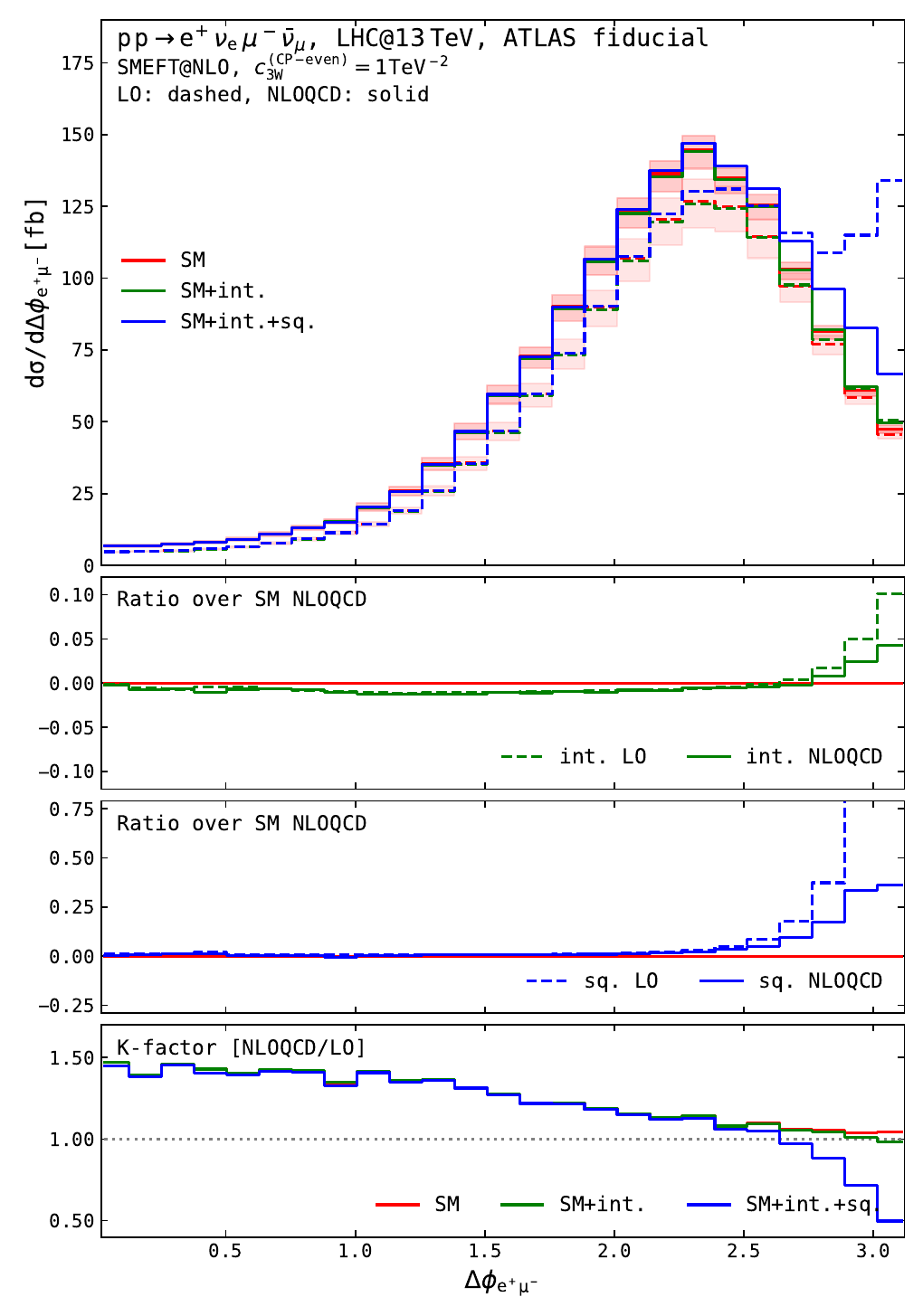}}
  \caption{Distributions in the azimuthal distance between the charged leptons in $\PW^+\PW^-$ production at the LHC@13TeV, for the inclusive setup with jet vetoes only (left) and for the fiducial ATLAS setup (right): effect of the inclusion of the CP-even operator ${O}_{3{W}}$ with Wilson coefficient set to $1{\rm TeV}^{-2}$. Same structure as \Cref{fig:m4lrec}.
  }\label{fig:WWeven_dphi}
\end{figure}

For the $\PW^+\PW^-$ process, we show distributions in \Cref{fig:WWeven_mll} for the invariant mass of the charged-lepton pair and in \Cref{fig:WWeven_dphi} for the azimuthal separation between the two charged leptons. These two observables do not require any reconstruction of the two W bosons.  
We find that EFT effects are less pronounced than for the observables studied in WZ production. Similarly to WZ, fiducial cuts increase the deviation from the SM predictions, mainly for the quadratic contributions. We find that the interference effects remain at the per cent level even in the case of fiducial cuts. We observe K-factors, which get reduced in the tail of the distribution; in particular, for the quadratic contributions, the NLO results are significantly smaller than the LO ones. This is due to the jet veto, which significantly suppresses the NLO real emissions.

The azimuthal separation of the two charged leptons shows a small impact of the EFT limited to the region close to $\pi$ for the inclusive setup. In the presence of fiducial cuts, the relative importance of the EFT effects is enhanced in the region of large separation. We find that, whilst, at LO, a large fraction of events survive in the large-separation region, these are rejected at NLO QCD, resulting in a much less pronounced deviation from the SM predictions. For this observable, employing a LO simulation for the EFT would lead to a significantly overestimated constraint on the EFT coefficient. 

Although not shown in the figures, the region $\Delta\phi_{\Pe^+\mu^-}\approx \pi$ is subject to similar effects in the case of the quadratic contribution from the CP-odd operator, though with diminished impact compared with the CP-even one. On the contrary, the interference contribution from the CP-odd operator is entirely negligible relative to the NLO QCD results in the SM.

\section{Boost asymmetries}
\label{sec:asym}
In this section, we evaluate the impact of SMEFT operators on asymmetries suitable for $\PW\PW$ and other diboson processes.
Owing to the different parton densities in the proton of up-like and down-like quarks, it is expected that in $\PW^+\PW^-$ inclusive production, $\PW^+$ bosons are typically more forward than $\PW^-$ ones \cite{Re:2018vac}. A similar reasoning applies to $\PW\gamma$ and $\PW\PZ$ \cite{Yang:2022bxv}. 
All asymmetries considered in this work can be written as follows,
\beq\label{eq:charge_Asy_general}
\mc A(i,j) = \frac{\rd \sigma (|y_{i}|>|y_{j}|)-\rd \sigma (|y_{i}|<|y_{j}|)}{\rd \sigma (|y_{i}|>|y_{j}|)+\rd \sigma (|y_{i}|<|y_{j}|)}\,.
\eeq
In $\PW^+\PW^-$, the asymmetry $\mc A(\PW^+,\PW^-)$ is highly sensitive to the polarisation state of the bosons and, therefore, could be noticeably affected by SMEFT effects. 
However, owing to the presence of two neutrinos preventing the reconstruction of individual $\PW$ bosons, it has been proposed \cite{Re:2018vac} to define a proxy of $\mc A(\PW^+,\PW^-)$ based on the decay-lepton kinematics, namely $\mc A(\ell^+,\ell^-)$.
In the production of $\PW\PZ$ pairs, one can construct an asymmetry \cite{Yang:2022bxv} relying on the reconstructed single-neutrino kinematics for the identification of boson moment, $\mc A(\PW^{\rm (rec)},\PZ)$, or the analogous proxy based on lepton kinematics, $\mc A({\ell^+_{\PW},\ell^-_{\PZ}})$.

Similarly to what we have done for the angular coefficients in \Cref{sec:angcoeff}, we parametrise the asymmetries in two ways,
\beqn\label{eq:asymm_par}
\mc A^{(1)}(\lambda) &=& \frac{\mc A^{\rm SM} + \lambda \,{\mc A^{\rm int}}\,\kappa^{\rm int} }{1 + \lambda \kappa^{\rm int}}\,,\qquad
\mc A^{(2)}(\lambda) = \frac{\mc A^{\rm SM} + \lambda \,{\mc A^{\rm int}}\,\kappa^{\rm int} +  {\lambda^2 \mc A^{\rm sq}} \kappa^{\rm sq}}{1 + \lambda \kappa^{\rm int} +  \lambda^2 \kappa^{\rm sq}}\,,
\eeqn
where $\kappa^{\rm int},\,\kappa^{\rm sq}$ and $\lambda$ are defined in Eq.~\ref{eq:kappadef}.

In \Cref{tab:AsymmWZ}, we show the NLO QCD results for W$^+$Z asymmetries in the fiducial ATLAS setup \cite{ATLAS:2022oge}, considering the SMEFT effects and assuming $\lambda=1$TeV$^{-2}$ in Eq.~\ref{eq:asymm_par}.
\begin{table}[t]
    \centering
    \begin{tabular}{lcccc}
    \hline
                    & \multicolumn{2}{c}{$\mc A({\ell^+_{\PW},\ell^-_{\PZ}})$}                                                        & \multicolumn{2}{c}{$\mc A({{\PW},{\PZ}}) $}                                                      \\[0.1cm]
\hline
                    &   LO                              &   NLO QCD                               &    LO                               & NLO QCD                             \\
\hline
SM                  & $ 0.1213 ( 1  )^{+2.2 \%}_ {-1.8 \%} $ &   $ 0.1048 ( 3  )^{+1.1 \%}_ {-1.0 \%} $  &  $ -0.0548 ( 1  )_{-3.1 \%}^ {+2.6 \%} $  &  $ -0.0697 ( 3  )_{-1.5 \%}^ {+1.3 \%} $ \\
CP-even,  $\mc A^{(1)}$    & $ 0.1139 ( 1  )^{+2.4 \%}_ {-2.0 \%} $ &   $ 0.1019 ( 3  )^{+1.0 \%}_ {-0.8 \%} $   &  $ -0.0627 ( 1  )_{-2.7 \%}^ {+2.3 \%} $  &  $ -0.0740 ( 3  )_{-1.3 \%}^ {+1.0 \%} $\\
CP-even,  $\mc A^{(2)}$ & $ 0.0880 ( 1  )^{+0.1 \%}_ {-0.4 \%} $ &   $ 0.0864 ( 3  )^{+0.4 \%}_ {-0.3 \%} $    &  $ -0.0695 ( 1  )_{-1.6 \%}^ {+1.4 \%} $  &  $ -0.0777 ( 4  )_{-0.9 \%}^ {+0.8 \%} $  \\
CP-odd,  $\mc A^{(1)}$    & $ 0.1223 ( 1  )^{+2.1 \%}_ {-1.8 \%} $ &   $ 0.1042 ( 2  )^{+1.2 \%}_ {-1.1 \%} $    &   $ -0.0546 ( 1  )_{-3.1 \%}^ {+2.6 \%} $  &  $ -0.0705 ( 2  )_{-1.5 \%}^ {+1.4 \%} $\\
CP-odd,  $\mc A^{(2)}$ & $ 0.0937 ( 1  )^{+0.2 \%}_ {-0.5 \%} $ &   $ 0.0883 ( 3  )^{+0.7 \%}_ {-0.8 \%} $    &  $ -0.0639 ( 1  )_{-1.5 \%}^ {+1.4 \%} $  &  $ -0.0750 ( 3  )_{-1.1 \%}^ {+1.2 \%} $ \\
\hline
    \end{tabular}
    \caption{Effect of the inclusion of the CP-even and CP-odd operators (${O}_{3{W}}$ and $O_{3\widetilde{W}}$) on the asymmetries $\mc A({\ell^+_{\PW},\ell^-_{\PZ}})$ and $\mc A({\PW,{\PZ}})$, defined according to Eq.~\ref{eq:charge_Asy_general} and the SMEFT parametrisations of Eq.~\ref{eq:asymm_par} with $\lambda = 1$TeV$^{-2}$.  Results are shown for $\PW^+\PZ$ production at the LHC@13TeV at LO and NLO QCD in the fiducial ATLAS setup. The uncertainties from correlated 9-point QCD-scale variations are shown in percentages. The MC uncertainties are shown in parentheses.}
    \label{tab:AsymmWZ}
\end{table}
The QCD-scale uncertainties are at the few-percent level, as they come from correlated scale variations of the numerator and denominator appearing in Eq.~\ref{eq:charge_Asy_general}. More conservative QCD uncertainties may be obtained with uncorrelated scale variations \cite{Re:2018vac}.
The mostly left-handed polarisation of the W boson causes the charged lepton from its decay to be produced more forward
than the leptons from the Z boson. Therefore, the leptonic asymmetry is positive. This effect is diminished in the presence of SMEFT CP-even effects, which, as we have shown in the discussion of \Cref{tab:Acoeff_phi}, increase the right-hand polarisation of the W boson. A similar effect is found for the CP-odd operator at the squared level, while its effect at the linear level is negligible. 
The bosonic asymmetry is instead negative and somewhat affected by the neutrino reconstruction, which typically leads to more central W-rapidity distributions compared to the MC-truth expectations \cite{Denner:2020eck}. This makes the asymmetry artificially marked but does not change its sign. The SMEFT effects further increase (in absolute value) the SM asymmetry by 5-to-10\% for the chosen Wilson coefficient (1~TeV$^{-2}$).

The numerical results for the leptonic asymmetry in $\PW^+\PW^-$ are shown in \Cref{tab:AsymmWW_new}, assuming the SMEFT parametrisations of Eq.~\ref{eq:asymm_par} with $\lambda=1$TeV$^{-2}$.
\begin{table}[t]
    \centering
    \begin{tabular}{lcccc}
    \hline
        & \multicolumn{4}{c}{$\mc A({\ell^+,\ell^-})$}                                         \\[0.1cm]
\hline\\[-0.4cm]
  accuracy                 &   LO             &   NLO QCD        &   NLO QCD        &   NLO QCD       \\[0.1cm]
\hline\\[-0.4cm]
  setup                &  fully inclusive &  fully inclusive &  jet-vetoes only & ATLAS fiducial   \\[0.1cm]
\hline\\[-0.4cm]
SM                      &  $ -0.0529 ( 2  )_{-3.2 \%}^ {+2.7 \%} $ & $ -0.0185 ( 3  )_{-9.6 \%}^{+9.8 \%} $& $ -0.0436 ( 4  )_{-2.7 \%}^ {+2.7 \%} $ & $ -0.022 ( 1  )_{-5.5 \%}^ {+3.8 \%} $\\ CP-even, $\mc A^{(1)}$         & $ -0.0574 ( 2  )_{-2.9 \%}^ {+2.4 \%} $ & $ -0.0212 ( 3  )_{-9.2 \%}^ {+8.7 \%} $ & $ -0.0474 ( 4  )_{-2.5 \%}^ {+2.5 \%} $ &  $ -0.024 ( 1  )_{-5.4 \%}^ {+3.9 \%} $\\ 
CP-even, $\mc A^{(2)}$      & $ -0.0557 ( 2  )_{-3.2 \%}^ {+2.7 \%} $    & $ -0.0212 ( 3  )_{-9.0 \%}^ {+8.4 \%} $ & $ -0.0473 ( 4  )_{-2.0 \%}^ {+2.1 \%} $  & $ -0.024 ( 1  )_{-5.4 \%}^ {+4.1 \%} $\\ 
CP-odd, $\mc A^{(1)}$           & $ -0.0528 ( 3  )_{-3.2 \%}^ {+2.7 \%} $& $ -0.0184 ( 4  )_{-9.7 \%}^ {+9.7 \%} $ & $ -0.0430 ( 6  )_{-2.6 \%}^ {+2.6 \%} $     &$ -0.022 ( 1  )_{-3.8 \%}^ {+2.7 \%} $\\ 
CP-odd, $\mc A^{(2)}$        & $ -0.0516 ( 3  )_{-3.4 \%}^ {+2.9 \%} $   &  $ -0.0181 ( 4  )_{-9.7 \%}^ {+9.9 \%} $ & $ -0.0426 ( 7  )_{-2.3 \%}^ {+2.4 \%} $ & $ -0.021 ( 1  )_{-3.3 \%}^ {+3.4 \%} $ \\ 
\hline
\end{tabular}
    \caption{Effect of the inclusion of the CP-even and CP-odd operators (${O}_{3{W}}$ and $O_{3\widetilde{W}}$) on the asymmetry $\mc A({\ell^+,\ell^-})$ defined according to Eq.~\ref{eq:charge_Asy_general} and the SMEFT parametrisations of Eq.~\ref{eq:asymm_par} with $\lambda = 1$TeV$^{-2}$. 
    Results are shown for $\PW^+\PW^-$ production at the LHC@13TeV in three setups: fully inclusive (LO, NLO QCD), inclusive with jet vetoes only (NLO QCD), fiducial ATLAS setup (NLO QCD). The uncertainties from correlated 9-point QCD-scale variations are shown in percentages. The MC uncertainties are shown in parentheses.}
    \label{tab:AsymmWW_new}
\end{table}
The $\mc A({\ell^+,\ell^-})$ is negative, and its size is maximal at LO in the SM. The inclusion of QCD corrections notably reduces its value, while applying jet vetoes makes Born-like topologies dominant again, giving similar results as at LO. The fiducial ATLAS selections lead to a realistic SM estimate of the asymmetry of about $-0.022$ with 5\%-wide QCD-scale bands. The CP-even linear term enhances the size of the (negative) SM asymmetry at LO and NLO QCD. The inclusion of the quadratic CP-even term does not change sizeably the asymmetries. As for integrated cross-sections (see \Cref{tab:sigmasWW}), the CP-odd effects are negligible also for the boost asymmetry.  

To further understand the actual impact of SMEFT effects on asymmetries, it is important to study their dependence on the WCs of the CP-even and CP-odd operators. 
\begin{figure}
  \centering
  \subfigure[$\mc A({\ell_{\PW},\ell^-_{\PZ}})$, CP even \label{fig:parasym_1}]{\includegraphics[scale=0.43,page=1]{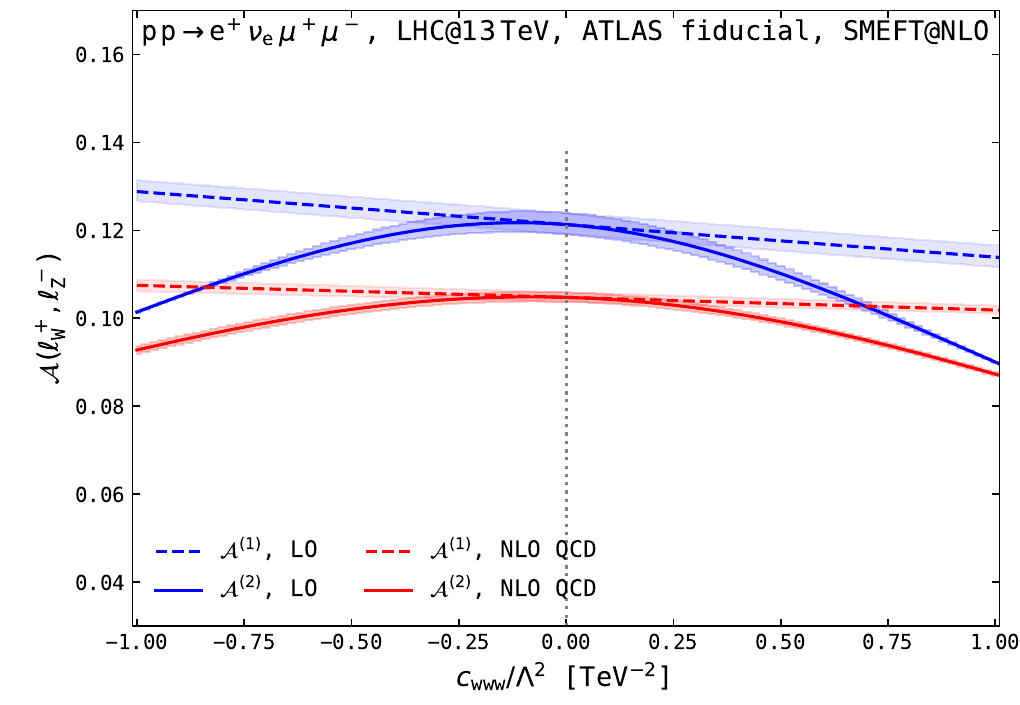}}
  \subfigure[$\mc A({\ell_{\PW},\ell^-_{\PZ}})$, CP odd \label{fig:parasym_2}]{\includegraphics[scale=0.43,page=3]{fig/parametricAsym.pdf}}
  \subfigure[$\mc A({{\PW},{\PZ}})$, CP even \label{fig:parasym_3}]{\includegraphics[scale=0.43,page=2]{fig/parametricAsym.pdf}}
  \subfigure[$\mc A({{\PW},{\PZ}})$, CP odd \label{fig:parasym_4}]{\includegraphics[scale=0.43,page=4]{fig/parametricAsym.pdf}}
\subfigure[$\mc A(\ell^+,\ell^-)$, CP even \label{fig:parasymW_1}]{\includegraphics[scale=0.43,page=1]{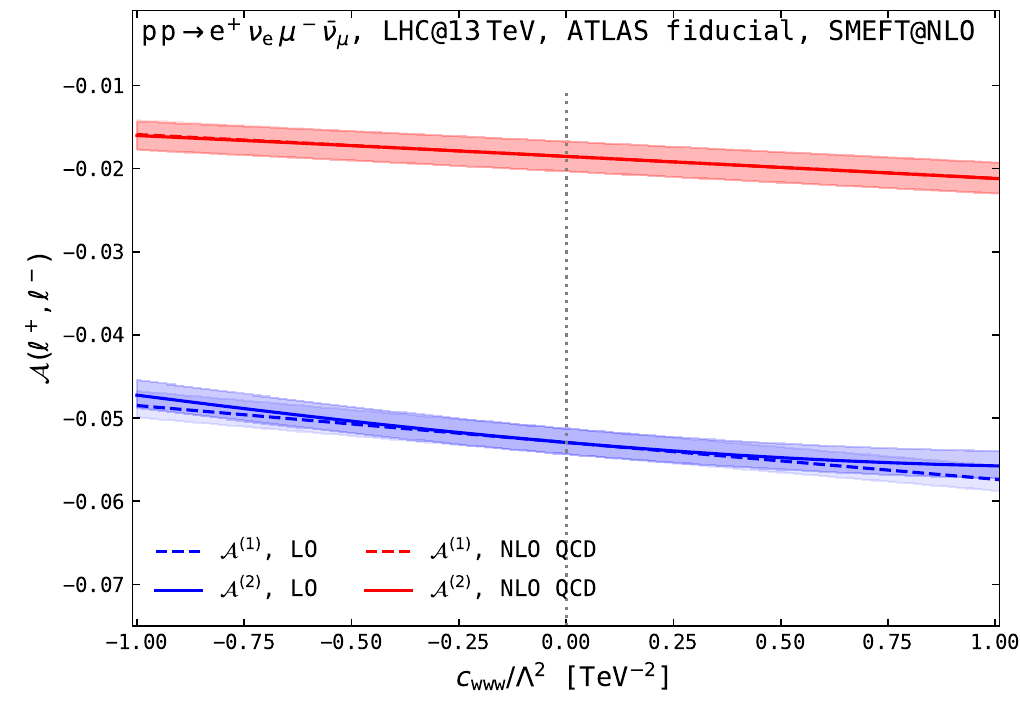}}
  \subfigure[$\mc A(\ell^+,\ell^-)$, CP odd \label{fig:parasymW_2}]{\includegraphics[scale=0.43,page=2]{fig/parametricAsymWW.pdf}}
  \caption{Effect of the inclusion of the CP-even operator ${O}_{3{W}}$ (left) and of the CP-odd one $O_{3\widetilde{W}}$ (right) on the asymmetries for W$^+$Z (\ref{fig:parasym_1}--\ref{fig:parasym_4}) and W$^+$W$^-$ (\ref{fig:parasymW_1}--\ref{fig:parasymW_2}) production at the LHC@13TeV, in the corresponding fiducial ATLAS setups \cite{ATLAS:2022oge,ATLAS:2019rob}. The asymmetry parametrisations are defined in Eq.~\ref{eq:asymm_par}. Shaded bands at LO (blue) and NLO QCD (red) come from 9-point QCD-scale variations.\label{fig:AsymmParametric}}
\end{figure}
Results are shown in \Cref{fig:AsymmParametric}
for the bosonic and leptonic asymmetries in W$^+$Z and the leptonic one in W$^+$W$^-$, respectively.
For both processes, the fiducial ATLAS selections are assumed. The parametrisations introduced in Eq.~\ref{eq:asymm_par} are considered, with $\lambda$ varying between $-1$ and $+1$~TeV$^{-2}$.
In WZ, the slope of the linear parametrisation of the asymmetries almost vanishes for the CP-odd operator (both at LO and NLO QCD). In contrast, the effects of the CP-even operator are more visible despite a slope reduction due to NLO QCD corrections. Quadratic effects are significant (thus unavoidable) for both the leptonic and the bosonic asymmetry.
At variance with WZ, the leptonic-asymmetry dependence on the WCs in WW is entirely dominated by the linear term (especially at NLO QCD), despite a sizeable quadratic contribution to the fiducial cross-section.
The quadratic parametrisation of the asymmetry lies well within the QCD-scale uncertainties of the linear parametrisation. In addition, the CP-odd effects are negligible over the whole scanned range of the Wilson coefficient. The leptonic asymmetry observable for WW at the inclusive level is not promising for constraining the WCs. 

\begin{figure}[h]
  \centering
  \subfigure[$\mc A({\ell^+_{\PW},\ell^-_{\PZ}})$, CP-even \label{fig:asym_1}]{\includegraphics[scale=0.43,page=1]{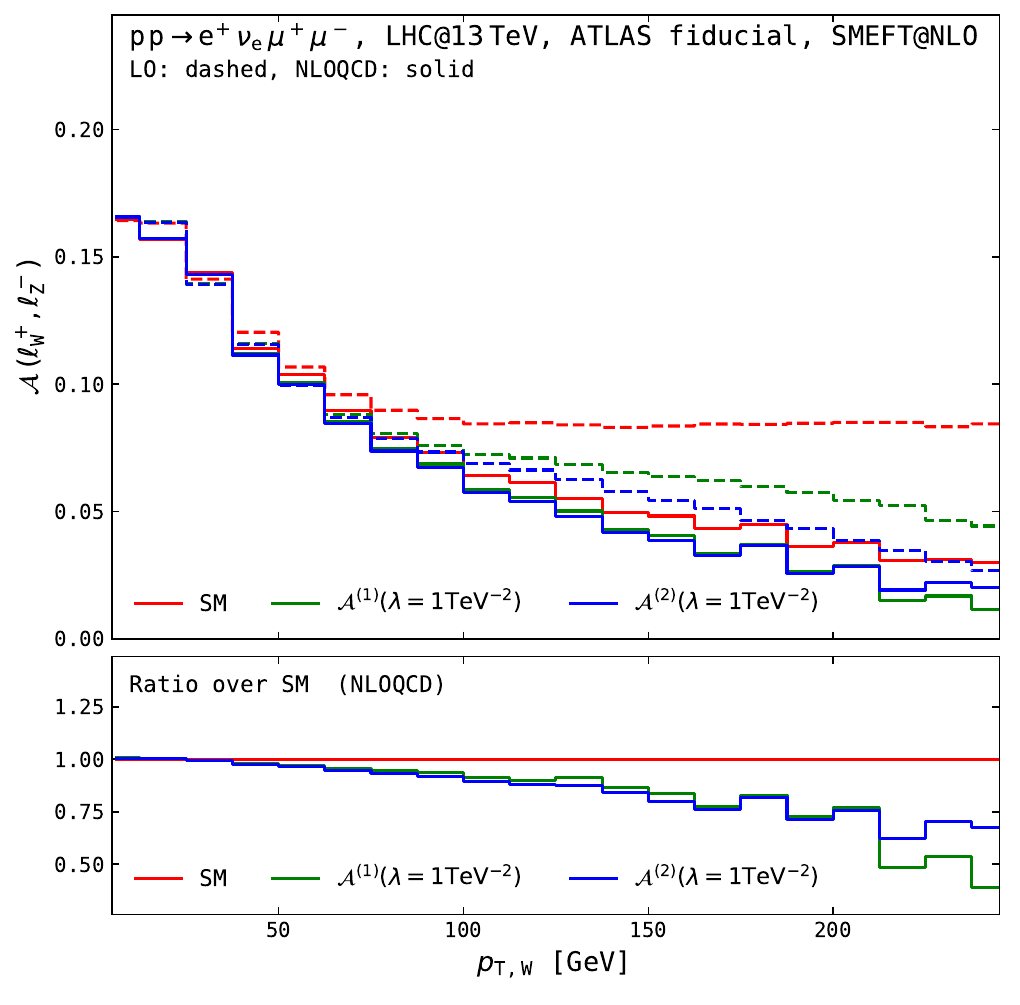}}
  \subfigure[$\mc A({{\PW},{\PZ}})$, CP-even\label{fig:asym_2}]{\includegraphics[scale=0.43,page=2]{fig/WZ_asym_ptW_dep.pdf}}
  \subfigure[$\mc A({\ell^+_{\PW},\ell^-_{\PZ}})$, CP-odd \label{fig:asym_3}]{\includegraphics[scale=0.43,page=3]{fig/WZ_asym_ptW_dep.pdf}}
  \subfigure[$\mc A({{\PW},{\PZ}})$, CP-odd\label{fig:asym_4}]{\includegraphics[scale=0.43,page=4]{fig/WZ_asym_ptW_dep.pdf}}
  \caption{Effect of the inclusion of the CP-even (a,b) and CP-odd (c,d) operators (${O}_{3{W}}$ and $O_{3\widetilde{W}}$) on the asymmetries $\mc A({\ell^+_{\PW},\ell^-_{\PZ}})$ (a,c) and $\mc A({{\PW},{\PZ}})$ (b,d), differentially in the transverse momentum of the $\PW$ boson for $\PW^+\PZ$ production at the LHC@13TeV, in the fiducial ATLAS setup \cite{ATLAS:2022oge}. The asymmetries are parametrised according to Eq.~\ref{eq:asymm_par}, with $\lambda = 1$~TeV$^{-2}$. 
  }\label{fig:Asymm_lep_pTW_CPeven}
\end{figure}

Finally, in \Cref{fig:Asymm_lep_pTW_CPeven}, we explore the dependence of the leptonic and bosonic asymmetries as a function of the transverse momentum of the W boson in WZ production. 
The leptonic asymmetry is significantly reduced at high $p_{\rm T,W}$. As expected, the deviation from the SM due to CP-even EFT effects is enhanced in the tail, with the EFT leading to smaller asymmetries. Similarly, the bosonic asymmetry is modified more at large transverse momentum, reaching 20\% in the tail. We also note that the difference between the SM and the EFT is reduced at NLO QCD in both cases. 
At variance with the LO picture, at NLO QCD, the deviation from the SM is entirely dominated by the linear SMEFT term, even in the distribution's tail.
The effect of the squared CP-odd term, though notably more minor compared to the CP-even one, can be appreciated at moderate transverse momentum, leading to a 10\% decrease in the NLO QCD $\mc A({\ell^+_{\PW},\ell^-_{\PZ}})$ asymmetry for $p_{\rm T,W}>150\GeV$.

Owing to the partial cancellation of QCD-scale uncertainties, and given the apparent deviation from the SM predictions due to the EFT contributions in the moderate-to-large transverse-momentum region, the differential measurement of boost asymmetries as defined in Eq.~\ref{eq:charge_Asy_general} represents a promising avenue towards tighter constraints of the EFT WCs. 

\section{Sensitivity study}
\label{sec:sens_study}
Using \(\chi^2\)-based statistical analysis, we now assess the agreement between experimentally measured cross-sections and SMEFT predictions, thereby setting constraints on WCs. 
Our analysis considers the complete off-shell modelling of diboson processes and incorporates differential information into the \(\chi^2\) statistics across various kinematic regions and observables. 
Compared to previous literature results \cite{Baglio:2018bkm,Baglio:2019uty}, we include several angular observables sensitive to the polarisation structure of the underlying diboson processes.
The squared differences between observed and expected values are normalised by the total error defined as the quadrature sum of absolute uncertainties from theoretical and experimental sources. Experimental uncertainties are considered, including uncorrelated and correlated systematics and statistical errors. Although ideally, the correlations between these errors should be accounted for, we assume they are uncorrelated and conservatively introduce an additional 5\% systematic uncertainty bin-by-bin. The total theoretical uncertainty is identified through the maximum uncertainties arising from the choices of the renormalisation and factorisation scales ($\mu_{\rm R}$, $\mu_{\rm F}$), as well as MC simulation errors, the latter being predominantly significant in the linear-only fit.

We utilise the SM predictions employed in the ATLAS experimental analysis,
namely NNLO QCD for WZ \cite{Grazzini:2016ctr} and NNLO QCD + NLO EW for WW \cite{Grazzini:2017ckn,Grazzini:2019jkl}. We notice that in the fiducial WZ region considered here \cite{ATLAS:2019bsc,ATLAS:2022oge}, the NLO EW corrections account for $-3\%$ compared to the NNLO QCD results \cite{Grazzini:2019jkl,Le:2022lrp}, therefore their inclusion is not expected to change sizeably the fit results. These SM predictions are extracted by digitising the plots in the aforementioned experimental publications, as numerical data is not provided in the HEPData repository \cite{Maguire:2017ypu}. Having established our SM predictions at NNLO accuracy, the \(\chi^2\) statistic varies by exploiting only the EFT information at LO and NLO from both CP-even and CP-odd operators.

By performing a \(\chi^2\) fit to determine 95\% confidence level (CL) intervals for the WCs, we quantify the compatibility of SMEFT with the experimental data. In obtaining these bounds, we apply no additional phase space cuts beyond those defined by the experimental fiducial regions and fix \(\Lambda\) to \(1 \, \text{TeV}\). Our bounds can be appropriately re-scaled for other values of $\Lambda$. We present limits for both \(\PW^{+}\PZ\) and \(\PW^{+}\PW^{-}\) processes separately. We display the individual bounds for each operator at LO and NLO QCD. This includes scenarios with only linear SMEFT terms and those incorporating the full SMEFT contributions, i.e. including linear and quadratic terms. Furthermore, we show the limits derived from inclusive cross-sections and differential distributions. 
As expected, there is no sensitivity to the CP-odd coefficients at the linear level, similar to the conclusions reached in Ref.~\cite{Azatov:2019xxn}; therefore, we refrain from presenting these bounds.
Moreover, it is worth reminding the reader, as pointed out in Ref.~\cite{Azatov:2019xxn}, that bounds on the CP-odd coefficients from the non-observation of electron and neutron electric dipole moments (EDM) are very stringent. In particular, when considering bounds from the electron EDM~\cite{ParticleDataGroup:2022pth}, the current constraints on the CP-odd coefficient could be even beyond the reach of the HL-LHC. At the same time, we emphasise that as Refs.~\cite{Dekens:2013zca,Cirigliano:2019vfc} suggest, low-energy experiments and LHC searches can feature some complementarity in constraining CP-violating interactions in the context of global studies.

\subsection{$\PW\PZ$ fit results}
For the \( \PW\PZ \) production process, we utilise the fiducial differential cross-sections from the latest ATLAS measurements~\cite{ATLAS:2019bsc, ATLAS:2022oge}. The reported differential data encompasses the following observables,
\beq
p_{\rm T, Z},\quad
p_{\rm T,W},\quad
p_{\rm T,mis},\quad
M_{\rm T,WZ},\quad
\Delta\phi_{\rm WZ},\quad
|\Delta y_{\rm Z,\ell_{\rm W}}|,\quad
\cos\theta^*_{\Pe^+}\,.
\eeq
 For the polar decay angle ($\cos\theta^*_{\Pe^+}$), we employ the EFT predictions for its reconstructed variant, relying on the neutrino-reconstruction strategy employed by ATLAS \cite{ATLAS:2019bsc}.
 We note that all differential data, except for $\cos\theta^*_{\Pe^+}$, are extracted from Ref.~\cite{ATLAS:2019bsc}, which utilises a dataset with an integrated luminosity of 36.1 fb$^{-1}$. The polar decay angle, however, is extracted from Ref.~\cite{ATLAS:2022oge}, which uses a dataset with an integrated luminosity of 139 fb$^{-1}$.
 
\begin{figure}[h]
  \includegraphics[scale=0.55]{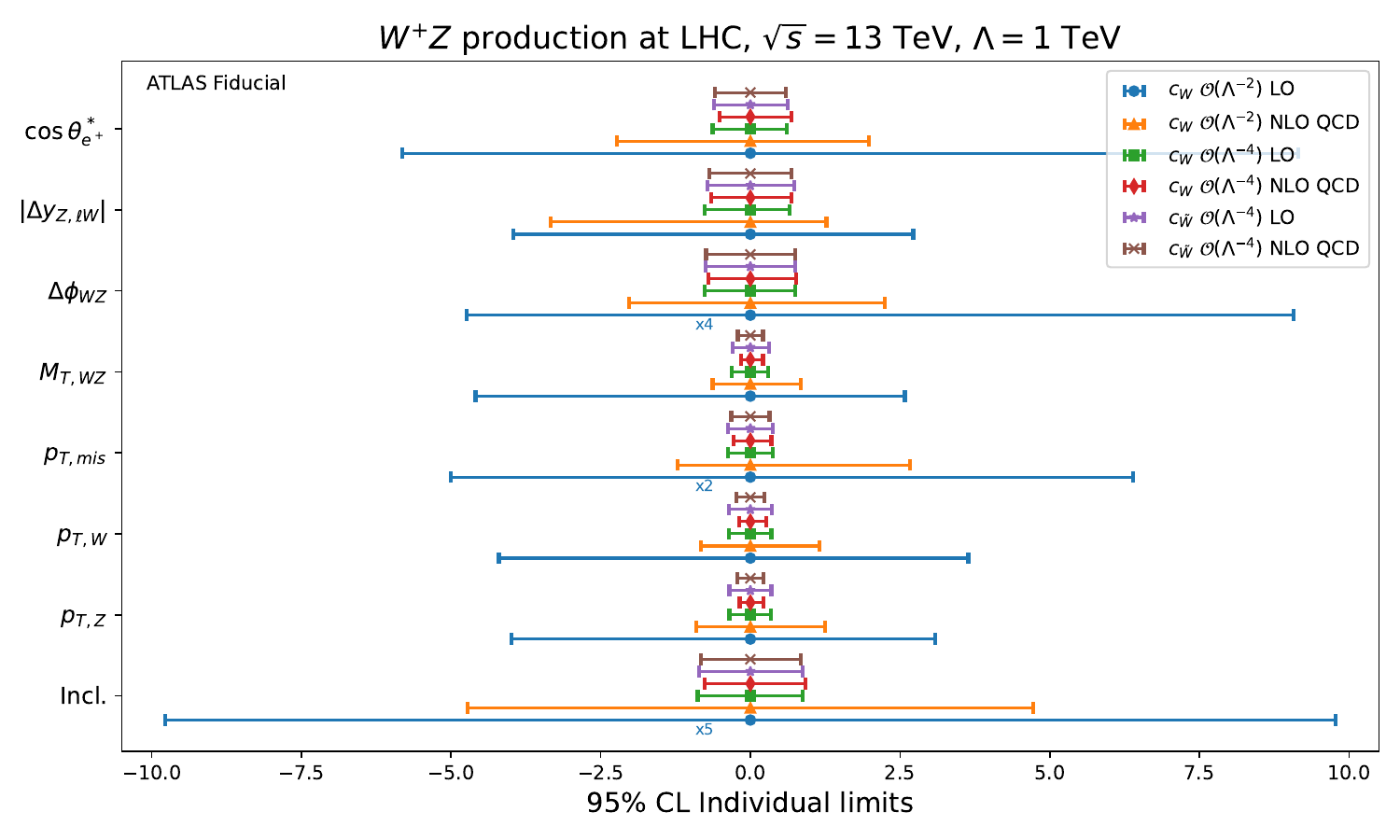}
  \caption{Individual 95\% CL bounds on CP-even coefficients from only SMEFT linear terms, and from both CP-even and CP-odd coefficients when all EFT terms are included, for $\PW^{+}\PZ$ production in the fiducial ATLAS region at LO and NLO. Each row provides limits obtained through differential information from the specified observable. The last row presents limits derived exclusively from the inclusive cross-section. Bounds with a quoted factor beneath are divided by that factor to fit in the plot.}
  \label{fig:WZ_fit}
\end{figure}

In \Cref{fig:WZ_fit}, we present the limits derived from solely the linear SMEFT terms and those considering quadratic contributions. The corresponding numerical values are given in \Cref{app:fit} (see  \Cref{tab:wz_bounds_all}).
We note an order of magnitude improvement in the bounds across all observables when including quadratic contributions: approximately \(\mathcal{O}(10)\) at LO and \(\mathcal{O}(5)\) at NLO, compared to bounds from linear terms alone. Additionally, the bounds derived from the linear NLO interference terms consistently exhibit an improvement over their LO counterparts, owing to the greater-than-unity EFT linear K-factor. 
The quadratic bounds at NLO are generally similar to their LO counterparts, except for transverse quantities for which the NLO constraints are more stringent.
We observe that dimensionful observables such as transverse mass and momenta yield more stringent bounds than {angular} measurements for this process. In particular, the transverse mass of the WZ system provides the most stringent bounds, as has also been observed in Refs.~\cite{Azatov:2019xxn,Baglio:2019uty,Baglio:2020oqu}\footnote{As mentioned, the bounds on the WCs are obtained by setting $\Lambda=1$ TeV but can be straightforwardly re-scaled for other values of $\Lambda$.}. Polarisation-sensitive observables, notably \( \cos\theta^*_{\Pe^+} \), demonstrate competitive bounds among dimensionless observables.

Finally, as expected, the limits on the CP-odd coefficients become non-trivial only when including quadratic contributions, which nearly align the CP-odd and CP-even coefficient bounds. We note the similar bounds obtained for the CP-even coefficient in the recent global fit by the SMEFit collaboration~\cite{Celada:2024mcf}.

\subsection{$\PW\PW$ fit results}
\begin{figure}[h]
  \includegraphics[scale=0.55]{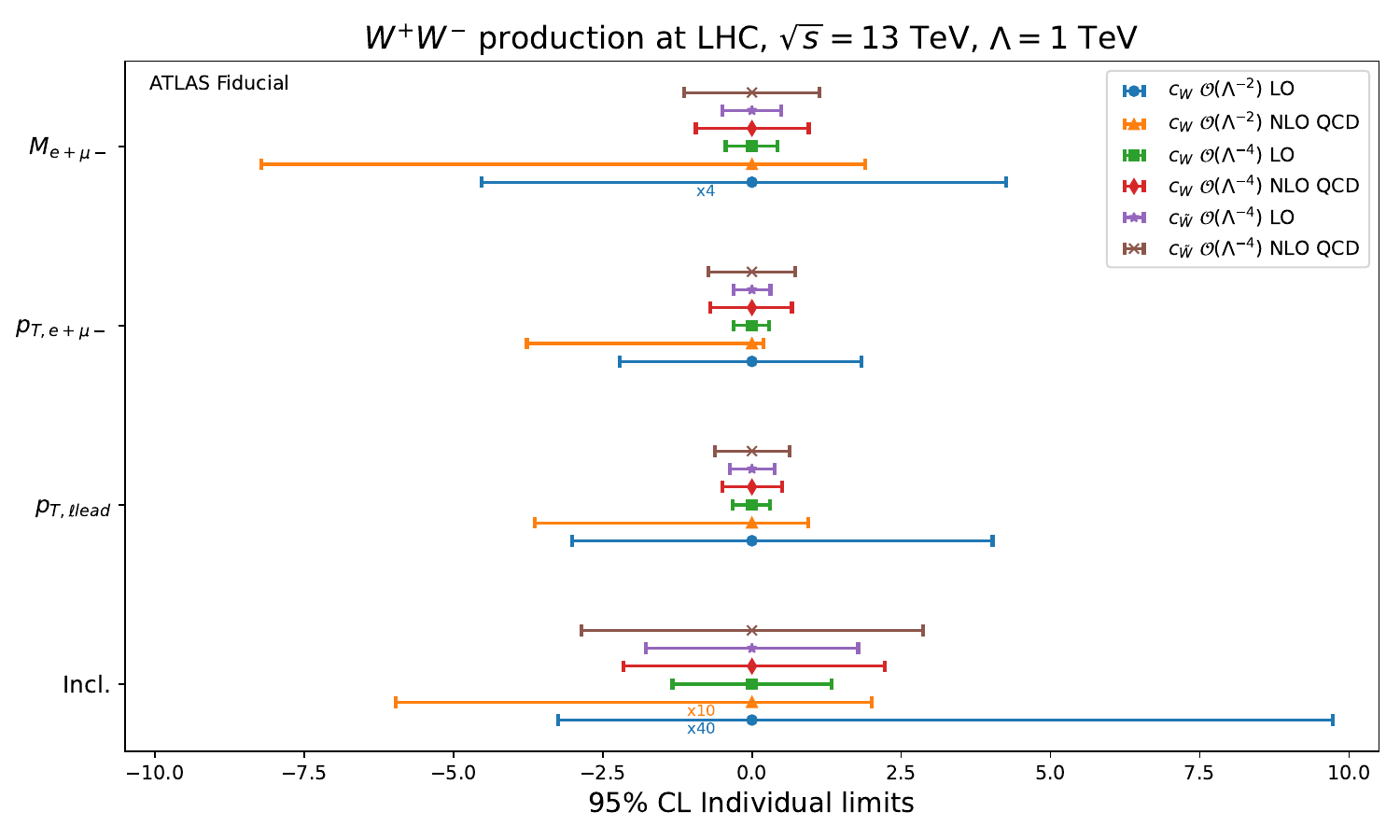}
  \caption{Same as~\Cref{fig:WZ_fit} but for $\PW^{+}\PW^{-}$.}
  \label{fig:WW_fit}
\end{figure}

For the \(\PW^+\PW^-\) analysis, we utilise differential information from the ATLAS measurement~\cite{ATLAS:2019rob}. The reported differential data includes various distributions,
\beq
p_{\rm T,\ell_{\rm lead}},\quad
p_{\rm T,\Pe^+\mu^-},\quad
M_{\Pe^+\mu^-},\quad
\Delta\phi_{\Pe^+\mu-},\quad
|y_{\Pe^+\mu^-}|,\quad
\left|\tanh\frac{\Delta\eta_{\Pe^+\mu^-}}2\right|,
\eeq
the first three of which we report the derived bounds in~\Cref{fig:WW_fit} (for the corresponding numerical values we refer to \Cref{tab:ww_bounds_all} in \Cref{app:fit}). The latter three variables feature consistent mis-modelling, as reported in the experimental publication~\cite{ATLAS:2019rob}, with predictions underestimating the measured fiducial cross-sections by 15-20\%. Consequently, we believe that the bounds derived from these variables do not accurately capture any SMEFT effects, and therefore, we refrained from presenting those results.

Our conclusions from the $\PW\PW$ analysis are twofold: \textit{(i)} Similar to the $\PW\PZ$ case, the transverse dimensionful observables, i.e. $p_{\rm T,\ell_{\rm lead}}$ and $p_{\rm T,\Pe^+\mu^-}$, provide a more stringent handle on the derived bounds. \textit{(ii)} In contrast to $\PW\PZ$, in the case of $\PW\PW$ production when including the full EFT contributions, \emph{i.e.}, considering the quadratic terms, the bounds derived with LO EFT predictions are more stringent than those from NLO. This characteristic corroborates the less-than-unity K-factor in $\PW\PW$ production. This can already be inferred from examining the differential results shown in~\Cref{fig:WWeven_mll_fid}. Such an observation emphasises the need to employ NLO EFT predictions for reliably constraining WCs. Finally, we note that the limits derived from WZ production dominate over those from WW production due to the former's higher sensitivity. Therefore, combining both datasets will not improve the bounds beyond those obtained in WZ, and thus, we refrain from doing so.

We conclude the sensitivity study by commenting that the bounds obtained from WZ and WW behave differently at NLO QCD accuracy because of the different EFT K-factors, as can be expected from the last insets of~\Cref{fig:ptZ_2,fig:WWeven_mll_fid}.

\section{Conclusions}
\label{sec:conclusions}
In this paper, we have presented the complete calculation of $\PW^+\PZ$ and $\PW^+\PW^-$
production at the LHC at NLO in QCD, in the fully leptonic decay channel. The SMEFT effects from two dimension-six operators (the CP-even $O_{3W}$ and the CP-odd $O_{3\widetilde{W}}$) modifying the triple-gauge couplings are computed at NLO in QCD. Complete off-shell effects and spin correlations are accounted for at NLO in the SM and SMEFT.

We have shown how the application of fiducial cuts leads to non-trivial effects in the relative size of the EFT contributions with respect to the SM predictions. This has been explored by computing the differential distributions of various kinematic variables. We have demonstrated that QCD corrections affect the SM and EFT contributions differently, and thus, NLO-accurate modelling for the EFT is necessary.

By studying observables such as the invariant mass of the four-lepton system and transverse momentum of the Z boson, we observe that the EFT contributions are more pronounced in the high-energy tails with both NLO corrections and fiducial cuts significantly increasing the interference between the CP-even operator and the SM. 

To further improve sensitivity to EFT operators, we considered polarisation-sensitive observables. We have quantified the impact of the EFT operators on the angular coefficients extracted from the decay-angle distributions in an inclusive setup. As these are meaningful only in the absence of selection cuts, we have then focused on the decay-angle distributions in realistic fiducial setups, finding that the interference of the SM and CP-even and CP-odd operators lead to different modulations over the SM and can thus be considered to distinguish between CP-odd and CP-even contributions. 

We explored WW production and WZ production in the presence of the same set of operators. In general, WW production is significantly less sensitive to EFT effects, and thus, sensitivity to the triple gauge interactions arises predominantly from the WZ process.

Beyond integrated cross-sections and differential distributions, we have investigated boost asymmetries, which benefit from reduced systematic uncertainties. These further enhanced sensitivity to CP-even linear and quadratic effects, especially when considered differentially in WZ production.

To conclude, we have explored the sensitivity reached by current LHC diboson measurements. We have assessed various observables and found that most sensitivity originates from the high-energy tails of distributions, which are dominated by the quadratic contributions. When including quadratic contributions, constraints are similar for CP-even and CP-odd effects. In this respect, precise measurements of angular polarisation-sensitive observables can further constrain the triple gauge operator coefficients and distinguish between their CP properties.

This work paves the way for future studies of angular observables in diboson processes at the LHC. Specifically, it points towards conducting a two-dimensional fit of CP-even and CP-odd effects and including the complete set of dimension-six operators affecting off-shell processes. This comprehensive approach will allow for the full exploitation of diboson processes and the consequential implications for global EFT fits. Additionally, the strong interference suppression observed in WW production, which is more pronounced than in WZ production, warrants further investigation, namely identifying suitable observables can enhance the detection of these subtle EFT effects.

\section*{Acknowledgements}
HF thanks Rafael Aoude, Eugenia Celada, Alejo Rossia, and Marion Thomas for the insightful conversations. GP is grateful to Giulia Zanderighi, Uli Haisch, and Jakob Linder for their useful discussions. The work of HF and EV is supported by the European Research Council (ERC) under the European Union’s Horizon 2020 research and innovation programme (Grant Agreement No. 949451) and by a Royal Society University Research Fellowship (Grant URF/R1/201553). The authors also acknowledge support from the COMETA COST Action CA22130.

\appendix
\section{Numerical data for fit results}\label{app:fit}
\begin{table}[h]
\centering
\small
\begin{tabular}{
  >{\raggedright\arraybackslash}p{1.2cm}| 
  R{0.8cm} @{${},{}$} L{0.8cm}
  R{0.8cm} @{${},{}$} L{0.8cm}
  R{0.8cm} @{${},{}$} L{0.8cm}
  R{0.8cm} @{${},{}$} L{0.8cm}
  R{0.8cm} @{${},{}$} L{0.8cm}
  R{0.8cm} @{${},{}$} L{0.8cm}
}
\toprule
\multicolumn{1}{>{\raggedright\arraybackslash}p{1.2cm}}{} & \multicolumn{4}{c}{$c_{W}$ $\mathcal{O}(\Lambda^{-2}$)}   
& \multicolumn{4}{c}{$c_{W}$ $\mathcal{O}(\Lambda^{-4})$} 
& \multicolumn{4}{c}{$c_{\tilde{W}}$ $\mathcal{O}(\Lambda^{-4})$} \\
\cmidrule(lr){2-5} \cmidrule(lr){6-9} \cmidrule(lr){10-13}
\multicolumn{1}{>{\raggedright\arraybackslash}p{1.2cm}}{} & \multicolumn{2}{c}{LO}                             
& \multicolumn{2}{c}{NLO QCD}                               
& \multicolumn{2}{c}{LO}                               
& \multicolumn{2}{c}{NLO QCD}    
& \multicolumn{2}{c}{LO}                               
& \multicolumn{2}{c}{NLO QCD}                            
\\
\midrule
Incl.                  
& \multicolumn{2}{c}{[-48.89, 48.89]} 
& \multicolumn{2}{c}{[-4.72, 4.72]}  
& \multicolumn{2}{c}{[-0.88, 0.87]}  
& \multicolumn{2}{c}{[-0.77, 0.92]}    
& \multicolumn{2}{c}{[-0.86, 0.87]}  
& \multicolumn{2}{c}{[-0.83, 0.84]} 
\\
$p_{\rm T, Z}$                  
& \multicolumn{2}{c}{[-3.99, 3.09]} 
& \multicolumn{2}{c}{[-0.90, 1.25]}  
& \multicolumn{2}{c}{[-0.35, 0.34]}  
& \multicolumn{2}{c}{[-0.18, 0.22]}    
& \multicolumn{2}{c}{[-0.35, 0.35]}  
& \multicolumn{2}{c}{[-0.22, 0.22]}
\\
$p_{\rm T, W}$  
& \multicolumn{2}{c}{[-4.20, 3.64]} 
& \multicolumn{2}{c}{[-0.83, 1.16]}   
& \multicolumn{2}{c}{[-0.36, 0.35]}  
& \multicolumn{2}{c}{[-0.19, 0.26]}    
& \multicolumn{2}{c}{[-0.36, 0.36]}  
& \multicolumn{2}{c}{[-0.24, 0.24]} 
\\
$p_{\rm T, mis}$   
& \multicolumn{2}{c}{[-10.01, 12.79]} 
& \multicolumn{2}{c}{[-1.22, 2.66]}  
& \multicolumn{2}{c}{[-0.38, 0.38]}  
& \multicolumn{2}{c}{[-0.28, 0.35]}    
& \multicolumn{2}{c}{[-0.38, 0.38]}  
& \multicolumn{2}{c}{[-0.32, 0.32]} 
\\
$m_{\rm T, WZ}$   
& \multicolumn{2}{c}{[-4.59, 2.58]} 
& \multicolumn{2}{c}{[-0.63, 0.84]}
& \multicolumn{2}{c}{[-0.31, 0.30]}  
& \multicolumn{2}{c}{[-0.16, 0.21]}    
& \multicolumn{2}{c}{[-0.30, 0.31]}  
& \multicolumn{2}{c}{[-0.21, 0.21]} 
\\
$\Delta\phi_{\rm WZ}$     
& \multicolumn{2}{c}{[-18.96, 36.28]} 
& \multicolumn{2}{c}{[-2.02, 2.25]}    
& \multicolumn{2}{c}{[-0.76, 0.75]}  
& \multicolumn{2}{c}{[-0.70, 0.76]}    
& \multicolumn{2}{c}{[-0.75, 0.75]}  
& \multicolumn{2}{c}{[-0.74, 0.75]} 
\\
$|\Delta y_{\rm Z,\ell_{\rm W}}|$   
& \multicolumn{2}{c}{[-3.96, 2.72]} 
& \multicolumn{2}{c}{[-3.33, 1.27]}    
& \multicolumn{2}{c}{[-0.77, 0.66]}  
& \multicolumn{2}{c}{[-0.66, 0.69]}    
& \multicolumn{2}{c}{[-0.72, 0.73]}  
& \multicolumn{2}{c}{[-0.69, 0.69]} 
\\
\( \cos\theta^*_{\Pe^+} \)  
& \multicolumn{2}{c}{[-5.82, 9.15]} 
& \multicolumn{2}{c}{[-2.23, 1.98]}  
& \multicolumn{2}{c}{[-0.63, 0.61]}  
& \multicolumn{2}{c}{[-0.52, 0.68]}    
& \multicolumn{2}{c}{[-0.61, 0.62]}  
& \multicolumn{2}{c}{[-0.59, 0.59]} 
\\
\bottomrule
\end{tabular}
\caption{Individual 95\% CL bounds on CP-even ($c_{W}$) coefficients from only SMEFT linear terms, and from both CP-even and CP-odd ($c_{\tilde{W}}$) coefficients when all EFT terms are included, for $\PW^{+}\PZ$ production in the fiducial ATLAS region at LO and NLO. The first row presents limits derived exclusively from the inclusive cross-section. The following rows provide limits obtained through differential information from the specified observables.}
\label{tab:wz_bounds_all}
\end{table}
\begin{table}[h]
\centering
\small
\begin{tabular}{
  >{\raggedright\arraybackslash}p{1.2cm}| 
  R{0.8cm} @{${},{}$} L{0.8cm}
  R{0.8cm} @{${},{}$} L{0.8cm}
  R{0.8cm} @{${},{}$} L{0.8cm}
  R{0.8cm} @{${},{}$} L{0.8cm}
  R{0.8cm} @{${},{}$} L{0.8cm}
  R{0.8cm} @{${},{}$} L{0.8cm}
}
\toprule
\multicolumn{1}{>{\raggedright\arraybackslash}p{1.2cm}}{} & \multicolumn{4}{c}{$c_{W}$ $\mathcal{O}(\Lambda^{-2})$}   
& \multicolumn{4}{c}{$c_{W}$ $\mathcal{O}(\Lambda^{-4})$} 
& \multicolumn{4}{c}{$c_{\tilde{W}}$ $\mathcal{O}(\Lambda^{-4})$} \\
\cmidrule(lr){2-5} \cmidrule(lr){6-9} \cmidrule(lr){10-13}
\multicolumn{1}{>{\raggedright\arraybackslash}p{1.2cm}}{} & \multicolumn{2}{c}{LO}                             
& \multicolumn{2}{c}{NLO QCD}                               
& \multicolumn{2}{c}{LO}                               
& \multicolumn{2}{c}{NLO QCD}    
& \multicolumn{2}{c}{LO}                               
& \multicolumn{2}{c}{NLO QCD}                            
\\
\midrule
Incl.                  
& \multicolumn{2}{c}{[-130, 389]}  
& \multicolumn{2}{c}{[-59.60, 20.00]} 
& \multicolumn{2}{c}{[-1.33, 1.33]}  
& \multicolumn{2}{c}{[-2.15, 2.23]} 
& \multicolumn{2}{c}{[-1.78, 1.78]}  
& \multicolumn{2}{c}{[-2.86, 2.86]} 
\\
$p_{\rm T, \ell_{\rm lead}}$        
& \multicolumn{2}{c}{[-3.01, 4.03]}  
& \multicolumn{2}{c}{[-3.64, 0.94]}   
& \multicolumn{2}{c}{[-0.32, 0.30]}  
& \multicolumn{2}{c}{[-0.50, 0.51]} 
& \multicolumn{2}{c}{[-0.37, 0.38]}  
& \multicolumn{2}{c}{[-0.62, 0.63]}   
\\
$p_{\rm T, \Pe^+\mu^-}$        
& \multicolumn{2}{c}{[-2.21, 1.83]}  
& \multicolumn{2}{c}{[-3.77, 0.19]} 
& \multicolumn{2}{c}{[-0.31, 0.29]}  
& \multicolumn{2}{c}{[-0.70, 0.67]} 
& \multicolumn{2}{c}{[-0.31, 0.31]}  
& \multicolumn{2}{c}{[-0.73, 0.73]}  
\\
$M_{\Pe^+\mu^-}$  
& \multicolumn{2}{c}{[-18.12, 17.02]}  
& \multicolumn{2}{c}{[-8.22, 1.90]} 
& \multicolumn{2}{c}{[-0.44, 0.43]}  
& \multicolumn{2}{c}{[-0.94, 0.95]} 
& \multicolumn{2}{c}{[-0.49, 0.49]}  
& \multicolumn{2}{c}{[-1.13, 1.13]}   
\\
\bottomrule
\end{tabular}
\caption{Same as~\Cref{tab:wz_bounds_all} but for $\PW^{+}\PW^{-}$.}
\label{tab:ww_bounds_all}
\end{table}
$\,$\\

\bibliography{bibliography}
\bibliographystyle{JHEP}
\end{document}